\documentclass[aps,prd,showpacs,notitlepage,nofootinbib,preprintnumbers,amsmath,amssymb]{revtex4-1}

\usepackage{graphics,graphicx}
\usepackage{dcolumn}
\usepackage{bm}
\usepackage{epsfig}
\usepackage[usenames]{color}
\usepackage{hyperref} 
\usepackage{mathbbol}
\usepackage{epstopdf}
\usepackage{simplewick}

\def\eq#1{{Eq.~(\ref{#1})}}
\def\fig#1{{Fig.~\ref{#1}}}
\newcommand{\ben}{\begin{eqnarray*}}
\newcommand{\een}{\end{eqnarray*}}

\newcommand{\as}{\alpha_s}

\newcommand{\dhd}{{\textstyle d}
\lower.03ex\hbox{\kern-0.38em$^{\scriptstyle-}$}\kern-0.05em{}}
\newcommand{\dbar}{{\textstyle \delta}
\lower.03ex\hbox{\kern-0.38em$^{\scriptstyle-}$}\kern-0.05em{}}

\newcommand{\sh}[1]{#1\hskip-7pt \diagup}

\begin{document}

\title{Classical Gluon Production Amplitude for Nucleus--Nucleus
  Collisions: \\ First Saturation Correction in the Projectile}

\author{Giovanni~A.~Chirilli,\footnote{chirilli.1@osu.edu}
  Yuri~V.~Kovchegov\footnote{kovchegov.1@osu.edu},
  Douglas~E.~Wertepny\footnote{wertepny.1@osu.edu}}

\affiliation{Department of Physics, The Ohio State University,
  Columbus, OH 43210, USA}

\begin{abstract}
  We calculate the classical single-gluon production amplitude in
  nucleus--nucleus collisions including the first saturation
  correction in one of the nuclei (the projectile) while keeping
  multiple-rescattering (saturation) corrections to all orders in the
  other nucleus (the target). In our approximation only two nucleons
  interact in the projectile nucleus: the single-gluon production
  amplitude we calculate is order-$g^3$ and is leading-order in the
  atomic number of the projectile, while resumming all order-one
  saturation corrections in the target nucleus. Our result is the
  first step towards obtaining an analytic expression for the first
  projectile saturation correction to the gluon production cross
  section in nucleus--nucleus collisions.
\end{abstract}

\pacs{12.38.-t, 12.38.Bx, 12.38.Cy}

\maketitle



\section{Introduction}

Complete understanding of heavy ion collisions is impossible without a
clear picture of the initial stages of the collision, i.e., the
processes leading to the creation of quarks and gluons which later on
thermalize forming quark-gluon plasma (QGP). The distribution of
quarks and gluons produced in these early-time processes, known as the
initial condition for the QGP formation, is the fundamental building
block of heavy ion theory. Questions concerning thermalization of the
quark-gluon medium and the determination of the initial conditions for
the hydrodynamic evolution of the QGP can not be answered in a fully
satisfactory manner without the qualitative and quantitative knowledge
of the production mechanism for the initial-state quarks and gluons in
a nuclear collision.

In the framework of saturation physics (see the reviews
\cite{Gribov:1984tu,Balitsky:2001gj,Iancu:2003xm,Jalilian-Marian:2005jf,Weigert:2005us,Gelis:2010nm,Albacete:2014fwa}
and the book \cite{KovchegovLevin}) the leading-order contribution to
gluon production is given by the classical gluon fields
\cite{Kovner:1995ts,Kovner:1995ja,Kovchegov:1996ty,Kovchegov:1997ke,Jalilian-Marian:1997xn}
of the McLerran--Venugopalan (MV) model
\cite{McLerran:1994vd,McLerran:1993ka,McLerran:1993ni}. Classical
gluon fields in heavy ion collisions resum powers of $\as^2 A_1^{1/3}$
and $\as^2 A_2^{1/3}$ \cite{Kovchegov:1997pc}, where $\as$ is the
strong coupling constant, while $A_1$ and $A_2$ are the atomic numbers
of the two nuclei (henceforth referred to as the projectile and the
target). Since the saturation scales squared of the two nuclei are
proportional to these parameters, $Q_{s1}^2 \sim \as^2 A_1^{1/3}$ and
$Q_{s2}^2 \sim \as^2 A_2^{1/3}$, we can write down the quasi-classical
single-gluon production cross section as
\begin{align}
  \label{eq:xsect1}
  \frac{d \sigma}{d^2 k \, d^2 B \, d^2 b} = \frac{1}{\as} \, f \left(
    \frac{Q_{s1}^2 ({\vec B}_\perp - {\vec b}_\perp)}{k_T^2} ,
    \frac{Q_{s2}^2 ({\vec b}_\perp)}{k_T^2} \right),
\end{align}
where ${\vec B}_\perp$ is the impact parameter between the two nuclei,
${\vec b}_\perp$ is the transverse position of the produced gluon with
respect to the center of the target nucleus, while ${\vec k}_\perp$ is
the transverse momentum of the produced gluon with $k_T = |{\vec
  k}_\perp|$. The expansion in the powers of $\as^2 A_1^{1/3}$ and
$\as^2 A_2^{1/3}$ corresponds to expansion in the powers of
$Q_{s1}^2/k_T^2$ and $Q_{s2}^2/k_T^2$,
\begin{align}
  \label{eq:xsect_exp}
  f \left( \frac{Q_{s1}^2}{k_T^2} , \frac{Q_{s2}^2}{k_T^2} \right) =
  \sum_{n,m =1}^\infty c_{n,m} \, \left( \frac{Q_{s1}^2}{k_T^2}
  \right)^n \, \left( \frac{Q_{s2}^2}{k_T^2} \right)^m.
\end{align}
Note that due to projectile-target symmetry (resulting in $Q_{s1}
\leftrightarrow Q_{s2}$ symmetry) we have $c_{n,m} = c_{m,n}$. (The
$k_T$-dependence on the right-hand side of \eq{eq:xsect1} also enters
through powers of $\ln (k_T/\Lambda)$ where $\Lambda$ is the infrared
(IR) cutoff of each nucleon: the powers of $\ln (k_T/\Lambda)$ are not
shown explicitly above. The coefficients $c_{n,m}$ from
\eq{eq:xsect_exp} are, in fact, polynomials in $\ln (k_T/\Lambda)$.)

At the moment we do not have an analytic expression for the function
$f(Q_{s1}^2/k_T^2, Q_{s2}^2/k_T^2)$ in \eq{eq:xsect1}. This function
was extensively studied numerically in
\cite{Krasnitz:1998ns,Krasnitz:1999wc,Krasnitz:2002mn,Krasnitz:2003jw,Krasnitz:2003nv,Lappi:2003bi,Blaizot:2010kh}. Still
it appears desirable to attain a better handle on the analytic form of
this function. Apart from the general advantage of having an analytic
solution, knowing the function $f(Q_{s1}^2/k_T^2, Q_{s2}^2/k_T^2)$
should greatly facilitate the inclusion of small-$x$ evolution
corrections
\cite{Kuraev:1977fs,Balitsky:1978ic,Balitsky:1996ub,Balitsky:1998ya,Kovchegov:1999yj,Kovchegov:1999ua,Jalilian-Marian:1997dw,Jalilian-Marian:1997gr,Iancu:2001ad,Iancu:2000hn}
along with the running-coupling corrections
\cite{Gardi:2006rp,Balitsky:2006wa,Kovchegov:2006vj,Albacete:2007yr,Horowitz:2010yg}
into the gluon production cross section.  These corrections are
essential for realistic phenomenological applications.

Let us briefly summarize what is known about the function
$f(Q_{s1}^2/k_T^2, Q_{s2}^2/k_T^2)$. The leading-order result (the
coefficient $c_{1,1}$ in \eq{eq:xsect_exp}) was obtained in
\cite{Kovner:1995ts,Kovner:1995ja,Kovchegov:1997ke}, reproducing the
earlier results of \cite{Fadin:1975cb,Kuraev:1977fs}. The case of
proton--nucleus ($p+A$) collisions, defined as the leading-order in
$Q_{s1}^2$ term in the expansion on the right-hand side of
\eq{eq:xsect_exp}, was solved in \cite{Kovchegov:1998bi} (see also
\cite{Kopeliovich:1998nw,Dumitru:2001ux}), yielding the coefficients
$c_{1,n}$ (and, due to target-projectile symmetry, $c_{n, 1}$) for any
positive integer $n$. This is all we presently know analytically about
the coefficients $c_{n,m}$.

An ansatz for the full solution of the classical gluon production
problem was proposed by one of the authors in
\cite{Kovchegov:2000hz}. A variational approach to the problem was
attempted in \cite{Blaizot:2008yb}.  While consistent with our
knowledge of coefficients $c_{1,n}$, neither of these results can be
verified further due to our lack of knowledge of the coefficients
$c_{n,m}$ for $n,m \ge 2$. In phenomenological applications one often
employs the $k_T$-factorization formula involving unintegrated gluon
distributions \cite{Kharzeev:2001gp,Kharzeev:2000ph,ALbacete:2010ad}:
while the gluon production cross section in $p+A$ collisions (the
lowest-order in $Q_{s1}^2$ terms in \eq{eq:xsect_exp}) does lead to
the $k_T$-factorization formula
\cite{Kovchegov:2001sc,Kharzeev:2003wz}, it is not clear whether
$k_T$-factorization holds beyond the $p+A$ approximation, again due to
our lack of knowledge of $c_{n,m}$ for $n,m \ge 2$. Moreover,
numerical simulations of the classical gluon production
\cite{Krasnitz:1998ns,Krasnitz:1999wc,Krasnitz:2002mn,Krasnitz:2003jw,Krasnitz:2003nv,Lappi:2003bi,Blaizot:2010kh}
for nucleus--nucleus ($A+A$) collisions appear to rule out the
$k_T$-factorization ansatz, suggesting that this factorized result is
only valid for $p+A$.
 
The goal of the current project is to determine the coefficients
$c_{2,n}$ (for $n \ge 2$). As one can see from \eq{eq:xsect_exp}, to
obtain $c_{2,n}$ coefficients one has to expand the gluon production
cross section to the second order in $Q_{s1}^2$ keeping all orders of
$Q_{s2}^2$. This means that one has to allow two nucleons in the
projectile nucleus to participate in the interaction, while allowing
all nucleons in the target nucleus to interact. Formally one can think
of this as working in the regime where $\as^2 \, A_2^{1/3} \sim 1$
while $\as^2 \, A_1^{1/3} \ll 1$: the goal is to calculate the ${\cal
  O} \left[ (\as^2 \, A_1^{1/3})^2 /\as \right] = {\cal O} \left[
  \as^3 \, A_1^{2/3} \right] $ correction to gluon production. Note
that one still has $A_1 \gg 1$ such that the ${\cal O} (\as^3)$
contribution to the cross section, where only one of the projectile
nucleons interacts, is small due to it being suppressed by a power of
$A_1^{1/3}$.

As we will detail below, having two nucleons interact in the
projectile nucleus results in the gluon production cross section
consisting of two contributions (see the top two diagrams in
\fig{xsect} below): (i) each of the two nucleons interacts both in the
amplitude and in the complex conjugate amplitude (or one nucleon
interacts only in the amplitude and another nucleon interacts only in
the complex conjugate amplitude), or (ii) one nucleon interacts both
in the amplitude and in the complex conjugate amplitude while the
other nucleon interacts only in the (complex conjugate) amplitude. In
the case (i) the scattering amplitude is ${\cal O} (g^3)$, such that
the contribution to the cross section is $\sim |g^3|^2 \sim \as^3$. In
the case (ii) the amplitude is ${\cal O} (g^5)$, while the complex
conjugate amplitude is ${\cal O} (g)$, such that the cross section
contribution is $\sim g^5 \, g \sim \as^3$. (As we will argue below,
the case where the amplitude is ${\cal O} (g^4)$ and the complex
conjugate amplitude is ${\cal O} (g^2)$ is included by using retarded
Green functions in the ${\cal O} (g^3)$ and ${\cal O} (g^5)$
amplitudes in this quasi-classical calculation.)

In the present paper we calculate the scattering amplitude in case
(i). While its square would give a contribution to the desired gluon
production cross section at order-$Q_{s1}^4$, the complete expression
for the cross section can only be obtained if one includes the
contribution of case (ii) as well. Calculation of the contribution
(ii) would involve the ${\cal O} (g^5)$ amplitude, which appears to be
significantly more involved and is left for future work. Our
calculation is performed in the light-cone gauge of the projectile,
with the final results given in Eqs.~\eqref{eq:ABCsum_coord} and
\eqref{eq:Dall_coord}.

The paper is structured as follows. In Sec.~\ref{sec:setup} we set up
the problem by first reproducing the lowest-order ($p+A$) gluon
production calculation and then by outlining the main ingredients
needed to complete the calculation of the coefficients $c_{2,n}$ in
\eq{eq:xsect_exp}. The elements of the calculation, along with the
main results, are presented in Sec.~\ref{sec:calc}. We conclude in
Sec.~\ref{sec:outlook}.


\section{The setup}
\label{sec:setup}

Consider the high-energy scattering of a projectile nucleus on a
target nucleus. Defining the light-cone variables as $v^\pm = (v^0 \pm
v^3)/\sqrt{2}$ with the $x^3$-axis being the collision axis, we choose
the projectile to be moving along the $x^+$ light-cone direction, and
the target moving along the $x^-$ direction. All of our calculation
will be performed in $A^+=0$ light-cone gauge. Transverse plane
vectors are denoted as ${\vec v}_\perp = (v^1, v^2)$, such that the
full 4-vector is $v^\mu = (v^+, v^-, {\vec v}_\perp)$ in the
$(+,-,\perp)$ notation.


\subsection{Classical gluon production in the $p+A$ approximation}

\begin{figure}[ht]
\begin{center}
\includegraphics[width=0.6 \textwidth]{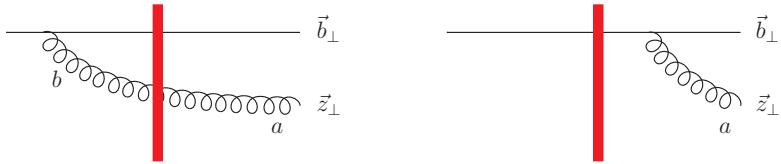} 
\caption{Diagrams contributing to the single-gluon production
  amplitude in $p+A$ collisions.}
\label{1gluon}
\end{center}
\end{figure}

To introduce our formalism, let us begin by outlining the calculation
of the single gluon production cross section in $p+A$ collisions in
the quasi-classical approximation. Working in the approximation where
$\as^2 \, A_2^{1/3} \sim 1$ and $\as^2 \, A_1^{1/3} \ll 1$, in this
Section we calculate the ${\cal O} \left[ \as^2 \, A_1^{1/3} /\as
\right] = {\cal O} \left[ \as \, A_1^{1/3} \right]$
contribution. Since only one power of $A_1^{1/3}$ is required we only
need to include diagrams where one nucleon from the projectile
interacts with the target. The diagrams contributing to the gluon
production cross section in $p+A$ are shown in \fig{1gluon}. The
target nucleus moving along the $x^-$-axis generates the shock wave,
shown in this paper as a vertical band. Diagrammatically this vertical
band represents multiple interactions with the field of the shock
wave, which happen over a very short time interval around $x^+
=0$. Since only one nucleon in the projectile is involved in the
interaction, we model it with a single quark, shown in \fig{1gluon},
by a horizontal solid straight line. The spectator quarks are not
shown explicitly in \fig{1gluon} (along with the spectator nucleons if
the projectile is a nucleus): below, for other processes, we also do
not show the spectators explicitly. The produced gluon can be emitted
either before the interaction with the shock wave (left diagram in
\fig{1gluon}) or after the interaction (right diagram in
\fig{1gluon}). Emission during the passage of the projectile through
the shock wave is suppressed by a power of center-of-mass
energy. Gluon emission from within the shock wave is also suppressed
in the $A^+=0$ light-cone gauge we are using.

Normalizing the incoming quark in the projectile as a Wilson line, we
write the following expression for the scattering amplitude resulting
from the diagrams in \fig{1gluon}:
\begin{align}
  \label{eq:1gluon}
  A ({\vec z}_\perp , {\vec b}_\perp) = 2 g \, \int \frac{d^2 k}{(2
    \pi)^2} \, e^{i {\vec k}_\perp \cdot ({\vec z}_\perp - {\vec
      b}_\perp)} \, \frac{{\vec \epsilon}_\perp^{\, \lambda *} \cdot
    {\vec k}_\perp}{k_\perp^2} \, \left[ U^{ab}_{{\vec z}_\perp} -
    U^{ab}_{{\vec b}_\perp} \right] \, \left( V_{{\vec b}_\perp} t^b
  \right).
\end{align}
The amplitude in \eq{eq:1gluon} includes a Fourier-transform from
transverse momentum space to transverse coordinate space. It also
involves the adjoint
\begin{align}
  \label{eq:Wadjoint}
  U^{ab}_{{\vec z}_\perp} = \left( \mbox{P} \exp \left\{ i \, g \,
    \int\limits_{-\infty}^\infty d x^+ T^c \, A^{c \, -} (x^+, z^-=0,
    {\vec z}_\perp) \right\} \right)^{ab}
\end{align}
and fundamental 
\begin{align}
  \label{eq:Wfund}
  V_{{\vec b}_\perp} = \mbox{P} \exp \left\{ i \, g \,
    \int\limits_{-\infty}^\infty d x^+ t^a \, A^{a \, -} (x^+, b^-=0,
    {\vec b}_\perp) \right\}
\end{align}
Wilson lines describing respectively the propagation of a high-energy
gluon and quark through the target gluon field $A^\mu$. Here $T^a$ and
$t^a$ are correspondingly the adjoint and fundamental SU($N_c$)
generators.  In arriving at \eq{eq:1gluon} we used the Fierz identity
$U^{ab} \left( V t^b \right) = t^a \, V$ when evaluating the right
diagram in \fig{1gluon} in order to put the contributions of both
diagrams into a similar form.

Performing the Fourier transform in \eq{eq:1gluon} we obtain
\begin{align}
  \label{eq:1gluon_coord}
  A ({\vec z}_\perp , {\vec b}_\perp) = \frac{i \, g}{\pi} \,
  \frac{{\vec \epsilon}_\perp^{\, \lambda *} \cdot ({\vec z}_\perp -
    {\vec b}_\perp)}{|{\vec z}_\perp - {\vec b}_\perp|^2} \, \left[
    U^{ab}_{{\vec z}_\perp} - U^{ab}_{{\vec b}_\perp} \right] \,
  \left( V_{{\vec b}_\perp} t^b \right).
\end{align}

The gluon production cross section is given by (see
e.g. \cite{KovchegovLevin})
\begin{align}
  \label{eq:xsect}
  \frac{d \sigma}{d^2 k_T \, dy} = \frac{1}{2 \, (2 \pi)^3} \, \int
  d^2 z \, d^2 z' \, d^2 b \, e^{-i {\vec k}_\perp \cdot ({\vec
      z}_\perp - {\vec z}^{\, \prime}_\perp)} \, \left\langle A ({\vec
      z}_\perp , {\vec b}_\perp) \, A^* ({\vec z}^{\, \prime}_\perp ,
    {\vec b}_\perp) \right\rangle,
\end{align}
where the summation over colors and polarizations of the final-state
particles along with the averaging over the quantum numbers of the
initial-state particles are implicitly implied. The angle brackets
$\langle \ldots \rangle$ denote averaging in the target nucleus wave
function.

Substituting the amplitude \eq{eq:1gluon_coord} into \eq{eq:xsect}
yields \cite{Kovchegov:1998bi}
\begin{align}
  \label{eq:pA_xsect}
  \frac{d \sigma}{d^2 k_T \, dy} = \frac{\as \, C_F}{4 \, \pi^4} \,
  \int d^2 z \, d^2 z' \, d^2 b \, e^{-i {\vec k}_\perp \cdot ({\vec
      z}_\perp - {\vec z}^{\, \prime}_\perp)} \, \frac{{\vec z}_\perp
    - {\vec b}_\perp}{|{\vec z}_\perp - {\vec b}_\perp|^2} \cdot
  \frac{{\vec z}^{\, \prime}_\perp - {\vec b}_\perp}{|{\vec z}^{\,
      \prime}_\perp - {\vec b}_\perp|^2} \notag \\ \times \, \left[
    S_G ({\vec z}_\perp , {\vec z}^{\, \prime}_\perp) - S_G ({\vec
      b}_\perp , {\vec z}^{\, \prime}_\perp) - S_G ({\vec z}_\perp ,
    {\vec b}_\perp) + 1 \right]
\end{align}
where the gluon dipole $S$-matrix is defined by
\begin{align}
  \label{eq:Gdipole}
  S_G ({\vec x}_\perp , {\vec y}_\perp) = \frac{1}{N_c^2 -1} \,
  \left\langle U^{ab}_{{\vec x}_\perp} \, U^{\dagger \, ba}_{{\vec
        y}_\perp} \right\rangle = \frac{1}{N_c^2 -1} \, \left\langle
    \mbox{Tr} \left[ U_{{\vec x}_\perp} \, U^{\dagger}_{{\vec
          y}_\perp} \right] \right\rangle.
\end{align}
In the quasi-classical MV/Glauber--Mueller (GM) approximation it is
given by \cite{Mueller:1989st}
\begin{align}
  \label{eq:SG}
  S_G ({\vec x}_\perp , {\vec y}_\perp) = \exp \left[ - \frac{1}{4} \,
    ({\vec x}_\perp - {\vec y}_\perp)^2 \, Q_{sG}^2 \left( \frac{{\vec
          x}_\perp + {\vec y}_\perp}{2} \right) \, \ln \frac{1}{|{\vec
        x}_\perp - {\vec y}_\perp| \, \Lambda} \right]
\end{align}
where $Q_{sG}^2 ({\vec b}_\perp) = 4 \pi \as^2 \, T({\vec b}_\perp)$
is the square of the gluon saturation scale with $T({\vec b}_\perp)$
the nuclear profile function and $\Lambda$ the IR cutoff of each
individual nucleon ($\Lambda \sim \Lambda_{QCD}$). \eq{eq:SG} resums
all the multiple rescatterings in the target nucleus: thus, in the MV
model, it resums all-order saturation corrections in the target.

Note a particular convenience of the form of the amplitude given in
\eq{eq:1gluon_coord}, with the fundamental Wilson line $V$ placed to
the left of the fundamental SU($N_c$) generator in both terms there:
when we square the amplitude, the fundamental Wilson line $V$ is
multiplied by its hermitean conjugate $V^\dagger$ giving an
identity. Below when calculating diagrams we will always cast them in
the same form, which would make all fundamental Wilson lines vanish
when we square the amplitude.


\subsection{First Saturation Correction in the Projectile: Diagram
  Types and Calculational Simplifications}

\begin{figure}[ht]
\begin{center}
\includegraphics[width=0.9 \textwidth]{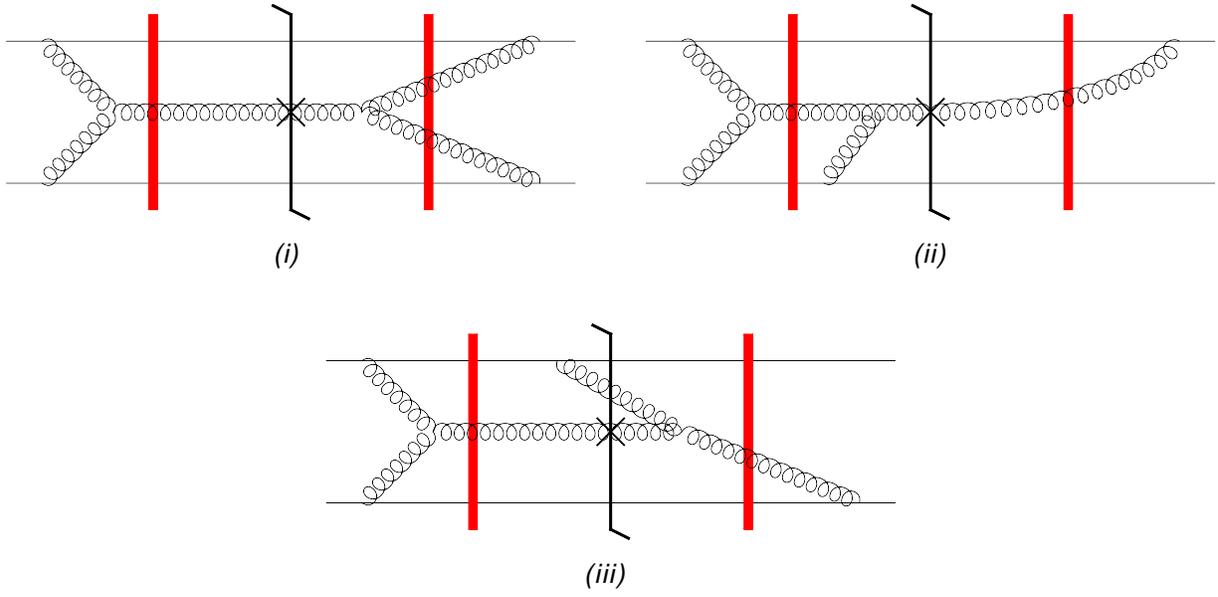} 
\caption{The diagrams representing different types of saturation
  corrections in the projectile nucleus to the gluon production cross
  section.}
\label{xsect}
\end{center}
\end{figure}

The goal of this project is to calculate the first projectile
saturation correction to the cross section in \eq{eq:pA_xsect}. That
is, we need to find the order-$\as^2 \, A_1^{1/3}$ correction to
\eqref{eq:pA_xsect}: this is the leading in $A_1$ part of the
order-$\as^2$ correction. To get the leading-$A_1$ contribution, the
correction must include an interaction with another nucleon in the
projectile nucleus. Hence we see that we need to find order-$\as^2$
correction to \eqref{eq:pA_xsect} involving one other nucleon in the
projectile. The main types of the diagrams we need to calculate are
shown in \fig{xsect}. There by two horizontal solid straight lines we
show two quarks from two different nucleons in the projectile nucleus,
again suppressing the spectators. The diagrams in \fig{xsect}
represent the amplitude squared which contributes to the cross
section: the solid vertical line denotes the final-state cut. The
cross labels the measured gluon.

The three types of projectile saturation corrections to gluon
production shown in \fig{xsect} are (i) the square of the order-$g^3$
amplitude; (ii) the interference between the order-$g^5$ amplitude and
the leading-order (order-$g$) amplitude from Eq.
\eqref{eq:1gluon_coord}; and (iii) the interference between the
order-$g^4$ amplitude and the order-$g^2$ amplitude. (In this power
counting we are assuming that the interaction with the shock wave is
order-$1$, that is $\as^2 \, A_2^{1/3} \sim 1$, and are counting only
powers of $g$ arising from the vertices shown explicitly in
\fig{xsect}.) The order-$\as^3$ diagrams in which one of the nucleons
is a spectator and does not participate in the interactions are
suppressed by $A_1^{1/3}$ and are neglected in our analysis. Note that
diagram (iii) has two gluons in the final state: since we are
calculating the single-inclusive production cross section, having more
than the measured gluon in the final state is allowed.

Below we calculate the order-$g^3$ amplitude that enters diagram (i)
in \fig{xsect}, thus constructing the order-$\as$ correction to the
amplitude in Eq. \eqref{eq:1gluon_coord} which is enhanced by the
leading power of $A_1^{1/3}$. The order-$g^5$ amplitude from diagram
(ii) is left for future work.

The situation with the diagram (iii) is more subtle. First let us note
that, since we are working in the classical MV model, the diagrams we
consider for gluon production also correspond to the diagrams
contributing to the classical gluon field with the two colliding
nuclei providing the source current
\cite{McLerran:1994vd,McLerran:1993ka,McLerran:1993ni,Kovner:1995ts,Kovner:1995ja,Kovchegov:1997ke}. Hence
one can think of these diagrams as graphically representing the
solution of the Yang--Mills equations
\cite{Kovchegov:1997pc,Kovchegov:1997ke}. In such case each Feynman
propagator would be replaced by the retarded Green function. While
such a replacement is relatively straightforward for gluon
propagators, it is less clear what this prescription means for quark
propagators in the time-ordered picture we employ here: for instance,
the solution of Yang--Mills equations can only have the produced gluon
emitted by the projectile quark {\sl after} the interaction with the
shock wave (see e.g. \cite{Kovchegov:1997ke}), whereas a naive
application of perturbation theory allows one to draw a number of {\sl
  a priori} non-zero graphs with the produced gluon emitted before the
shock wave interaction (see \fig{Bgraphs} below).

\begin{figure}[hb]
\begin{center}
\includegraphics[width=0.9 \textwidth]{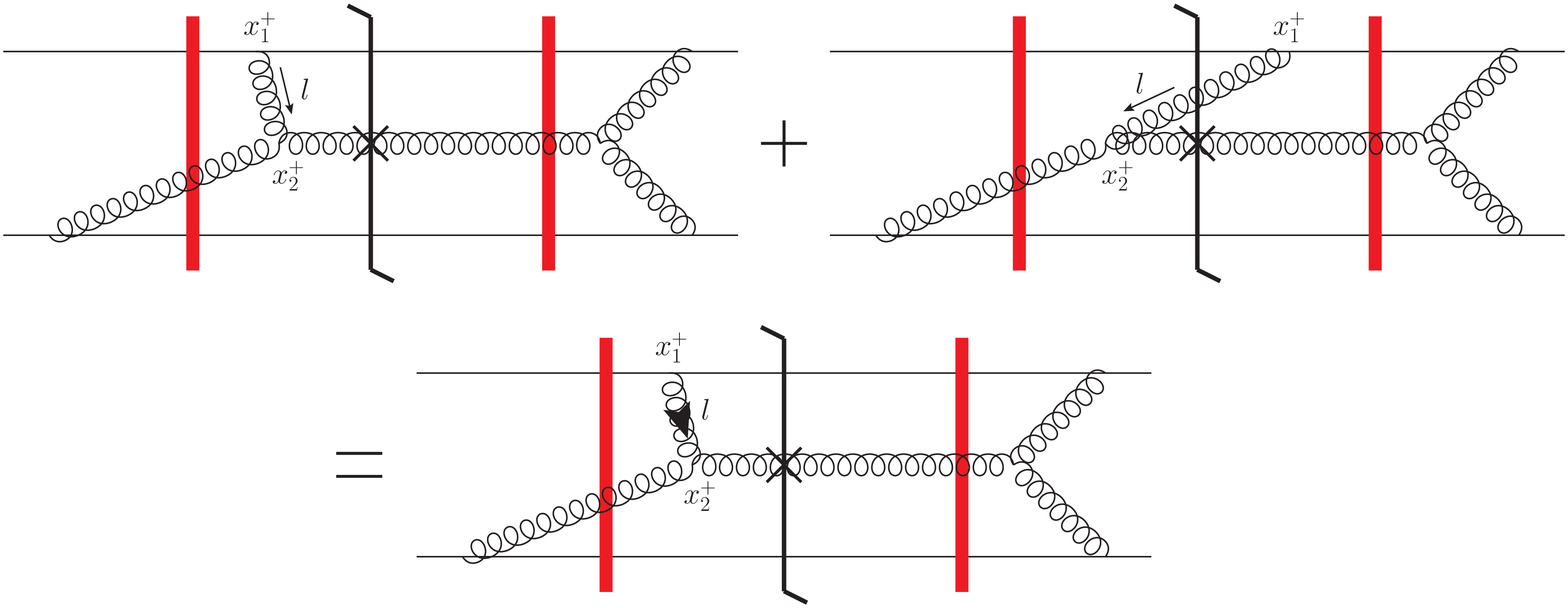} 
\caption{An example of diagrams which add up to convert a gluon
  Feynman propagator into a retarded Green function. The bold arrow on
  the gluon line in the last diagram denotes a retarded gluon Green
  function, with the arrow's direction pointing in the direction of
  light-cone time flow, that is, indicating that $x_2^+ > x_1^+$.}
\label{retardation}
\end{center}
\end{figure}

To clarify this issue we need to perform a detailed diagrammatic
analysis. First let us review how the gluon Feynman propagators become
retarded Green functions. As an example consider the top two diagrams
shown in \fig{retardation} where regular Feynman propagators are
implied for uncut lines. The difference between the top two diagrams
in \fig{retardation} is due to one of the vertices involving the gluon
carrying momentum $l$ being moved across the cut in the right
diagram. The rest of the diagram is the same in both
cases. Concentrating on the propagator of the gluon carrying momentum
$l$ and suppressing the rest of the diagrams' contribution we see that
adding the two graphs gives \cite{Balitsky:2004rr}
\begin{align}
  \label{eq:retarded}
  \frac{- i \, D_{\mu\nu} (l)}{l^2 + i \, \epsilon} + 2 \pi \, \theta
  (-l^+) \, \delta (l^2) \, D_{\mu\nu} (l) = \frac{- i \, D_{\mu\nu}
    (l)}{l^2 + i \, \epsilon \, l^+},
\end{align}
where 
\begin{align}
  \label{eq:Dmunu}
  D_{\mu\nu} (l) = g_{\mu\nu} - \frac{\eta_\mu \, l_\nu + \eta_\nu \,
    l_\mu}{\eta \cdot l} = - \sum_{\lambda = \pm 1}
  \epsilon_\mu^\lambda (l) \, \epsilon_\nu^{\lambda *} (l) - \eta^\mu
  \, \eta^\nu \, \frac{l^2}{(\eta \cdot l)^2}
\end{align}
is the numerator of the light-cone gauge gluon propagator with
$\eta^\mu = (0,1,0_\perp)$ in the $(+,-,\perp)$ notation, such that
$A^+ = \eta \cdot A =0$ is the gauge condition. For an on-shell ($l^2
=0$) gluon, the numerator of the gluon propagator in \eq{eq:Dmunu} can
be written as a sum over physical (transverse) gluon polarizations
with the polarization 4-vector
\begin{align}
  \label{eq:pol}
  \epsilon^\mu (l) = \left( 0, \frac{{\vec \epsilon}_\perp^{\,
        \lambda} \cdot {\vec l}_\perp}{l^+}, {\vec \epsilon}_\perp^{\,
      \lambda} \right)
\end{align}
in the light-cone gauge and ${\vec \epsilon}_\perp^{\, \lambda} =
-(1/\sqrt{2}) (\lambda, i)$. (Note that the last term on the right
hand side of \eqref{eq:Dmunu} vanishes if $l^2 =0$.) In arriving at
\eq{eq:retarded} we have also used the fact that the quark-gluon
vertex changes sign when carried across the cut due to complex
conjugation.

From \eq{eq:retarded} we conclude that contributions of top two
diagrams in \fig{retardation} can be found by only calculating the
left diagram with the retarded gluon Green function: this conclusion
is illustrated in the bottom diagram in \fig{retardation}, with the
retarded gluon Green function denoted by a bold arrow on the gluon
line. Note also that the retarded Green function arising in
\eq{eq:retarded} implies that $x_2^+ > x_1^+$ in the notation shown in
\fig{retardation}, such that the gluon is first emitted by the quark
line, and then it enters the triple gluon vertex leading to the
production of the measured gluon. The arrow in the last diagram of
\fig{retardation} indicates this direction of time flow.

So far we have considered an example of just two diagrams which
combine to give us a retarded gluon Green function. However, the
statement that by moving an end of a gluon line across the cut and by
adding that contribution to the original diagram we would obtain a
retarded Green function for that gluon is valid in general, at this
classical level. To demonstrate this explicitly we need to consider
many different types of diagrams: this is done in
Appendix~\ref{A}. Note also that our ability to use a retarded gluon
Green function in calculating the amplitude should not depend on what
goes on in the complex conjugate amplitude: while naively our argument
in \fig{retardation} seems to depend on the absence of final-state
(post-shock wave) interactions in the complex conjugate amplitude, the
argument is in fact true in general, as long as we are working at the
classical level, that is, calculating order-$\as^3$ cross-section
contribution involving two interacting projectile nucleons
(quarks). We refer the reader to Appendix~\ref{A} for details.

The diagram in the top left panel of \fig{retardation} is in the
(i)-class by the classification presented in \fig{xsect}, while the
diagram in the top right panel of \fig{retardation} is in the
(iii)-class. We see that the contribution of the type-(iii) diagram is
included in the type-(i) diagram by using a retarded gluon propagator
in the latter. Again this conclusion is true in general: type-(iii)
diagrams are included in the calculation of type-(i) and type-(ii)
diagrams if we use retarded gluon propagators.

\begin{figure}[ht]
\begin{center}
\includegraphics[width=0.9 \textwidth]{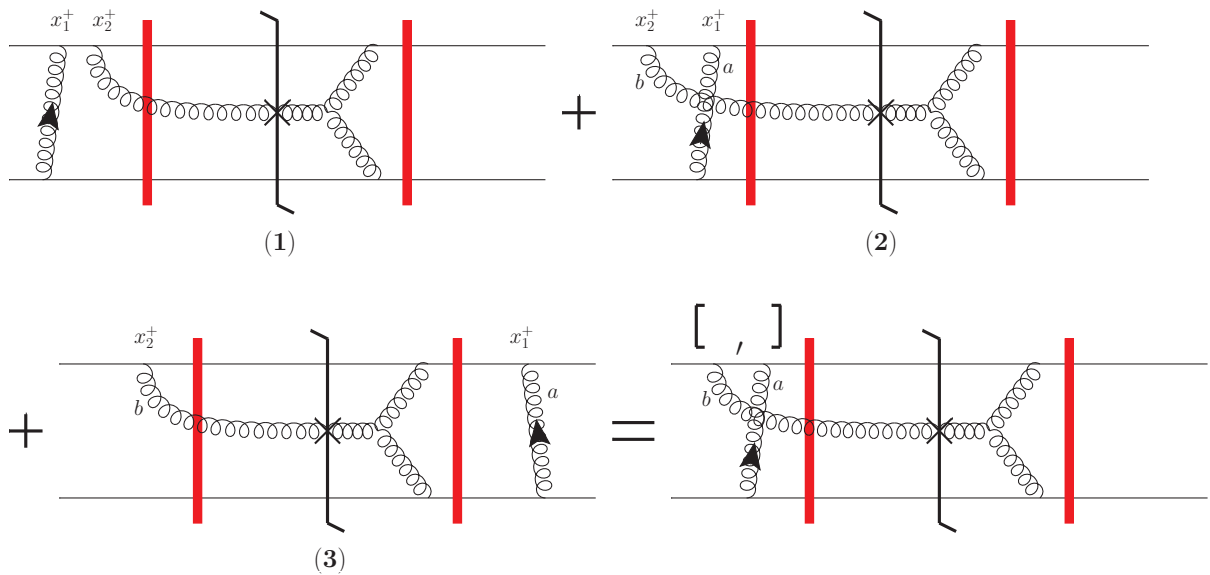} 
\caption{Sample diagrams with a gluon exchange between the projectile quarks.}
\label{commutators}
\end{center}
\end{figure}

Now let us consider diagrams where the projectile quarks exchange a
gluon with each other (the gluon may or may not go through the shock
wave). An example of several such diagrams is given in
\fig{commutators}, where graphs labeled (1), (2) and (3) are different
from each other only by the placement of the gluon exchanged between
the projectile quarks. Once again, the arrow on this gluon line
denotes a retarded gluon Green function, with the direction of the
arrow indicating the direction of the resulting (light-cone) time
ordering. We see that each diagram in \fig{commutators} represents a
sum of two diagrams (akin to \fig{retardation}): each graph of
\fig{commutators} implies a sum of itself (with the Feynman gluon
propagator for the gluon with the arrow on it) and the contribution
where the lower quark-gluon vertex of the same gluon is carried across
the cut (to the region before the shock wave interaction in the
complex conjugate amplitude), resulting in the retarded Green function
for the gluon marked by an arrow. Note that the time-ordering of the
retarded Green function for the gluon in the graph of
\fig{commutators} again points towards the produced gluon, similar to
the case illustrated in \fig{retardation}.

\begin{figure}[ht]
\begin{center}
\includegraphics[width=0.9 \textwidth]{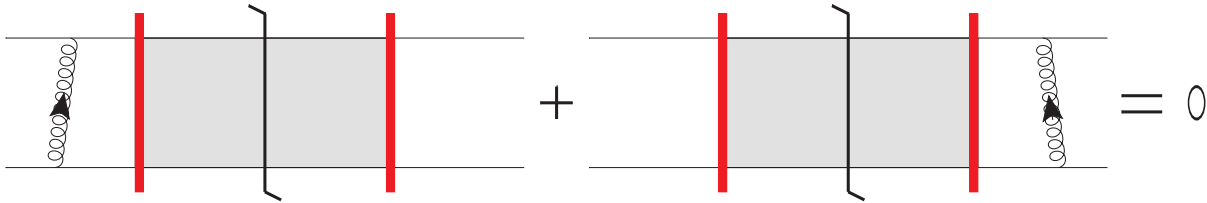} 
\caption{An example of diagrams cancellation resulting from moving a
  retarded gluon propagator (denoted by the arrow indicating the
  direction of time-ordering) across the final-state cut. Any possible
  interaction may happen at $x^+ >0$ both in the amplitude and in the
  complex conjugate amplitude, as indicated by a shaded region.}
\label{cancellation}
\end{center}
\end{figure}

To analyze the diagrams in \fig{commutators} we need the following
additional observation: moving a retarded gluon Green function across
the cut flips the sign of the contribution. This is illustrated by a
simple example in \fig{cancellation} (see Appendix~\ref{A} for more
details), where the cancellation is valid independent of the
interactions which may happen after the shock wave interaction (i.e.,
for $x^+ >0$ shown by the shaded region in \fig{cancellation}) on both
sides of the cut.

Employing the cancellation of \fig{commutators}, we see that diagram
(1) in \fig{commutators} is canceled by the contribution to diagram
(3) coming from the $x_2^+ > x_1^+$ region,
\begin{align}
  \label{eq:cancel1}
  (1) + (3)_{x_2^+ > x_1^+} = 0.
\end{align}
Due to the difference in color factors, such cancellation does not
happen between the diagram (2) in \fig{commutators} and the diagram
(3) with $x_2^+ < x_1^+$: instead, if we write the contribution of
diagram (2) factoring out the color factor as $(2) = t^a \, t^b \, M$,
we get
\begin{align}
  \label{eq:cancel2}
  (2) + (3)_{x_2^+ > x_1^+} = [t^a, t^b] \, M = (2) \ \mbox{with} \
  [t^a, t^b].
\end{align}
This conclusion is illustrated in the lower right diagram of
\fig{commutators}: the sum of diagrams (1), (2) and (3) is simply
diagram (2) with the commutator instead of the product of the
fundamental generators in the amplitude. Once again, while we have
illustrated the point by a simple example, it is valid in all cases:
contributions of diagrams with the single or double gluon exchange
between the two valence quarks can all be found by simply calculating
a small subset of those graphs replacing products of $t^a$'s in them
by commutators. The details of how this happens and which diagrams
survive in the end are given in Appendix~\ref{A}. At order-$g^3$ the
surviving diagrams are $B_2$, $B_9$ and $B_{11}$ from \fig{Bgraphs}
below (indeed diagram $B_2$ is the amplitude to the left of the cut in
the graph (2) of \fig{commutators}), which have to be calculated with
the commutators instead of the products of fundamental generators. As
we will see below, $B_2 =0$ (even with the commutator), such that only
produced gluon emissions by projectile quarks after the shock wave
interaction, that is for $x^+ >0$, survive. This observation completes
the analogy between the Feynman diagrams in the shock wave formalism
and the diagrams representing the classical gluon fields in the MV
model.

We would like to stress that to establish this diagrams versus
classical fields analogy, and simplify our calculations in the
process, we had to utilize the fact that the amplitude we are
calculating would have to be squared to obtain the cross
section. Hence we have absorbed some contributions from the complex
conjugate amplitude into our amplitude by employing the retarded Green
functions and, for some graphs, commutators. Therefore,
strictly-speaking, below we will not be calculating a standard
Feynman-diagram amplitude, but a somewhat modified amplitude with
retarded ``propagators'' and commutators: the square of this amplitude
would still give the gluon production cross section.

Let us summarize the main conclusions of this Section. First of all,
to find the first saturation correction in the projectile, we only
need to calculate ${\cal O} (g^3)$ and ${\cal O} (g^5)$ gluon
production amplitudes with the retarded gluon Green functions. The
time-ordering of the retarded ``propagators'' should be such that the
gluon would be emitted first, and then would participate in an
interaction ultimately leading to the production of the tagged
gluon. Secondly, the contributions of the diagrams with the gluon
exchanges between the projectile nucleons (quarks) can be more
economically constructed by calculating a subset of those graph with
the commutators of fundamental color matrices.


\section{Gluon production amplitude at order-$g^3$}
\label{sec:calc}


\subsection{Graphs A, B, and C}

\begin{figure}[ht]
\begin{center}
\includegraphics[width=0.9 \textwidth]{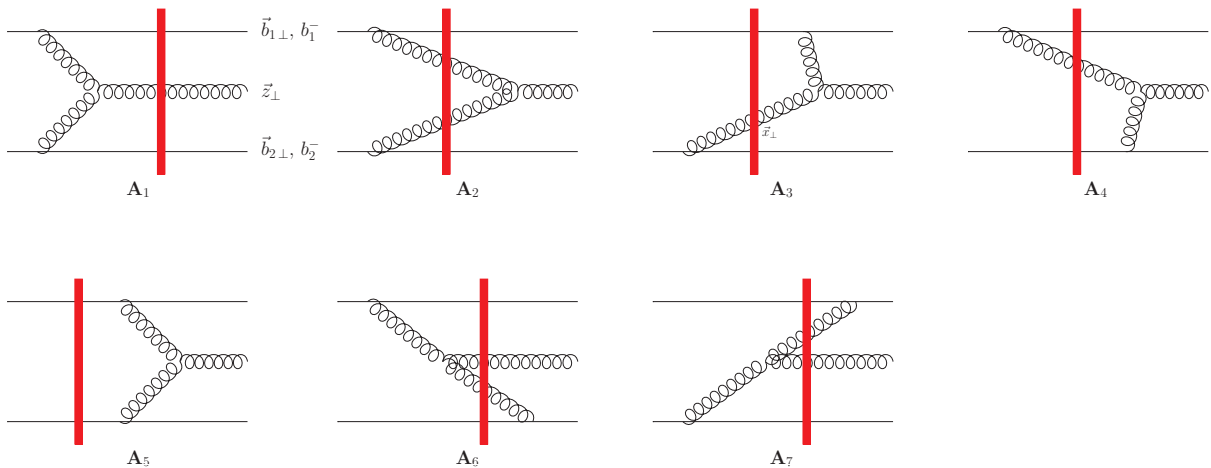} 
\caption{The diagrams containing a triple-gluon vertex. Horizontal
  solid lines depict two Wilson lines representing two quarks coming
  from two different nucleons in the projectile. All gluon propagators
  are retarded, with the time flowing in the direction of the produced
  gluon.}
\label{Agraphs}
\end{center}
\end{figure}

\begin{figure}[ht]
\begin{center}
\includegraphics[width=0.9 \textwidth]{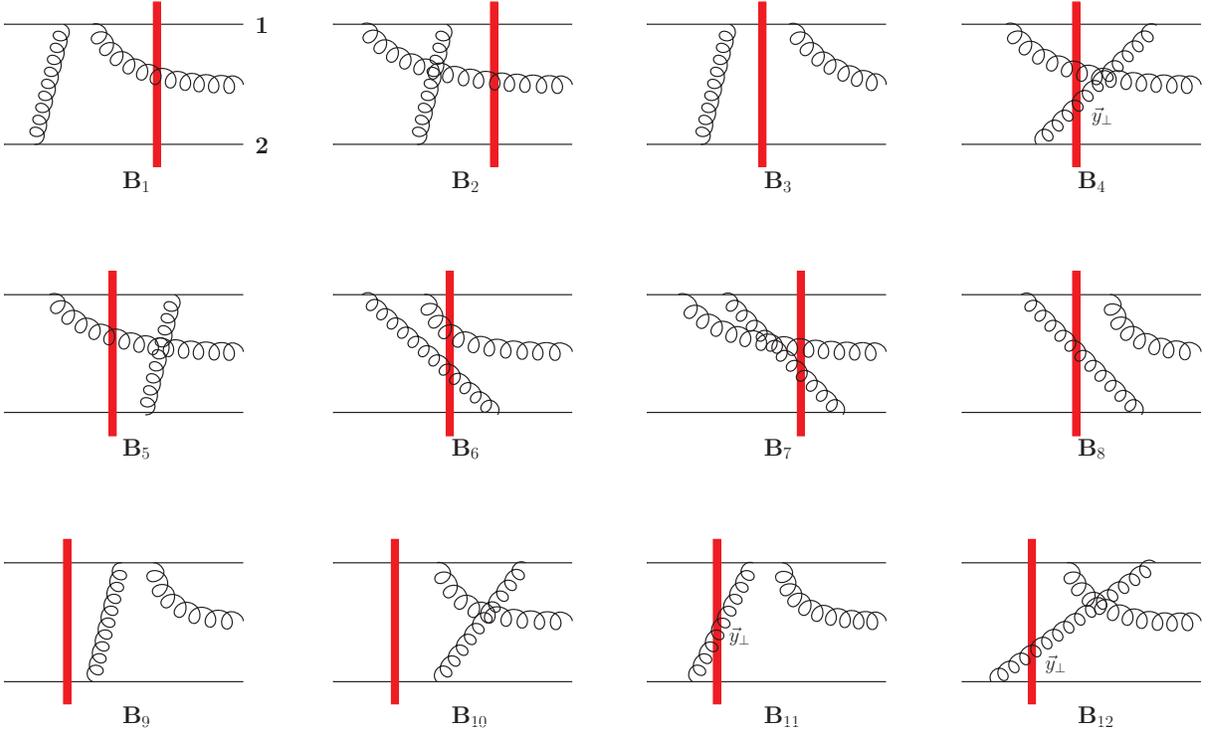} 
\caption{The connected diagrams without a triple-gluon vertex. The
  remaining twelve graphs of this type are obtained by swapping the
  Wilson lines in the diagrams shown, $C_i (b_1, b_2; 1, 2) = B_i
  (b_2, b_1; 2, 1)$. All gluon propagators are retarded, with the time
  flowing in the direction of the produced gluon.}
\label{Bgraphs}
\end{center}
\end{figure}

The diagrams contributing to the gluon production amplitude at ${\cal
  O} (g^3)$ that include interactions with both nucleons are shown in
Figs.~\ref{Agraphs} and \ref{Bgraphs}. The diagrams from \fig{Agraphs}
are labeled $A_i$ with $i = 1, \ldots , 7$ and will be referred to as
the $A$-graphs. Similarly the graphs in \fig{Bgraphs} are labeled
$B_i$ with $i=1, \ldots , 12$ and will be called the $B$-graphs. In
addition to all the $B$-graphs one has to consider the diagrams
similar to those in \fig{Bgraphs} but with the two projectile quarks
interchanged: these will be labeled $C_i$ with $i=1, \ldots , 12$ and
referred to as $C$-graphs.

To calculate the diagrams in Figs.~\ref{Agraphs} and \ref{Bgraphs} we
will follow the rules established in the previous Section. All gluon
propagators are retarded, such that time flows towards the measured
gluon in the diagram. This means that the retarded propagator for the
gluon exchanged between the two quarks in \fig{Bgraphs} is such that
the lower quark-gluon vertex comes earlier than the upper one. The
projectile quark lines are normalizes as Wilson lines (that is, each
line is divided by $2 P^+$ where $P^+$ is the large ``$+$'' momentum
in the line).

Our ultimate goal is to obtain an amplitude in coordinate space, in a
form similar to \eq{eq:1gluon_coord}. The reason for that is the
relative ease with which the correlators of Wilson lines are
calculated in the transverse coordinate space, particularly in the MV
model. For the diagrams in Figs.~\ref{Agraphs} and \ref{Bgraphs} we
would have to perform Fourier transforms into transverse coordinate
space, similar to the transition from \eq{eq:1gluon} to
\eq{eq:1gluon_coord} above. Our standard notation will be to label the
projectile quarks $1$ and $2$, such that their transverse positions
are ${\vec b}_{1 \perp}$ and ${\vec b}_{2 \perp}$. Note also that {\sl
  a priori} the two quarks in the $x^+$-direction moving projectile
have different $x^-$ positions, labeled here as $b_1^-$ and $b_2^-$
(see diagram $A_1$ in \fig{Agraphs}). Since the projectile is also a
large nucleus, the difference $b_1^- - b_2^-$, while suppressed by a
power of energy, is enhanced by a power of the projectile atomic
number $A_1$
\cite{Kovchegov:1996ty,Jalilian-Marian:1997xn,Kovchegov:1997pc}: in
the following we will keep the difference $b_1^- - b_2^-$ non-zero
throughout the calculation while remembering that it is sub-eikonally
small, and will put it to zero at the end of the calculation where
allowed. The difference $b_1^- - b_2^-$ will serve as a regulator in
the longitudinal Fourier transform. This is in accordance to the
standard regularization of singularities in the MV model
\cite{Kovchegov:1996ty,Jalilian-Marian:1997xn,Kovchegov:1997pc}.

To obtain an amplitude dependent on $b_1^-$ and $b_2^-$ we need to
perform a longitudinal Fourier transform integrating over the ``$+$''
component of the momentum exchanged between the projectile quarks in
the diagrams of Figs.~\ref{Agraphs} and \ref{Bgraphs}. Since we are
working in the $A^+=0$ light-cone gauge with the numerator of the
gluon propagator given in \eq{eq:Dmunu}, we have to specify the
regularization of the light-cone poles at $l^+ =0$ in order to perform
this longitudinal Fourier transform. Each way of regulating the
light-cone pole is equivalent to specifying a particular sub-gauge
within the light-cone gauge. In our calculation we will use the
principal value (PV) prescription, where
\begin{align}
  \label{eq:DmunuPV}
  D_{\mu\nu} (l) = g_{\mu\nu} - \left[ \eta_\mu \, l_\nu + \eta_\nu \,
    l_\mu \right] \, \mbox{PV} \left( \frac{1}{l^+} \right)
\end{align}
and
\begin{align}
  \label{eq:polPV}
  \epsilon^\mu (l) = \left[ 0, {\vec \epsilon}_\perp^{\, \lambda}
    \cdot {\vec l}_\perp \,\mbox{PV} \left( \frac{1}{l^+} \right),
    {\vec \epsilon}_\perp^{\, \lambda} \right].
\end{align}
The PV prescription corresponds to the field of the projectile nucleus
moving in the $x^+$ direction satisfying the ${\vec A}_\perp (x^- \to
+ \infty) = - {\vec A}_\perp (x^- \to - \infty)$ sub-gauge condition
(see e.g. \cite{Belitsky:2002sm}).

Other possible sub-gauge choices include requiring that the field of
the projectile nucleus obeys the ${\vec A}_\perp (x^- \to + \infty) =
0$ condition \cite{Belitsky:2002sm,Balitsky:2004rr}. This corresponds
to \cite{Kovchegov:1997pc}
\begin{align}
  \label{eq:Dmunu_me}
  D_{\mu\nu} (l) = g_{\mu\nu} - \frac{\eta_\mu \, l_\nu}{l^+ - i
    \epsilon} - \frac{\eta_\nu \, l_\mu}{l^+ + i \epsilon}
\end{align}
if the momentum $l$ flows from the index $\mu$ to the index $\nu$
along the gluon line. The corresponding polarization 4-vector is
\begin{align}
  \label{eq:pol_me}
  \epsilon^\mu (l) = \left( 0, \frac{{\vec \epsilon}_\perp^{\,
        \lambda} \cdot {\vec l}_\perp}{l^+ + i \epsilon} , {\vec
      \epsilon}_\perp^{\, \lambda} \right).
\end{align}
One may also employ the ${\vec A}_\perp (x^- \to - \infty) = 0$
sub-gauge condition, which results in reversing the sign of $i
\epsilon$ in Eqs.~\eqref{eq:Dmunu_me} and \eqref{eq:pol_me}.

The reason we are going to use the PV sub-gauge is explained in detail
in Appendix~\ref{B}. It turns out that while the ${\vec A}_\perp (x^-
\to + \infty) = 0$ and ${\vec A}_\perp (x^- \to - \infty) = 0$
sub-gauges are very useful for many classical gluon field calculations
\cite{Kovchegov:1996ty,Kovchegov:1997pc,Kovchegov:1998bi,Liou:2012xy,Liou:2013qya},
in our shock-wave target setup the PV prescription is more
economical. Namely, as detailed in Appendix~\ref{B}, using the ${\vec
  A}_\perp (x^- \to + \infty) = 0$ or ${\vec A}_\perp (x^- \to -
\infty) = 0$ gauge choices leads to the need to include new diagrams,
which are not included in Figs.~\ref{Agraphs} and \ref{Bgraphs} (and
are not part of the $C$-graphs). See for instance the first two
diagrams in \fig{shock_graphs} of Appendix~\ref{B}. These diagrams
were considered before in \cite{Kovchegov:1998bi}. In the shock-wave
formalism these diagrams are probably classified as higher-order
corrections to the interactions with the shock wave: namely, if one
takes the leading-order gluon production diagrams from \fig{1gluon},
and considers an order-$\as$ correction to the interactions of either
the quark or the gluon with the shock wave, one would then obtain an
order-$g^3$ contribution to the gluon production amplitude. Indeed
such corrections are not going to be enhanced by an extra $A_1^{1/3}$
(in the cross section): however, shock wave interaction corrections to
the lowest-order graphs from \fig{1gluon} involving a quark from
another projectile nucleon would give an order-$g^3$ contribution
enhanced by a power of $A_1$, that is, they would be comparable to the
$A$, $B$ and $C$ graphs.  In Appendix~\ref{B} we show that while in
the ${\vec A}_\perp (x^- \to + \infty) = 0$ and ${\vec A}_\perp (x^-
\to - \infty) = 0$ sub-gauges such diagrams are important, most of
their contributions are zero in the PV sub-gauge. The remaining shock
wave corrections contributions which are non-zero in the PV sub-gauge
cancel in the amplitude squared, and, hence, can also be neglected in
the calculation. We will therefore proceed by calculating the Fourier
transform for all diagrams using the PV sub-gauge of the light-cone
gauge.

We performed all diagram calculations treating the projectile quark
lines both as the regular high-energy quarks and as the light-cone
Wilson lines from \eq{eq:Wfund} (only with non-zero $b^-$). In the PV
sub-gauge both results are identically equal. However, in the ${\vec
  A}_\perp (x^- \to + \infty) = 0$ and ${\vec A}_\perp (x^- \to -
\infty) = 0$ sub-gauges we found that the Wilson-line approximation
does not give the right answer for the $B$ and $C$ graphs. For more
details we refer the reader to Appendix~\ref{B}.


\subsubsection{Sample Diagram Calculation}

\begin{figure}[ht]
\begin{center}
\includegraphics[width=0.5 \textwidth]{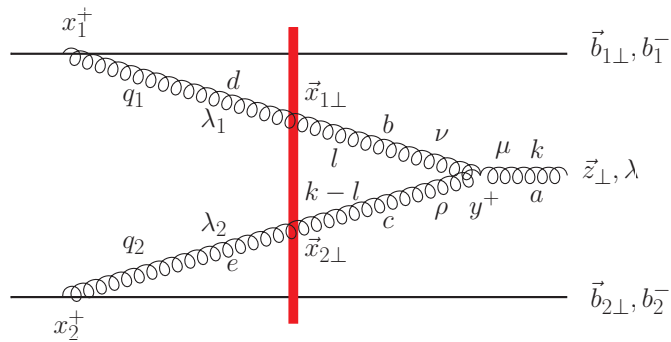} 
\caption{A more detailed rendering of the diagram $A_2$.}
\label{A2}
\end{center}
\end{figure}

Since presenting a detailed calculation of each diagram is rather
tedious, and would make the paper difficult to read, we will first
work out one sample diagram, and then state the answers for the rest
of the graphs. It appears that the diagram whose calculation
illustrates most of the issues relevant to computing $A$, $B$ and $C$
graphs is $A_2$. The diagrams is shown with all the momentum,
coordinate, polarization, Lorentz-index and color labels in \fig{A2}.

Using \fig{A2} and treating the projectile quarks as light-cone Wilson
lines we write the contribution of this diagram as
\begin{align}
  A_2 & = (i g)^2 \, \int\limits_{-\infty}^0 d x_1^+ \, e^{\epsilon \,
    x_1^+} \, \int\limits_{-\infty}^0 d x_2^+ \, e^{\epsilon \, x_2^+}
  \, \int\limits_0^{\infty} d y^+ \, e^{-\epsilon \, y^+} \, \int d^2
  x_1 \, d^2 x_2 \, \int\limits_{-\infty}^{\infty} \frac{d l^+}{2 \pi}
  \, e^{- i \, l^+ \, (b_2^- - b_1^-)}
  \notag \\
  & \times \, \int\limits_{-\infty}^{\infty} \frac{d q_1^-}{2 \pi} \,
  e^{- i \, q_1^- \, (0 - x_1^+)} \ \frac{d q_2^-}{2 \pi} \, e^{- i \,
    q_2^- \, (0 - x_2^+)} \ \frac{d l^-}{2 \pi} \, e^{- i \, l^- \,
    (y^+ -0)} \ \frac{d (k-l)^-}{2 \pi} \, e^{- i \, (k-l)^- \, (y^+
    -0)} \
  e^{i \, k^- \, y^+} \notag \\
  & \times \, \int \frac{d^2 q_1}{(2 \pi)^2} \, e^{i {\vec q}_{1\perp}
    \cdot ({\vec x}_{1\perp} - {\vec b}_{1\perp})} \ \frac{d^2 q_2}{(2
    \pi)^2} \, e^{i {\vec q}_{2\perp} \cdot ({\vec x}_{2\perp} - {\vec
      b}_{2\perp})} \ \frac{d^2 l}{(2 \pi)^2} \, e^{i {\vec l}_{\perp}
    \cdot ({\vec x}_{2\perp} - {\vec x}_{1\perp})} \, \frac{d^2 k}{(2
    \pi)^2} \, e^{i {\vec k}_{\perp} \cdot ({\vec z}_{\perp} - {\vec
      x}_{2\perp})} \notag \\
  & \times \, \frac{-i}{2 \, l^+ \, q_1^- - {q}_{1\perp}^{\, 2} + i \,
    \epsilon \, l^+} \, \frac{-i}{2 \, (k^+ - l^+) \, q_2^- -
    {q}_{2\perp}^{\, 2} + i \, \epsilon \, (k^+ - l^+)} \, \frac{-i}{2
    \, l^+ \, l^- - {l}_\perp^{\, 2} + i \, \epsilon \,
    l^+}  \notag \\
  & \times \, \frac{-i}{2 \, (k^+ - l^+) \, (k-l)^- - ({\vec k}_\perp
    - {\vec l}_\perp)^{\, 2} + i \, \epsilon \, (k^+ - l^+)} \,
  \sum_{\lambda_1} \epsilon_{\lambda_1}^{- \, *} (q_1) \,
  \epsilon_{\lambda_1}^{\nu} (l) \, \sum_{\lambda_2}
  \epsilon_{\lambda_2}^{- \, *} (q_2) \, \epsilon_{\lambda_2}^{\rho}
  (k-l) \notag \\
  & \times \, g \, f^{abc} \, \left[ (2 l - k)_\mu \, g_{\nu\rho} -
    (k+l)_\rho \, g_{\nu\mu} + (2 k -l)_\nu \, g_{\mu\rho} \right] \,
  \epsilon_\lambda^{\mu \, *} (k)  \notag \\
  & \times \, (2 l^+) \, U^{bd}_{{\vec x}_{1\perp}} \, 2 (k^+ - l^+)
  \, U^{ce}_{{\vec x}_{2\perp}} \, \left( V_{{\vec b}_{1\perp}} t^d
  \right)_1 \, \left( V_{{\vec b}_{2\perp}} t^e \right)_2. \label{eq:A2_1}
\end{align}
The integrals over $x_1^+$ and $x_2^+$, coming from the projectile
quarks Wilson lines, along with the integral over $y^+$, are regulated
by the exponentials like $e^{\pm \epsilon \, x^+}$ to make them
convergent at their respective infinities. This regularization is
consistent with the regularization of Feynman propagators. Note that
in \eq{eq:A2_1} we are first Fourier-transforming the diagram into
coordinate space, then integrating over $x_1^+$, $x_2^+$ and $y^+$
with the proper light-cone time ordering and over ${\vec x}_{1\perp}$
and ${\vec x}_{2\perp}$. The shock wave is at $x^+ =0$. Due to the
nature of the Fourier transform, the ``$+$'' components of momentum
are conserved in all interaction vertices, including the interactions
with the shock wave. Conversely, the ``$-$'' momentum component is not
conserved at any of the vertices, such that e.g. $(k-l)^- \neq k^- -
l^-$, just like in light-cone perturbation theory (LCPT)
\cite{Lepage:1980fj}, though our internal lines are not on mass
shell. The outgoing gluon brings in a Fourier factor of $e^{i \, k^-
  \, y^+}$, where $k^- = k_\perp^2/(2 k^+)$ since it is on mass
shell. Note that $k^+ >0$. The gluon interaction with the shock wave
brings in a factor of $(2 k^+) \, U^{ab}_{{\vec x}_{\perp}} \,
g^{\mu\nu}$ for a gluon line with light-cone momentum $k^+$ and
transverse coordinate ${\vec x}_{\perp}$, according to the standard
rules of the eikonal approximation \cite{Balitsky:1996ub}. Similarly,
eikonal interaction of a quark with a shock wave yields $(2 P^+) \,
V_{{\vec b}_{\perp}}$ for a quark line with light-cone momentum $P^+$
and transverse coordinate ${\vec b}_{\perp}$: as we have mentioned
before, the factor of $(2 P^+)$ is removed by our normalization of the
external quark lines. To simplify the calculations, we have replaced
the numerators of the gluon propagators by polarization sums in
\eq{eq:A2_1}: this is justified since (cf. \eq{eq:Dmunu})
\begin{align}
  \label{eq:Dmunu2}
  D_{\mu\nu} (l) = - \sum_{\lambda = \pm 1} \epsilon_\mu^\lambda (l)
  \, \epsilon_\nu^{\lambda *} (l) - \eta^\mu \, \eta^\nu \,
  \frac{l^2}{(\eta \cdot l)^2}
\end{align}
where the contribution of the ``instantaneous'' last term on the right
vanishes in $A_2$. Note that we are using retarded Green function
regularization of all gluon propagators, in agreement with the
discussion above and the arguments presented in Appendix~\ref{A}.

Integrating \eq{eq:A2_1} over $x_1^+$, $x_2^+$ and $y^+$ and employing
\eq{eq:polPV} yields
\begin{align}
  \label{eq:A2_2}
  A_2 & = - 4 \, i \, g^3 \, f^{abc} \, \int d^2 x_1 \, d^2 x_2 \,
  U^{bd}_{{\vec x}_{1\perp}} \, U^{ce}_{{\vec x}_{2\perp}} \, \left(
    V_{{\vec b}_{1\perp}} t^d \right)_1 \, \left( V_{{\vec
        b}_{2\perp}} t^e \right)_2 \int\limits_{-\infty}^{\infty}
  \frac{d l^+}{2 \pi} \, e^{- i \, l^+ \, (b_2^- - b_1^-)}
  \notag \\
  & \times \, \int\limits_{-\infty}^{\infty} \frac{d q_1^-}{2 \pi} \,
  \frac{d q_2^-}{2 \pi} \, \frac{d l^-}{2 \pi} \, \frac{d (k-l)^-}{2
    \pi} \, \frac{1}{q_1^- - i \epsilon} \, \frac{1}{q_2^- - i
    \epsilon} \,
  \frac{1}{l^- + (k-l)^- - k^- - i \epsilon}  \notag \\
  & \times \, \int \frac{d^2 k}{(2 \pi)^2} \, \frac{d^2 l}{(2 \pi)^2}
  \, \frac{d^2 q_1}{(2 \pi)^2} \, \frac{d^2 q_2}{(2 \pi)^2} \, e^{i
    {\vec q}_{1\perp} \cdot ({\vec x}_{1\perp} - {\vec b}_{1 \perp}) +
    i {\vec q}_{2\perp} \cdot ({\vec x}_{2\perp} - {\vec b}_{2 \perp})
    + i {\vec l}_\perp \cdot ({\vec x}_{2\perp} - {\vec x}_{1 \perp})
    + i {\vec k}_\perp \cdot ({\vec z}_\perp - {\vec x}_{2 \perp})}
  \notag \\
  & \times \, \frac{1}{2 \, l^+ \, q_1^- - {q}_{1\perp}^{\, 2} + i \,
    \epsilon \, l^+} \, \frac{1}{2 \, (k^+ - l^+) \, q_2^- -
    {q}_{2\perp}^{\, 2} + i \, \epsilon \, (k^+ - l^+)} \, \frac{1}{2
    \, l^+ \, l^- - {l}_\perp^{\, 2} + i \, \epsilon \,
    l^+}  \notag \\
  & \times \, \frac{1}{2 \, (k^+ - l^+) \, (k-l)^- - ({\vec k}_\perp -
    {\vec l}_\perp)^{\, 2} + i \, \epsilon \, (k^+ - l^+)} \,
  \sum_{\lambda_1, \lambda_2} {\vec \epsilon}_{\perp}^{\, \lambda_1 *}
  \cdot {\vec q}_{1\perp} \ {\vec \epsilon}_{\perp}^{\, \lambda_2 *}
  \cdot {\vec
    q}_{2\perp}  \notag \\
  & \times \, \left[ 2 l \cdot \epsilon_\lambda^* (k) \
    \epsilon_{\lambda_1} (l) \cdot \epsilon_{\lambda_2} (k-l) - (k+l)
    \cdot \epsilon_{\lambda_2} (k-l) \ \epsilon_\lambda^* (k) \cdot
    \epsilon_{\lambda_1} (l) + (2 k -l) \cdot \epsilon_{\lambda_1} (l)
    \ \epsilon_\lambda^* (k) \cdot \epsilon_{\lambda_2} (k-l) \right].
\end{align}
Since the expression in the square brackets of \eq{eq:A2_2} is
independent of the ``$-$" components of momenta, we can integrate over
$q_1^-$, $q_2^-$, $l^-$ and $(k-l)^-$ obtaining
\begin{align}
  \label{eq:A2_3}
  A_2 & = - 4 \, i \, g^3 \, f^{abc} \, \int d^2 x_1 \, d^2 x_2 \,
  U^{bd}_{{\vec x}_{1\perp}} \, U^{ce}_{{\vec x}_{2\perp}} \, \left(
    V_{{\vec b}_{1\perp}} t^d \right)_1 \, \left( V_{{\vec
        b}_{2\perp}} t^e \right)_2 \int\limits_{-\infty}^{\infty}
  \frac{d \alpha}{2 \pi} \, e^{- i \, \alpha \, k^+ \, (b_2^- -
    b_1^-)}
  \notag \\
  & \times \, \int \frac{d^2 k}{(2 \pi)^2} \, \frac{d^2 l}{(2 \pi)^2}
  \, \frac{d^2 q_1}{(2 \pi)^2} \, \frac{d^2 q_2}{(2 \pi)^2} \, e^{i
    {\vec q}_{1\perp} \cdot ({\vec x}_{1\perp} - {\vec b}_{1 \perp}) +
    i {\vec q}_{2\perp} \cdot ({\vec x}_{2\perp} - {\vec b}_{2 \perp})
    + i {\vec l}_\perp \cdot ({\vec x}_{2\perp} - {\vec x}_{1 \perp})
    + i {\vec k}_\perp \cdot ({\vec z}_\perp - {\vec x}_{2 \perp})} \,
  \frac{1}{{q}_{1\perp}^{2} \, {q}_{2\perp}^{2}
    \, ({\vec l}_\perp - \alpha \, {\vec k}_\perp )^{2} } \notag \\
  & \times\, \left[ ({\vec l}_\perp - \alpha \, {\vec k}_\perp ) \cdot
    {\vec \epsilon}_{\perp}^{\, \lambda *} \ {\vec q}_{1\perp} \cdot
    {\vec q}_{2\perp} - \mbox{PV} \frac{1}{1-\alpha} \ {\vec
      q}_{2\perp} \cdot ( {\vec l}_\perp - \alpha \, {\vec k}_\perp )
    \ {\vec \epsilon}_{\perp}^{\, \lambda *} \cdot {\vec q}_{1\perp} -
    \mbox{PV} \frac{1}{\alpha} \ {\vec q}_{1\perp} \cdot ({\vec
      l}_\perp - \alpha \, {\vec k}_\perp) \ {\vec
      \epsilon}_{\perp}^{\, \lambda *} \cdot {\vec q}_{2\perp}
  \right],
\end{align}
where we have also simplified the expression in the square brackets
using \eq{eq:polPV} again and defined $\alpha = l^+/k^+$.

Integration over $\alpha$ in \eq{eq:A2_3}, while a little lengthy, can
be straightforwardly performed using residues. In the spirit of
regulating the $\alpha$-integral with $b_2^- - b_1^-$, we will use
$b_2^- - b_1^-$ to tell us which way to close the contour for
divergent terms in the $\alpha$-integral, and drop it in the exponent
afterwards. For the convergent terms in the $\alpha$-integral we can
put $b_2^- - b_1^-$ to zero from the start. To be more specific,
consider the following integral (not present in \eq{eq:A2_3}):
\begin{align}
  \label{eq:int_sample}
  \int\limits_{-\infty}^{\infty} \frac{d \alpha}{2 \pi} \, e^{- i \,
    \alpha \, k^+ \, (b_2^- - b_1^-)} \, \mbox{PV} \frac{1}{1-\alpha}
  = \frac{i}{2} \, e^{- i \, k^+ \, (b_2^- - b_1^-)} \, \mbox{Sign}
  (b_2^- - b_1^-) \approx \frac{i}{2} \, \mbox{Sign} (b_2^- - b_1^-).
\end{align}
First of all, without the exponential containing $b_2^- - b_1^-$ the
integral is divergent, and it is regulated by the exponential. In the
last step in \eq{eq:int_sample} we drop $b_2^- - b_1^-$ after doing
the integral. Indeed we can not drop $b_2^- - b_1^-$ within the Sign
function, so we leave it as is. The above is the standard procedure
when calculating observables in the MV model
\cite{Kovchegov:1996ty,Jalilian-Marian:1997xn,Kovchegov:1997pc}. After
squaring the amplitude we would have to average the resulting cross
section in the projectile nucleus wave function, which would include
integrating over all $b_1^-$ and $b_2^-$. This is left for future work
since, as explained above, we are not ready to calculate the gluon
production cross section having constructed only the ${\cal O} (g^3)$
amplitude in this paper.

After carrying out the $\alpha$-integral \eq{eq:A2_3} reduces to
\begin{align}
  \label{eq:A2_4}
  A_2 & = - 2 \, i \, g^3 \, f^{abc} \, \int d^2 x_1 \, d^2 x_2 \,
  U^{bd}_{{\vec x}_{1\perp}} \, U^{ce}_{{\vec x}_{2\perp}} \, \left(
    V_{{\vec b}_{1\perp}} t^d \right)_1 \, \left( V_{{\vec
        b}_{2\perp}} t^e \right)_2
  \notag \\
  & \times \, \int \frac{d^2 k}{(2 \pi)^2} \, \frac{d^2 l}{(2 \pi)^2}
  \, \frac{d^2 q_1}{(2 \pi)^2} \, \frac{d^2 q_2}{(2 \pi)^2} \, e^{i
    {\vec q}_{1\perp} \cdot ({\vec x}_{1\perp} - {\vec b}_{1 \perp}) +
    i {\vec q}_{2\perp} \cdot ({\vec x}_{2\perp} - {\vec b}_{2 \perp})
    + i {\vec l}_\perp \cdot ({\vec x}_{2\perp} - {\vec x}_{1 \perp})
    + i {\vec k}_\perp \cdot ({\vec z}_\perp - {\vec x}_{2 \perp})} \,
  \frac{1}{{q}_{1\perp}^{2} \, {q}_{2\perp}^{2}
    \, |{\vec k}_\perp \times {\vec l}_\perp| } \notag \\
  & \times\, \left\{ {\vec q}_{1\perp} \cdot {\vec q}_{2\perp} \,
    \frac{{\vec \epsilon}_{\perp}^{\, \lambda *} \cdot \left[
        k_\perp^2 \, {\vec l}_\perp - ({\vec k}_\perp \cdot {\vec
          l}_\perp) \, {\vec k}_\perp \right]}{k_\perp^2} + {\vec
      \epsilon}_{\perp}^{\, \lambda *} \cdot {\vec q}_{1\perp} \,
    \frac{{\vec q}_{2\perp} \cdot \left[ {\vec k}_\perp \ {\vec
          l}_\perp \cdot ({\vec k}_\perp - {\vec l}_\perp) - {\vec
          l}_\perp \ {\vec k}_\perp \cdot ({\vec k}_\perp - {\vec
          l}_\perp) \right]}{({\vec k}_\perp - {\vec l}_\perp)^2}
  \right. \notag \\ & \left. + {\vec \epsilon}_{\perp}^{\, \lambda *}
    \cdot {\vec q}_{2\perp} \, \frac{{\vec q}_{1\perp} \cdot \left[
        {\vec k}_\perp \, {l}_\perp^2 - {\vec l}_\perp \, ({\vec
          k}_\perp \cdot {\vec l}_\perp) \right]}{l_\perp^2} + i \,
    \mbox{Sign} (b_2^- - b_1^-) \, |{\vec k}_\perp \times {\vec
      l}_\perp| \, {\vec q}_{1\perp} \cdot {\vec q}_{2\perp} \,
    \frac{{\vec \epsilon}_\perp^{\, \lambda *} \cdot {\vec
        k}_\perp}{k_\perp^2} \right\},
\end{align}
where we have defined
\begin{align}
  \label{eq:cross}
  {\vec k}_\perp \times {\vec l}_\perp \equiv k_x \, l_y - k_y \, l_x.
\end{align}
Finally noticing that
\begin{subequations}
\begin{align}
  \label{eq:identities}
  {\vec \epsilon}_{\perp}^{\, \lambda *} \cdot \left[ k_\perp^2 \,
    {\vec l}_\perp - ({\vec k}_\perp \cdot {\vec l}_\perp) \, {\vec
      k}_\perp \right] & = - {\vec \epsilon}_{\perp}^{\, \lambda *}
  \times {\vec k}_\perp \ {\vec k}_\perp \times {\vec l}_\perp \\
  {\vec q}_{2\perp} \cdot \left[ {\vec k}_\perp \ {\vec l}_\perp \cdot
    ({\vec k}_\perp - {\vec l}_\perp) - {\vec l}_\perp \ {\vec
      k}_\perp \cdot ({\vec k}_\perp - {\vec l}_\perp) \right] & =
  {\vec q}_{2\perp} \times ({\vec k}_\perp - {\vec l}_\perp) \ {\vec
    k}_\perp \times {\vec l}_\perp \\
  {\vec q}_{1\perp} \cdot \left[ {\vec k}_\perp \, {l}_\perp^2 - {\vec
      l}_\perp \, ({\vec k}_\perp \cdot {\vec l}_\perp) \right] & =
  {\vec q}_{1\perp} \times {\vec l}_\perp \ {\vec k}_\perp \times
  {\vec l}_\perp,
\end{align}
\end{subequations}
we can further simplify \eq{eq:A2_4} to the form given below in
\eq{A2mom}.


\subsubsection{Results}

Before performing the transverse Fourier transforms, let us first list
the expressions for the contributions of all the diagrams in
Figs.~\ref{Agraphs} and \ref{Bgraphs} in transverse momentum
space. All the diagram values below are given in the notation where
the fundamental Wilson lines are moved to the left of all the
$t^a$-matrices for the same nucleon, similar to \eqref{eq:1gluon}:
this way all the $V$'s will vanish when we square the amplitude.

Using the PV prescription to regulate the light-cone poles, the
$A$-diagram contributions to the amplitude are
\begin{subequations}\label{eq:A}
\begin{align}
  A_1 = & \, 2 \, g^3 \, \int \frac{d^2 k}{(2 \pi)^2} \, \frac{d^2
    l}{(2 \pi)^2} \, e^{i {\vec l}_\perp \cdot ({\vec b}_{2\perp} -
    {\vec b}_{1 \perp}) + i {\vec k}_\perp \cdot ({\vec z}_\perp -
    {\vec b}_{2 \perp})} \, U^{ad}_{{\vec z}_\perp} \, f^{dbc} \,
  \left( V_{{\vec b}_{1 \perp}} t^b \right)_1 \, \left( V_{{\vec b}_{2
        \perp}} t^c \right)_2 \notag \\ & \times \frac{{\vec
      \epsilon}_\perp^{\, \lambda *} \cdot {\vec k}_\perp \, {\vec
      l}_\perp \cdot ({\vec l}_\perp - {\vec k}_\perp)}{k_\perp^2 \,
    l_\perp^2 \, ({\vec k}_\perp - {\vec l}_\perp)^2} \,
  \mbox{Sign} (b_2^- - b_1^-) , \\
  A_2 = & - 2 \, i \, g^3 \, \int d^2 x_1 \, d^2 x_2 \, \int \frac{d^2
    k}{(2 \pi)^2} \, \frac{d^2 l}{(2 \pi)^2} \, \frac{d^2 q_1}{(2
    \pi)^2} \, \frac{d^2 q_2}{(2 \pi)^2} \, e^{i {\vec q}_{1\perp}
    \cdot ({\vec x}_{1\perp} - {\vec b}_{1 \perp}) + i {\vec
      q}_{2\perp} \cdot ({\vec x}_{2\perp} - {\vec b}_{2 \perp}) + i
    {\vec l}_\perp \cdot ({\vec x}_{2\perp} - {\vec x}_{1 \perp}) + i
    {\vec k}_\perp \cdot ({\vec z}_\perp - {\vec x}_{2 \perp})} \notag
  \\ & \times \frac{1}{q_{1\perp}^2 \, q_{2\perp}^2} \, \Bigg[ \left(
    - {\vec q}_{1\perp} \cdot {\vec q}_{2\perp} \, \frac{{\vec
        \epsilon}_\perp^{\, \lambda *} \times {\vec
        k}_\perp}{k_\perp^2} + {\vec \epsilon}_\perp^{\, \lambda *}
    \cdot {\vec q}_{1\perp} \, \frac{{\vec q}_{2\perp} \times ({\vec
        k}_\perp - {\vec l}_\perp)}{({\vec k}_\perp - {\vec
        l}_\perp)^2} + {\vec \epsilon}_\perp^{\, \lambda *} \cdot
    {\vec q}_{2\perp} \, \frac{{\vec q}_{1\perp} \times {\vec
        l}_\perp}{{l}_\perp^2} \right) \, \mbox{Sign} ({\vec k}_\perp
  \times {\vec l}_\perp) \notag \\ & + i \, {\vec q}_{1\perp} \cdot
  {\vec q}_{2\perp} \, \frac{{\vec \epsilon}_\perp^{\, \lambda *}
    \cdot {\vec k}_\perp}{k_\perp^2} \,  \mbox{Sign} (b_2^- - b_1^-) 
  \Bigg] \, f^{abc}
  \, U^{bd}_{{\vec x}_{1\perp}} \, U^{ce}_{{\vec x}_{2\perp}} \,
  \left( V_{{\vec b}_{1 \perp}} t^d \right)_1 \, \left( V_{{\vec
        b}_{2 \perp}} t^e \right)_2 , \label{A2mom} \\
  A_3 = & - 2 \, i \, g^3 \, \int d^2 x \, \int \frac{d^2 k}{(2
    \pi)^2} \, \frac{d^2 q}{(2 \pi)^2} \, \frac{d^2 l}{(2 \pi)^2} \,
  e^{i {\vec q}_\perp \cdot ({\vec x}_{\perp} - {\vec b}_{2 \perp}) +
    i {\vec l}_\perp \cdot ({\vec x}_\perp - {\vec b}_{1 \perp}) + i
    {\vec k}_\perp \cdot ({\vec z}_\perp - {\vec x}_{\perp})} \,
  f^{abc} \, U^{bd}_{{\vec b}_{1\perp}} \, U^{ce}_{{\vec x}_{\perp}}
  \, \left( V_{{\vec b}_{1 \perp}} t^d \right)_1 \, \left( V_{{\vec
        b}_{2 \perp}} t^e \right)_2 \notag \\ & \times
  \frac{1}{l_{\perp}^2 \, q_\perp^2} \, \Bigg[ \left( \frac{{\vec
        \epsilon}_\perp^{\, \lambda *} \times {\vec
        k}_\perp}{k_\perp^2} \, {\vec l}_\perp \cdot {\vec q}_\perp -
    {\vec \epsilon}_\perp^{\, \lambda *} \cdot {\vec l}_\perp \,
    \frac{{\vec q}_\perp \times ({\vec k}_\perp - {\vec
        l}_\perp)}{({\vec k}_\perp - {\vec l}_\perp)^2} \right) \,
  \mbox{Sign} ({\vec k}_\perp \times {\vec l}_\perp) \notag \\ & - i
  \left( \frac{{\vec \epsilon}_\perp^{\, \lambda *} \cdot {\vec
        k}_\perp}{k_\perp^2} \, {\vec l}_\perp \cdot {\vec q}_\perp -
    {\vec \epsilon}_\perp^{\, \lambda *} \cdot {\vec l}_\perp \,
    \frac{({\vec k}_\perp - {\vec l}_\perp) \cdot {\vec
        q}_\perp}{({\vec k}_\perp - {\vec l}_\perp)^2} -
    \frac{1}{2} \, {\vec \epsilon}_\perp^{\, \lambda *} \cdot {\vec q}_\perp \right) \Bigg], \\
  A_4  (b_1 & \, , b_2;  1, 2) = A_3 (b_2, b_1; 2, 1), \\
  A_5 = & - 2 \, i \, g^3 \, \int \frac{d^2 k}{(2 \pi)^2} \, \frac{d^2
    l}{(2 \pi)^2} \, e^{i {\vec l}_\perp \cdot ({\vec b}_{2\perp} -
    {\vec b}_{1 \perp}) + i {\vec k}_\perp \cdot ({\vec z}_\perp -
    {\vec b}_{2 \perp})} \, \frac{1}{k_\perp^2 \, l_\perp^2 \, ({\vec
      k}_\perp - {\vec l}_\perp)^2} \, f^{abc} \, U^{bd}_{{\vec
      b}_{1\perp}} \, U^{ce}_{{\vec b}_{2\perp}} \, \left( V_{{\vec
        b}_{1 \perp}} t^d \right)_1 \, \left( V_{{\vec b}_{2 \perp}}
    t^e \right)_2 \notag \\ & \times \left[ \frac{i}{2} \, {\vec
      \epsilon}_\perp^{\, \lambda *} \cdot {\vec k}_\perp \,
    {k}_\perp^2 - i \, {\vec \epsilon}_\perp^{\, \lambda *} \cdot
    {\vec l}_\perp \, k_\perp^2 - {\vec l}_\perp \cdot ({\vec k}_\perp
    - {\vec l}_\perp) \, {\vec \epsilon}_\perp^{\, \lambda *} \times
    {\vec k}_\perp \,
    \mbox{Sign} ({\vec k}_\perp \times {\vec l}_\perp) \right], \\
  A_6 = & \, 0, \\ A_7 = & \, 0.
\end{align}
\end{subequations}

The $B$-diagram contributions are
\begin{subequations}\label{eq:B}
\begin{align}
  B_1 = & - i g^3 \, \int \frac{d^2 k}{(2 \pi)^2} \, \frac{d^2 l}{(2
    \pi)^2} \, e^{i {\vec l}_\perp \cdot ({\vec b}_{2\perp} - {\vec
      b}_{1 \perp}) + i {\vec k}_\perp \cdot ({\vec z}_\perp - {\vec
      b}_{2 \perp})} \, \frac{{\vec \epsilon}_\perp^{\, \lambda *}
    \cdot {\vec k}_\perp}{k_\perp^2 \, ({\vec k}_\perp - {\vec
      l}_\perp)^2} \, U^{ad}_{{\vec z}_\perp} \, \left( V_{{\vec b}_{1
        \perp}} t^d t^b \right)_1 \, \left( V_{{\vec b}_{2 \perp}} t^b
  \right)_2 \, \mbox{Sign} (b_2^- - b_1^-), \\
  B_2 = & \, 0, \\
  B_3 = & \, i g^3 \, \int \frac{d^2 k}{(2 \pi)^2} \, \frac{d^2 l}{(2
    \pi)^2} \, e^{i {\vec l}_\perp \cdot ({\vec b}_{2\perp} - {\vec
      b}_{1 \perp}) + i {\vec k}_\perp \cdot ({\vec z}_\perp - {\vec
      b}_{2 \perp})} \, \frac{{\vec \epsilon}_\perp^{\, \lambda *}
    \cdot {\vec k}_\perp}{k_\perp^2 \, ({\vec k}_\perp - {\vec
      l}_\perp)^2} \, U^{ad}_{{\vec b}_{1\perp}} \, \left( V_{{\vec
        b}_{1 \perp}} t^d t^b \right)_1 \, \left( V_{{\vec b}_{2
        \perp}} t^b \right)_2 \,
  \mbox{Sign} (b_2^- - b_1^-),  \label{B3} \\
  B_4 = & \, 2 i g^3 \, \int d^2 y \, \int \frac{d^2 k}{(2 \pi)^2} \,
  \frac{d^2 q_1}{(2 \pi)^2} \, \frac{d^2 q_2}{(2 \pi)^2} \, e^{i {\vec
      q}_{1\perp} \cdot ({\vec b}_{1\perp} - {\vec y}) + i {\vec
      q}_{2\perp} \cdot ({\vec y}_\perp - {\vec b}_{2\perp}) + i {\vec
      k}_\perp \cdot ({\vec z}_\perp - {\vec b}_{1 \perp})} \,
  \frac{{\vec \epsilon}_\perp^{\, \lambda *} \cdot {\vec
      k}_\perp}{k_\perp^2} \, \frac{{\vec q}_{1\perp} \cdot {\vec
      q}_{2\perp}}{q_{1\perp}^2 \, q_{2\perp}^2} \, U^{ad}_{{\vec
      z}_{\perp}} \, U^{be}_{{\vec b}_{1\perp}} \notag \\ & \times \,
  \left( V_{{\vec b}_{1 \perp}} t^e t^d \right)_1 \, U^{bc}_{{\vec
      y}_\perp} \, \left( V_{{\vec b}_{2
        \perp}} t^c \right)_2 \, \mbox{Sign} (b_2^- - b_1^-), \\
  B_5 = & - i g^3 \, \int \frac{d^2 k}{(2 \pi)^2} \, \frac{d^2 l}{(2
    \pi)^2} \, e^{i {\vec l}_\perp \cdot ({\vec b}_{2\perp} - {\vec
      b}_{1 \perp}) + i {\vec k}_\perp \cdot ({\vec z}_\perp - {\vec
      b}_{2 \perp})} \, \frac{{\vec \epsilon}_\perp^{\, \lambda *}
    \cdot {\vec k}_\perp}{k_\perp^2 \, ({\vec k}_\perp - {\vec
      l}_\perp)^2} \, U^{ad}_{{\vec z}_{\perp}} \, U^{bc}_{{\vec
      b}_{1\perp}} \, \left( V_{{\vec b}_{1 \perp}} t^c t^d \right)_1
  \notag \\ & \times \, U^{be}_{{\vec b}_{2\perp}} \, \left( V_{{\vec
        b}_{2 \perp}} t^e
  \right)_2  \, \mbox{Sign} (b_2^- - b_1^-), \\
  B_6 = & \, 0, \\
  B_7 = & \, 0,  \\
  B_8 = & \, 0, \\
  B_9 = & \, 2 i g^3 \, \int \frac{d^2 k}{(2 \pi)^2} \, \frac{d^2
    l}{(2 \pi)^2} \, e^{i {\vec l}_\perp \cdot ({\vec b}_{2\perp} -
    {\vec b}_{1 \perp}) + i {\vec k}_\perp \cdot ({\vec z}_\perp -
    {\vec b}_{2 \perp})} \, \frac{{\vec \epsilon}_\perp^{\, \lambda *}
    \cdot {\vec k}_\perp}{k_\perp^2 \, ({\vec k}_\perp - {\vec
      l}_\perp)^2} \, U^{ad}_{{\vec b}_{1\perp}} \, U^{bc}_{{\vec
      b}_{1\perp}} \, \left( V_{{\vec b}_{1 \perp}} t^d t^c \right)_1
  \, U^{be}_{{\vec
      b}_{2\perp}} \, \left( V_{{\vec b}_{2 \perp}} t^e \right)_2 , \label{B9} \\
  B_{10} = & - i g^3 \, \int \frac{d^2 k}{(2 \pi)^2} \, \frac{d^2
    l}{(2 \pi)^2} \, e^{i {\vec l}_\perp \cdot ({\vec b}_{2\perp} -
    {\vec b}_{1 \perp}) + i {\vec k}_\perp \cdot ({\vec z}_\perp -
    {\vec b}_{2 \perp})} \, \frac{{\vec \epsilon}_\perp^{\, \lambda *}
    \cdot {\vec k}_\perp}{k_\perp^2 \, ({\vec k}_\perp - {\vec
      l}_\perp)^2} \, U^{ad}_{{\vec b}_{1\perp}} \, U^{bc}_{{\vec
      b}_{1\perp}} \, \left( V_{{\vec b}_{1 \perp}} t^c t^d \right)_1
  \, U^{be}_{{\vec b}_{2\perp}} \, \left( V_{{\vec b}_{2 \perp}} t^e
  \right)_2 \notag \\ & \times \, \left[2 - \mbox{Sign} (b_2^- - b_1^-) \right], \\
  B_{11} = & - 2 i g^3 \, \int d^2 y \, \int \frac{d^2 k}{(2 \pi)^2}
  \, \frac{d^2 q_1}{(2 \pi)^2} \, \frac{d^2 q_2}{(2 \pi)^2} \, e^{i
    {\vec q}_{1\perp} \cdot ({\vec b}_{1\perp} - {\vec y}_\perp) + i
    {\vec q}_{2\perp} \cdot ({\vec y}_\perp - {\vec b}_{2\perp}) + i
    {\vec k}_\perp \cdot ({\vec z}_\perp - {\vec b}_{1 \perp})} \,
  \frac{{\vec \epsilon}_\perp^{\, \lambda *} \cdot {\vec
      k}_\perp}{k_\perp^2} \, \frac{{\vec q}_{1\perp} \cdot {\vec
      q}_{2\perp}}{q_{1\perp}^2 \, q_{2\perp}^2} \notag \\ & \times \,
  U^{ad}_{{\vec b}_{1\perp}} \, U^{bc}_{{\vec b}_{1\perp}} \, \left(
    V_{{\vec b}_{1 \perp}} t^d t^c \right)_1 \, U^{be}_{{\vec
      y}_\perp} \, \left( V_{{\vec b}_{2
        \perp}} t^e \right)_2 , \label{B11} \\
  B_{12} = & \, 2 i g^3 \, \int d^2 y \, \int \frac{d^2 k}{(2 \pi)^2}
  \, \frac{d^2 q_1}{(2 \pi)^2} \, \frac{d^2 q_2}{(2 \pi)^2} \, e^{i
    {\vec q}_{1\perp} \cdot ({\vec b}_{1\perp} - {\vec y}_\perp) + i
    {\vec q}_{2\perp} \cdot ({\vec y}_\perp - {\vec b}_{2\perp}) + i
    {\vec k}_\perp \cdot ({\vec z}_\perp - {\vec b}_{1 \perp})} \,
  \frac{{\vec \epsilon}_\perp^{\, \lambda *} \cdot {\vec
      k}_\perp}{k_\perp^2} \, \frac{{\vec q}_{1\perp} \cdot {\vec
      q}_{2\perp}}{q_{1\perp}^2 \, q_{2\perp}^2} \notag \\ & \times \,
  U^{ad}_{{\vec b}_{1\perp}} \, U^{bc}_{{\vec b}_{1\perp}} \, \left(
    V_{{\vec b}_{1 \perp}} t^c t^d \right)_1 \, U^{be}_{{\vec
      y}_{\perp}} \, \left( V_{{\vec b}_{2 \perp}} t^e \right)_2 \,
  \left[ 1 - \mbox{Sign} (b_2^- - b_1^-) \right].
\end{align}
\end{subequations}

The $C$ diagrams are obtained by swapping the two projectile Wilson
lines in \fig{Bgraphs}, such that
\begin{align}
  \label{eq:C}
  C_i (b_1, b_2; 1, 2) = B_i (b_2, b_1; 2, 1).
\end{align}

As explained in Appendix~\ref{A}, if one includes the contributions
arising from moving the gluon exchanged between the two projectile
quarks in the $B$-graphs across the cut, only the diagrams $B_2$,
$B_9$ and $B_{11}$ need to be calculated with the commutators of
fundamental generators on the quark-$1$ line. Since $B_2 =0$ we only
need to take $B_9$ and $B_{11}$ from Eqs.~\eqref{B9} and \eqref{B11}
above and replace $t^d t^c \to [t^d , t^c]$ in them. The result is
\begin{align}
  \label{eq:Bsum_final}
  \sideset{}{'} \sum\limits_{i=1}^{12} \, B_i & \, = - 2 \, g^3 \,
  \int d^2 y \, \int \frac{d^2 k}{(2 \pi)^2} \, \frac{d^2 q_1}{(2
    \pi)^2} \, \frac{d^2 q_2}{(2 \pi)^2} \, e^{i {\vec q}_{1\perp}
    \cdot ({\vec b}_{1\perp} - {\vec y}_\perp) + i {\vec q}_{2\perp}
    \cdot ({\vec y}_\perp - {\vec b}_{2\perp}) + i {\vec k}_\perp
    \cdot ({\vec z}_\perp - {\vec b}_{1 \perp})} \, \frac{{\vec
      \epsilon}_\perp^{\, \lambda *} \cdot {\vec k}_\perp}{k_\perp^2}
  \, \frac{{\vec q}_{1\perp} \cdot {\vec q}_{2\perp}}{q_{1\perp}^2 \,
    q_{2\perp}^2} \, \left( V_{{\vec b}_{2 \perp}} t^e \right)_2
  \notag \\ & \times \, f^{cdh} \, U^{ad}_{{\vec b}_{1\perp}} \,
  U^{bc}_{{\vec b}_{1\perp}} \left[ U^{be}_{{\vec y}_{\perp}} -
    U^{be}_{{\vec b}_{2 \perp}} \right] \left( V_{{\vec b}_{1 \perp}}
    t^h \right)_1,
\end{align}
where the prime over the sum indicates that it is not a simple sum of
the contributions in Eqs.~\eqref{eq:B}, but a sum of all of those
graphs and the contributions with the exchanged gluon moved across the
cut. An expression similar to \eqref{eq:Bsum_final} can be obtained
using \eq{eq:C} for the sum of $C$ graphs (which would also include
contributions arising from moving the gluons across the cut).

Adding all the $A$, $B$ and $C$ graphs by using Eqs.~\eqref{eq:A},
\eqref{eq:Bsum_final} and \eqref{eq:C} yields
\begin{align}
  \label{eq:ABCsum}
  \sum\limits_{i=1}^{7} & \, A_i + \sideset{}{'}
  \sum\limits_{i=1}^{12} B_i + \sideset{}{'} \sum\limits_{i=1}^{12}
  C_i \notag \\ = & - 2 \, i \, g^3 \, \int d^2 x_1 \, d^2 x_2 \, \int
  \frac{d^2 k}{(2 \pi)^2} \, \frac{d^2 l}{(2 \pi)^2} \, \frac{d^2
    q_1}{(2 \pi)^2} \, \frac{d^2 q_2}{(2 \pi)^2} \, e^{i {\vec
      q}_{1\perp} \cdot ({\vec x}_{1\perp} - {\vec b}_{1 \perp}) + i
    {\vec q}_{2\perp} \cdot ({\vec x}_{2\perp} - {\vec b}_{2 \perp}) +
    i {\vec l}_\perp \cdot ({\vec x}_{2\perp} - {\vec x}_{1 \perp}) +
    i {\vec k}_\perp \cdot ({\vec z}_\perp - {\vec x}_{2 \perp})}
  \notag \\ & \times \frac{1}{q_{1\perp}^2 \, q_{2\perp}^2} \, \left(
    - {\vec q}_{1\perp} \cdot {\vec q}_{2\perp} \, \frac{{\vec
        \epsilon}_\perp^{\, \lambda *} \times {\vec
        k}_\perp}{k_\perp^2} + {\vec \epsilon}_\perp^{\, \lambda *}
    \cdot {\vec q}_{1\perp} \, \frac{{\vec q}_{2\perp} \times ({\vec
        k}_\perp - {\vec l}_\perp)}{({\vec k}_\perp - {\vec
        l}_\perp)^2} + {\vec \epsilon}_\perp^{\, \lambda *} \cdot
    {\vec q}_{2\perp} \, \frac{{\vec q}_{1\perp} \times {\vec
        l}_\perp}{{\vec l}_\perp^2} \right) \, \mbox{Sign} ({\vec
    k}_\perp \times {\vec l}_\perp) \notag \\ & \times \, f^{abc} \,
  \left[ U^{bd}_{{\vec x}_{1\perp}} - U^{bd}_{{\vec b}_{1\perp}}
  \right] \, \left[ U^{ce}_{{\vec x}_{2\perp}} - U^{ce}_{{\vec
        b}_{2\perp}} \right] \, \left( V_{{\vec b}_{1 \perp}} t^d
  \right)_1 \, \left( V_{{\vec b}_{2 \perp}} t^e \right)_2 \notag \\ &
  - 2 \, g^3 \, \int d^2 x \, \int \frac{d^2 k}{(2 \pi)^2} \,
  \frac{d^2 l}{(2 \pi)^2} \, \frac{d^2 q}{(2 \pi)^2} \, e^{i {\vec
      q}_{\perp} \cdot ({\vec x}_{\perp} - {\vec b}_{2 \perp}) + i
    {\vec l}_{\perp} \cdot ({\vec x}_{\perp} - {\vec b}_{1 \perp}) + i
    {\vec k}_\perp \cdot ({\vec z}_\perp - {\vec x}_{\perp})} \,
  f^{abc} \, \left( V_{{\vec b}_{1 \perp}} t^d \right)_1 \, \left(
    V_{{\vec b}_{2 \perp}} t^e \right)_2 \notag \\ & \times \, \left[
    U^{bd}_{{\vec b}_{1\perp}} \, \left( U^{ce}_{{\vec x}_{\perp}} -
      U^{ce}_{{\vec b}_{2\perp}} \right) \, \left( \frac{{\vec
          \epsilon}_\perp^{\, \lambda *} \cdot {\vec
          k}_\perp}{k_\perp^2} \, \frac{{\vec l}_\perp \cdot {\vec
          q}_\perp}{l_\perp^2 \, q_\perp^2} - \frac{{\vec
          \epsilon}_\perp^{\, \lambda *} \cdot {\vec
          l}_\perp}{l_\perp^2} \, \frac{({\vec k}_\perp - {\vec
          l}_\perp) \cdot {\vec q}_\perp}{({\vec k}_\perp - {\vec
          l}_\perp)^2 \, q_\perp^2} - \frac{{\vec \epsilon}_\perp^{\,
          \lambda *} \cdot {\vec q}_\perp}{2 \, l_\perp^2 \,
        q_\perp^2} + \frac{{\vec \epsilon}_\perp^{\, \lambda *} \cdot
        {\vec k}_\perp}{k_\perp^2} \, \frac{({\vec k}_\perp - {\vec
          l}_\perp) \cdot {\vec q}_\perp}{({\vec k}_\perp - {\vec
          l}_\perp)^2 \, q_\perp^2} \right) \right. \notag \\ & -
  \left.  \left( U^{bd}_{{\vec x}_{\perp}} - U^{bd}_{{\vec
          b}_{1\perp}} \right) \, U^{ce}_{{\vec b}_{2\perp}} \, \left(
      \frac{{\vec \epsilon}_\perp^{\, \lambda *} \cdot {\vec
          k}_\perp}{k_\perp^2} \, \frac{{\vec l}_\perp \cdot {\vec
          q}_\perp}{l_\perp^2 \, q_\perp^2} - \frac{{\vec
          \epsilon}_\perp^{\, \lambda *} \cdot {\vec
          q}_\perp}{q_\perp^2} \, \frac{({\vec k}_\perp - {\vec
          q}_\perp) \cdot {\vec l}_\perp}{({\vec k}_\perp - {\vec
          q}_\perp)^2 \, l_\perp^2} - \frac{{\vec \epsilon}_\perp^{\,
          \lambda *} \cdot {\vec l}_\perp}{2 \, l_\perp^2 \,
        q_\perp^2} + \frac{{\vec \epsilon}_\perp^{\, \lambda *} \cdot
        {\vec k}_\perp}{k_\perp^2} \, \frac{({\vec k}_\perp - {\vec
          q}_\perp) \cdot {\vec l}_\perp}{({\vec
          k}_\perp - {\vec q}_\perp)^2 \, l_\perp^2} \right) \right] \notag \\
  & + 2 g^3 \, \int d^2 x \, \int \frac{d^2 k}{(2 \pi)^2} \, \frac{d^2
    q_1}{(2 \pi)^2} \, \frac{d^2 q_2}{(2 \pi)^2} \, e^{i {\vec q}_{1
      \perp} \cdot ({\vec x}_{\perp} - {\vec b}_{1 \perp}) + i {\vec
      q}_{2 \perp} \cdot ({\vec x}_{\perp} - {\vec b}_{2 \perp}) + i
    {\vec k}_\perp \cdot ({\vec z}_\perp - {\vec x}_{\perp})} \,
  \frac{{\vec q}_{1 \perp} \cdot {\vec q}_{2 \perp}}{{q}^2_{1 \perp}
    \, {q}^2_{2 \perp}} \, \frac{{\vec \epsilon}_\perp^{\, \lambda *}
    \cdot {\vec k}_\perp}{k_\perp^2} \, \mbox{Sign} (b_2^- - b_1^-)
  \notag \\
  & \times \left[ U_{x_\perp}^{ab} - U_{z_\perp}^{ab} \right] \,
  f^{bde} \, \left( V_{{\vec b}_{1 \perp}} t^d \right)_1 \, \left(
    V_{{\vec b}_{2 \perp}} t^e \right)_2 .
\end{align}
This is our final answer for the sum of $A$, $B$ and $C$ graphs in
transverse momentum space.

Fourier transforming \eq{eq:ABCsum} into transverse coordinate space
is straightforward. We use 
\begin{align}
  \label{eq:Fourier1}
  \int \frac{d^2 q}{(2 \pi)^2} \, \frac{e^{i {\vec q}_\perp \cdot
      {\vec x}_\perp}}{q_\perp^2} = \frac{1}{2 \pi} \, \ln
  \frac{1}{x_\perp \, \Lambda}
\end{align}
along with 
\begin{align}
  \label{eq:Fourier2}
  \int \frac{d^2 q}{(2 \pi)^2} \, e^{i {\vec q}_\perp \cdot {\vec
      x}_\perp} \, \frac{{\vec q}_\perp}{q_\perp^2} = \frac{i}{2 \pi}
  \, \frac{{\vec x}_\perp}{x_\perp^2 }
\end{align}
and
\begin{align}
  \label{eq:FTSign}
  \int \frac{d^2 k}{(2 \pi)^2} \, \frac{d^2 l}{(2 \pi)^2} \, e^{i
    {\vec k}_\perp \cdot ({\vec z}_{\perp} - {\vec x}_{2\perp}) + i
    {\vec l}_\perp \cdot ({\vec x}_{2\perp} - {\vec x}_{1\perp})} \
  \frac{{\vec \epsilon}_\perp^{\, \lambda *} \times {\vec
      k}_\perp}{k^2_\perp} \, \mbox{Sign} ({\vec k}_\perp \times {\vec
    l}_\perp) \notag \\ = \frac{-i}{2 \, \pi^2} \, \frac{{\vec
      \epsilon}_\perp^{\, \lambda *} \cdot ({\vec x}_{2\perp} - {\vec
      x}_{1\perp})}{|{\vec x}_{2\perp} - {\vec x}_{1\perp}|^2} \,
  \delta [({\vec z}_{\perp} - {\vec x}_{1\perp}) \times ({\vec
    z}_{\perp} - {\vec x}_{2\perp})].
\end{align}
Here, again, $\Lambda \sim \Lambda_{QCD}$ is the IR cutoff in each
individual nucleon.

Performing the Fourier transform of \eq{eq:ABCsum} we obtain
\begin{align}
  \label{eq:ABCsum_coord}
  \sum\limits_{i=1}^{7} & \, A_i + \sideset{}{'}
  \sum\limits_{i=1}^{12} B_i + \sideset{}{'} \sum\limits_{i=1}^{12}
  C_i \notag \\ = & - \frac{g^3}{4 \, \pi^4} \int d^2 x_1 \, d^2 x_2
  \, \delta [({\vec z}_{\perp} - {\vec x}_{1\perp}) \times ({\vec
    z}_{\perp} - {\vec x}_{2\perp})] \left[ \frac{{\vec
        \epsilon}_\perp^{\, \lambda *} \cdot ({\vec x}_{2\perp} -
      {\vec x}_{1\perp}) }{|{\vec x}_{2\perp} - {\vec x}_{1\perp}|^2}
    \, \frac{{\vec x}_{1\perp} - {\vec b}_{1\perp}}{|{\vec x}_{1\perp}
      - {\vec b}_{1\perp}|^2} \cdot \frac{{\vec x}_{2\perp} - {\vec
        b}_{2\perp}}{|{\vec x}_{2\perp} - {\vec b}_{2\perp}|^2}
  \right. \notag \\ & \left. - \frac{{\vec \epsilon}_\perp^{\, \lambda
        *} \cdot ({\vec x}_{1\perp} - {\vec b}_{1\perp}) }{|{\vec
        x}_{1\perp} - {\vec b}_{1\perp}|^2} \, \frac{{\vec z}_{\perp}
      - {\vec x}_{1\perp}}{|{\vec z}_{\perp} - {\vec x}_{1\perp}|^2}
    \cdot \frac{{\vec x}_{2\perp} - {\vec b}_{2\perp}}{|{\vec
        x}_{2\perp} - {\vec b}_{2\perp}|^2} + \frac{{\vec
        \epsilon}_\perp^{\, \lambda *} \cdot ({\vec x}_{2\perp} -
      {\vec b}_{2\perp}) }{|{\vec x}_{2\perp} - {\vec b}_{2\perp}|^2}
    \, \frac{{\vec x}_{1\perp} - {\vec b}_{1\perp}}{|{\vec x}_{1\perp}
      - {\vec b}_{1\perp}|^2} \cdot \frac{{\vec z}_{\perp} - {\vec
        x}_{2\perp}}{|{\vec z}_{\perp} - {\vec x}_{2\perp}|^2} \right]
  \notag \\ & \times \, f^{abc} \, \left[ U^{bd}_{{\vec x}_{1\perp}} -
    U^{bd}_{{\vec b}_{1\perp}} \right] \, \left[ U^{ce}_{{\vec
        x}_{2\perp}} - U^{ce}_{{\vec b}_{2\perp}} \right] \, \left(
    V_{{\vec b}_{1 \perp}} t^d \right)_1 \, \left( V_{{\vec b}_{2
        \perp}} t^e \right)_2 \notag \\ & + \frac{i \, g^3}{4 \,
    \pi^3} \, f^{abc} \, \left( V_{{\vec b}_{1 \perp}} t^d \right)_1
  \, \left( V_{{\vec b}_{2 \perp}} t^e \right)_2 \int d^2 x \, \left[
    U^{bd}_{{\vec b}_{1\perp}} \, \left( U^{ce}_{{\vec x}_{\perp}} -
      U^{ce}_{{\vec b}_{2\perp}} \right) \, \left( \frac{{\vec
          \epsilon}_\perp^{\, \lambda *} \cdot ({\vec z}_{\perp} -
        {\vec x}_{\perp}) }{|{\vec z}_{\perp} - {\vec x}_{\perp}|^2}
      \, \frac{{\vec x}_{\perp} - {\vec b}_{1\perp}}{|{\vec x}_{\perp}
        - {\vec b}_{1\perp}|^2} \cdot \frac{{\vec x}_{\perp} - {\vec
          b}_{2\perp}}{|{\vec x}_{\perp} - {\vec b}_{2\perp}|^2}
    \right. \right. \notag \\ & \left. - \frac{{\vec
        \epsilon}_\perp^{\, \lambda *} \cdot ({\vec z}_{\perp} - {\vec
        b}_{1\perp}) }{|{\vec z}_{\perp} - {\vec b}_{1\perp}|^2} \,
    \frac{{\vec z}_{\perp} - {\vec x}_{\perp}}{|{\vec z}_{\perp} -
      {\vec x}_{\perp}|^2} \cdot \frac{{\vec x}_{\perp} - {\vec
        b}_{2\perp}}{|{\vec x}_{\perp} - {\vec b}_{2\perp}|^2} -
    \frac{{\vec \epsilon}_\perp^{\, \lambda *} \cdot ({\vec z}_{\perp}
      - {\vec b}_{1\perp}) }{|{\vec z}_{\perp} - {\vec b}_{1\perp}|^2}
    \, \frac{{\vec x}_{\perp} - {\vec b}_{1\perp}}{|{\vec x}_{\perp} -
      {\vec b}_{1\perp}|^2} \cdot \frac{{\vec x}_{\perp} - {\vec
        b}_{2\perp}}{|{\vec x}_{\perp} - {\vec b}_{2\perp}|^2} \right)
  \notag \\ & - \left( U^{bd}_{{\vec x}_{\perp}} - U^{bd}_{{\vec
        b}_{1\perp}} \right) \, U^{ce}_{{\vec b}_{2\perp}} \, \left(
    \frac{{\vec \epsilon}_\perp^{\, \lambda *} \cdot ({\vec z}_{\perp}
      - {\vec x}_{\perp}) }{|{\vec z}_{\perp} - {\vec x}_{\perp}|^2}
    \, \frac{{\vec x}_{\perp} - {\vec b}_{1\perp}}{|{\vec x}_{\perp} -
      {\vec b}_{1\perp}|^2} \cdot \frac{{\vec x}_{\perp} - {\vec
        b}_{2\perp}}{|{\vec x}_{\perp} - {\vec b}_{2\perp}|^2} -
    \frac{{\vec \epsilon}_\perp^{\, \lambda *} \cdot ({\vec z}_{\perp}
      - {\vec b}_{2\perp}) }{|{\vec z}_{\perp} - {\vec b}_{2\perp}|^2}
    \, \frac{{\vec z}_{\perp} - {\vec x}_{\perp}}{|{\vec z}_{\perp} -
      {\vec x}_{\perp}|^2} \cdot \frac{{\vec x}_{\perp} - {\vec
        b}_{1\perp}}{|{\vec x}_{\perp} - {\vec b}_{1\perp}|^2}
  \right. \notag \\ & \left. \left. - \frac{{\vec \epsilon}_\perp^{\,
          \lambda *} \cdot ({\vec z}_{\perp} - {\vec b}_{2\perp})
      }{|{\vec z}_{\perp} - {\vec b}_{2\perp}|^2} \, \frac{{\vec
          x}_{\perp} - {\vec b}_{1\perp}}{|{\vec x}_{\perp} - {\vec
          b}_{1\perp}|^2} \cdot \frac{{\vec x}_{\perp} - {\vec
          b}_{2\perp}}{|{\vec x}_{\perp} - {\vec b}_{2\perp}|^2}
    \right) \right] \notag \\
  & - \frac{i \, g^3}{4 \, \pi^2} \, f^{abc} \, \left( V_{{\vec b}_{1
        \perp}} t^d \right)_1 \, \left( V_{{\vec b}_{2 \perp}} t^e
  \right)_2 \notag \\ & \times \, \left[ (U^{bd}_{{\vec z}_\perp} -
    U^{bd}_{{\vec b}_{1 \perp}}) \, U^{ce}_{{\vec b}_{2 \perp}} \,
    \frac{{\vec \epsilon}_\perp^{\, \lambda *} \cdot ({\vec z}_{\perp}
      - {\vec b}_{1\perp})}{|{\vec z}_{\perp} - {\vec b}_{1\perp}|^2}
    \, \ln \frac{1}{|{\vec z}_{\perp} - {\vec b}_{2\perp}| \, \Lambda}
    - U^{bd}_{{\vec b}_{1 \perp}} \, (U^{ce}_{{\vec z}_{\perp}} -
    U^{ce}_{{\vec b}_{2 \perp}}) \, \frac{{\vec \epsilon}_\perp^{\,
        \lambda *} \cdot ({\vec z}_{\perp} - {\vec
        b}_{2\perp})}{|{\vec z}_{\perp} - {\vec b}_{2\perp}|^2} \, \ln
    \frac{1}{|{\vec z}_{\perp} - {\vec b}_{1\perp}| \, \Lambda} \right]  \notag \\
  & - \frac{i \, g^3}{4 \, \pi^3} \, \int d^2 x \, \left[
    U_{x_\perp}^{ab} - U_{z_\perp}^{ab} \right] \, f^{bde} \, \left(
    V_{{\vec b}_{1 \perp}} t^d \right)_1 \, \left( V_{{\vec b}_{2
        \perp}} t^e \right)_2 \notag \\ & \times \, \frac{{\vec
      \epsilon}_\perp^{\, \lambda *} \cdot ({\vec z}_{\perp} - {\vec
      x}_{\perp})}{|{\vec z}_{\perp} - {\vec x}_{\perp}|^2} \
  \frac{{\vec x}_{\perp} - {\vec b}_{1 \perp}}{|{\vec x}_{\perp} -
    {\vec b}_{1 \perp}|^2} \cdot \frac{{\vec x}_{\perp} - {\vec b}_{2
      \perp}}{|{\vec x}_{\perp} - {\vec b}_{2 \perp}|^2} \,
  \mbox{Sign} (b_2^- - b_1^-).
\end{align}
This is our final answer for the $A$, $B$ and $C$ graphs. 

Previously a calculation of a similar quantity, the order-$g^3$
classical gluon production amplitude, was carried out in
\cite{Balitsky:2004rr}. It would be desirable to compare the two
results. Unfortunately a direct comparison appears to be impossible at
the moment: first of all, the calculation in \cite{Balitsky:2004rr}
appears to have employed a different regularization of
$\alpha$-integrals from our use of the $b_2^- - b_1^-$ difference. The
result obtained in \cite{Balitsky:2004rr} does not depend on $b_2^- -
b_1^-$. In addition, the expression derived in \cite{Balitsky:2004rr}
in momentum space still contains an analogue of our $|{\vec k}_\perp
\times {\vec l}_\perp|$ in the denominator (see e.g. our \eq{eq:A2_4}
above), and is not simplified to cancel out this factor. It appears
that significant simplifications of the result in
\cite{Balitsky:2004rr} need to be performed before it could be
compared with our \eq{eq:ABCsum}. Furthermore, the calculation of
\cite{Balitsky:2004rr} appears to be done in a gauge where there are
no $B$ or $C$ graphs. Therefore, it is likely that the result of
\cite{Balitsky:2004rr} does not involve the calculational tricks of
moving the whole gluon across the cut in diagrams $B$ and $C$ detailed
above around \fig{commutators} and below in Appendix~\ref{A}. This
could be another difference between the two calculations.  Clearly
more work is needed to compare our results with those in
\cite{Balitsky:2004rr}.


\subsection{Graphs D and E}

\begin{figure}[ht]
\begin{center}
\includegraphics[width=0.45 \textwidth]{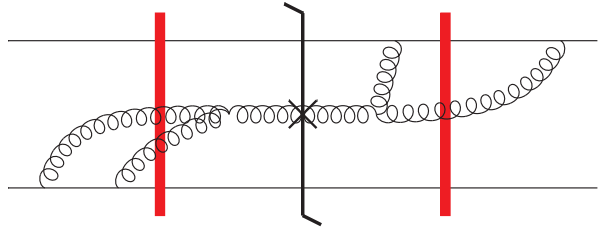} 
\caption{An example of a contribution to the amplitude squared, in
  which one of the projectile nucleons in the diagram to the left of
  the cut does not emit any $s$-channel gluons, while the other
  projectile nucleon does not emit any gluons in the complex conjugate
  amplitude.}
\label{Dsquared}
\end{center}
\end{figure}

There is a subset of type-(i) diagrams (in the classification of
\fig{xsect}) that we have not calculated yet. These are the diagrams
where one of the nucleons in the projectile is a spectator in the
amplitude, that is, it does not emit any $s$-channel gluons but may
interact with the shock wave, while another nucleon is a ``spectator''
in the complex-conjugate amplitude. An example of a diagram of this
type representing the amplitude squared is shown in \fig{Dsquared}.

\begin{figure}[ht]
\begin{center}
\includegraphics[width=0.7 \textwidth]{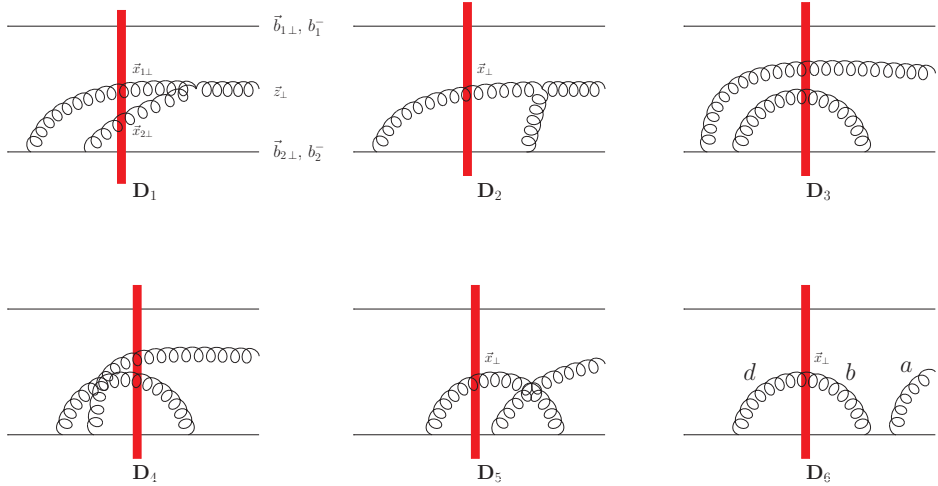} 
\caption{The ``disconnected'' diagrams, where one nucleon does not
  emit any gluons. The remaining graphs of this type are obtained by
  interchanging the quark lines in the diagrams shown, $E_i (b_1, b_2)
  = D_i (b_2, b_1)$.}
\label{Dgraphs}
\end{center}
\end{figure}

Clearly diagrams of the type shown to the left and right of the cut in
\fig{Dsquared} contribute at order-$g^3$ to the gluon production
amplitude and need to be resummed. (Since both projectile valence
quarks participate in the interaction in the amplitude squared, the
contribution in \fig{Dsquared} comes in with two powers of
$A_1^{1/3}$, just like squares of $A$, $B$ and $C$ graphs.) We will
label such diagrams as $D$-graphs: they are illustrated in
\fig{Dgraphs}. Let us stress that \fig{Dgraphs} only contains the
$D$-diagrams which are not zero due to color averaging: we have to
take a color trace (divided by $N_c$) for each projectile quark after
squaring the amplitude. Taking into account that there are no
emissions from quark $2$ on the other side of the cut implies that
many of the diagrams in this class give zero contribution to the
amplitude squared: those graphs are not shown.

There are also $E$-diagrams, which are obtained by swapping the two
projectile quarks (Wilson lines) in \fig{Dgraphs},
\begin{align}
  \label{eq:E}
  E_i (b_1, b_2; 1, 2) = D_i (b_2, b_1; 2, 1).
\end{align}

Just like for the $A$, $B$ and $C$ graphs, in summing $D$-graphs one
has to include the contribution where the gluons emitted by quark $2$
are either emitter, absorbed, or both emitted and absorbed on either
side of the cut. This again results in all gluon propagators being
retarded. In addition, as explained in Appendix~\ref{A}, the sum of
diagrams $D_5$ and $D_6$, along with the diagrams $D'_5, D''_5, D'_6,
D''_6$ where either part of or the whole gluon loop are in the complex
conjugate amplitude (see \fig{Dgraphs56}) is equal to diagram $D_6$
with the commutator of the color matrices $[t^a, t^b]$ instead of $t^a
\, t^b$. With this in mind, the values of the $D$ graphs are
\begin{subequations}\label{eq:D}
\begin{align}
  D_1 = & - i \, g^3 \, \int d^2 x_1 \, d^2 x_2 \, \int \frac{d^2
    k}{(2 \pi)^2} \, \frac{d^2 l}{(2 \pi)^2} \, \frac{d^2 q_1}{(2
    \pi)^2} \, \frac{d^2 q_2}{(2 \pi)^2} \, e^{i {\vec q}_{1\perp}
    \cdot ({\vec x}_{1\perp} - {\vec b}_{2 \perp}) + i {\vec
      q}_{2\perp} \cdot ({\vec x}_{2\perp} - {\vec b}_{2 \perp}) + i
    {\vec l}_\perp \cdot ({\vec x}_{2\perp} - {\vec x}_{1 \perp}) + i
    {\vec k}_\perp \cdot ({\vec z}_\perp - {\vec x}_{2 \perp})} \notag
  \\ & \times \frac{1}{q_{1\perp}^2 \, q_{2\perp}^2} \, \left( - {\vec
      q}_{1\perp} \cdot {\vec q}_{2\perp} \, \frac{{\vec
        \epsilon}_\perp^{\, \lambda *} \times {\vec
        k}_\perp}{k_\perp^2} + {\vec \epsilon}_\perp^{\, \lambda *}
    \cdot {\vec q}_{1\perp} \, \frac{{\vec q}_{2\perp} \times ({\vec
        k}_\perp - {\vec l}_\perp)}{({\vec k}_\perp - {\vec
        l}_\perp)^2} + {\vec \epsilon}_\perp^{\, \lambda *} \cdot
    {\vec q}_{2\perp} \, \frac{{\vec q}_{1\perp} \times {\vec
        l}_\perp}{{l}_\perp^2} \right) \, \mbox{Sign} ({\vec k}_\perp
  \times {\vec l}_\perp) \notag \\ & \times \, f^{abc} \,
  U^{bd}_{{\vec x}_{1\perp}} \, U^{ce}_{{\vec x}_{2\perp}} \, \left(
    V_{{\vec b}_{1 \perp}} \right)_1 \, \left( V_{{\vec
        b}_{2 \perp}} t^e \, t^d \right)_2 , \label{D1} \\
  D_2 = & - 2 \, i \, g^3 \, \int d^2 x \, \int \frac{d^2 k}{(2
    \pi)^2} \, \frac{d^2 q}{(2 \pi)^2} \, \frac{d^2 l}{(2 \pi)^2} \,
  e^{i {\vec q}_\perp \cdot ({\vec x}_{\perp} - {\vec b}_{2 \perp}) +
    i {\vec l}_\perp \cdot ({\vec x}_\perp - {\vec b}_{2 \perp}) + i
    {\vec k}_\perp \cdot ({\vec z}_\perp - {\vec x}_{\perp})} \,
  f^{abc} \, U^{be}_{{\vec b}_{2\perp}} \, U^{cd}_{{\vec x}_{\perp}}
  \, \left( V_{{\vec b}_{1 \perp}} \right)_1 \, \left( V_{{\vec b}_{2
        \perp}} t^e \, t^d \right)_2 \notag \\ & \times
  \frac{1}{l_{\perp}^2 \, q_\perp^2} \, \Bigg[ \left( \frac{{\vec
        \epsilon}_\perp^{\, \lambda *} \times {\vec
        k}_\perp}{k_\perp^2} \, {\vec l}_\perp \cdot {\vec q}_\perp -
    {\vec \epsilon}_\perp^{\, \lambda *} \cdot {\vec l}_\perp \,
    \frac{{\vec q}_\perp \times ({\vec k}_\perp - {\vec
        l}_\perp)}{({\vec k}_\perp - {\vec l}_\perp)^2} \right) \,
  \mbox{Sign} ({\vec k}_\perp \times {\vec l}_\perp) \notag \\ & - i
  \left( \frac{{\vec \epsilon}_\perp^{\, \lambda *} \cdot {\vec
        k}_\perp}{k_\perp^2} \, {\vec l}_\perp \cdot {\vec q}_\perp -
    {\vec \epsilon}_\perp^{\, \lambda *} \cdot {\vec l}_\perp \,
    \frac{({\vec k}_\perp - {\vec l}_\perp) \cdot {\vec
        q}_\perp}{({\vec k}_\perp - {\vec l}_\perp)^2} - \frac{1}{2}
    \, {\vec \epsilon}_\perp^{\, \lambda *} \cdot {\vec q}_\perp
  \right) \Bigg], \\ D_3 = & \, 0, \\ D_4 = & \, 0, \\ D_5 + & D_6 +
  D'_5 + D'_6 + D''_5 + D''_6 = 2 \, g^3 \, \int d^2 x \, \int
  \frac{d^2 k}{(2 \pi)^2} \, \frac{d^2 q_1}{(2 \pi)^2} \, \frac{d^2
    q_2}{(2 \pi)^2} \, e^{i {\vec q}_{1\perp} \cdot ({\vec b}_{2\perp}
    - {\vec x}_\perp) + i {\vec q}_{2\perp} \cdot ({\vec x}_\perp -
    {\vec b}_{2\perp}) + i {\vec k}_\perp \cdot ({\vec z}_\perp -
    {\vec b}_{2 \perp})} \notag \\ & \times \, \frac{{\vec
      \epsilon}_\perp^{\, \lambda *} \cdot {\vec k}_\perp}{k_\perp^2}
  \, \frac{{\vec q}_{1\perp} \cdot {\vec q}_{2\perp}}{q_{1\perp}^2 \,
    q_{2\perp}^2} \, f^{abc} \, U^{bd}_{{\vec x}_{\perp}} \,
  U^{ce}_{{\vec b}_{2\perp}} \, \left( V_{{\vec b}_{1 \perp}}
  \right)_1 \, \left( V_{{\vec b}_{2 \perp}} t^e \, t^d \right)_2 .
\end{align}
\end{subequations}
Note also that color-averaging is implied: for instance, the part of
the diagram $D_1$ contribution shown in \eq{D1} is the one that
survives a color-trace in the space of nucleon $2$ after the amplitude
is squared. More specifically when writing $(t^e t^d)_2$ in
Eqs.~\eqref{eq:D} we only keep the part of the expression which
survives the $(t^e t^d)_2 \to \delta^{ed}/(2 N_c)$ substitution.

The sum of all the $D$-graphs (including the contributions from moving
the gluons across the cut, as identified by the prime over the sum
sign) is
\begin{align}
  \label{eq:Dall2}
  \sideset{}{'} \sum_{i=1}^6 D_i = & - i \, g^3 \, \int d^2 x_1 \, d^2
  x_2 \, \int \frac{d^2 k}{(2 \pi)^2} \, \frac{d^2 l}{(2 \pi)^2} \,
  \frac{d^2 q_1}{(2 \pi)^2} \, \frac{d^2 q_2}{(2 \pi)^2} \, e^{i {\vec
      q}_{1\perp} \cdot ({\vec x}_{1\perp} - {\vec b}_{2 \perp}) + i
    {\vec q}_{2\perp} \cdot ({\vec x}_{2\perp} - {\vec b}_{2 \perp}) +
    i {\vec l}_\perp \cdot ({\vec x}_{2\perp} - {\vec x}_{1 \perp}) +
    i {\vec k}_\perp \cdot ({\vec z}_\perp - {\vec x}_{2 \perp})}
  \notag \\ & \times \frac{1}{q_{1\perp}^2 \, q_{2\perp}^2} \, \left(
    - {\vec q}_{1\perp} \cdot {\vec q}_{2\perp} \, \frac{{\vec
        \epsilon}_\perp^{\, \lambda *} \times {\vec
        k}_\perp}{k_\perp^2} + {\vec \epsilon}_\perp^{\, \lambda *}
    \cdot {\vec q}_{1\perp} \, \frac{{\vec q}_{2\perp} \times ({\vec
        k}_\perp - {\vec l}_\perp)}{({\vec k}_\perp - {\vec
        l}_\perp)^2} + {\vec \epsilon}_\perp^{\, \lambda *} \cdot
    {\vec q}_{2\perp} \, \frac{{\vec q}_{1\perp} \times {\vec
        l}_\perp}{{l}_\perp^2} \right) \, \mbox{Sign} ({\vec k}_\perp
  \times {\vec l}_\perp) \notag \\ & \times \, f^{abc} \, \left[
    U^{bd}_{{\vec x}_{1\perp}} - U^{bd}_{{\vec b}_{2\perp}} \right] \,
  \left[ U^{ce}_{{\vec x}_{2\perp}} - U^{ce}_{{\vec b}_{2\perp}}
  \right]\, \left( V_{{\vec b}_{1 \perp}} \right)_1 \, \left( V_{{\vec
        b}_{2 \perp}} t^e \, t^d \right)_2 - 2 \, g^3 \, \int d^2 x \,
  \int \frac{d^2 k}{(2 \pi)^2} \, \frac{d^2 q}{(2 \pi)^2} \, \frac{d^2
    l}{(2 \pi)^2} \notag \\ & \times \, e^{i {\vec q}_\perp \cdot
    ({\vec x}_{\perp} - {\vec b}_{2 \perp}) + i {\vec l}_\perp \cdot
    ({\vec x}_\perp - {\vec b}_{2 \perp}) + i {\vec k}_\perp \cdot
    ({\vec z}_\perp - {\vec x}_{\perp})} \, f^{abc} \, U^{bd}_{{\vec
      b}_{2\perp}} \, \left[ U^{ce}_{{\vec x}_{\perp}} - U^{ce}_{{\vec
        b}_{2\perp}} \right] \, \left( V_{{\vec b}_{1 \perp}}
  \right)_1 \, \left( V_{{\vec b}_{2 \perp}} t^e \, t^d \right)_2
  \notag \\ & \times \, \left( \frac{{\vec \epsilon}_\perp^{\, \lambda
        *} \cdot {\vec k}_\perp}{k_\perp^2} \, \frac{{\vec l}_\perp
      \cdot {\vec q}_\perp}{l_\perp^2 \, q_\perp^2} - \frac{{\vec
        \epsilon}_\perp^{\, \lambda *} \cdot {\vec
        l}_\perp}{l_\perp^2} \, \frac{({\vec k}_\perp - {\vec
        l}_\perp) \cdot {\vec q}_\perp}{({\vec k}_\perp - {\vec
        l}_\perp)^2 \, q_\perp^2} - \frac{{\vec \epsilon}_\perp^{\,
        \lambda *} \cdot {\vec q}_\perp}{2 \, l_\perp^2 \, q_\perp^2}
    + \frac{{\vec \epsilon}_\perp^{\, \lambda *} \cdot {\vec
        k}_\perp}{k_\perp^2} \, \frac{({\vec k}_\perp - {\vec
        l}_\perp) \cdot {\vec q}_\perp}{({\vec k}_\perp - {\vec
        l}_\perp)^2 \, q_\perp^2} \right).
\end{align}
This result, along with the sum of all the $E$ graphs, may also be
directly obtained from \eq{eq:ABCsum} by ``moving'' the quark-gluon
vertices on the line of quark $1$ to the line of quark $2$ in $A$, $B$
and $C$ graphs with an appropriate modification of color factors.

Fourier-transforming \eq{eq:Dall2} we obtain
\begin{align}
  \label{eq:Dall_coord}
  \sideset{}{'} \sum_{i=1}^6 D_i = & - \frac{g^3}{8 \, \pi^4} \, \int
  d^2 x_1 \, d^2 x_2 \, \delta [({\vec z}_{\perp} - {\vec x}_{1\perp})
  \times ({\vec z}_{\perp} - {\vec x}_{2\perp})] \left[ \frac{{\vec
        \epsilon}_\perp^{\, \lambda *} \cdot ({\vec x}_{2\perp} -
      {\vec x}_{1\perp}) }{|{\vec x}_{2\perp} - {\vec x}_{1\perp}|^2}
    \, \frac{{\vec x}_{1\perp} - {\vec b}_{2\perp}}{|{\vec x}_{1\perp}
      - {\vec b}_{2\perp}|^2} \cdot \frac{{\vec x}_{2\perp} - {\vec
        b}_{2\perp}}{|{\vec x}_{2\perp} - {\vec b}_{2\perp}|^2}
  \right. \notag \\ & \left. - \frac{{\vec \epsilon}_\perp^{\, \lambda
        *} \cdot ({\vec x}_{1\perp} - {\vec b}_{2\perp}) }{|{\vec
        x}_{1\perp} - {\vec b}_{2\perp}|^2} \, \frac{{\vec z}_{\perp}
      - {\vec x}_{1\perp}}{|{\vec z}_{\perp} - {\vec x}_{1\perp}|^2}
    \cdot \frac{{\vec x}_{2\perp} - {\vec b}_{2\perp}}{|{\vec
        x}_{2\perp} - {\vec b}_{2\perp}|^2} + \frac{{\vec
        \epsilon}_\perp^{\, \lambda *} \cdot ({\vec x}_{2\perp} -
      {\vec b}_{2\perp}) }{|{\vec x}_{2\perp} - {\vec b}_{2\perp}|^2}
    \, \frac{{\vec x}_{1\perp} - {\vec b}_{2\perp}}{|{\vec x}_{1\perp}
      - {\vec b}_{2\perp}|^2} \cdot \frac{{\vec z}_{\perp} - {\vec
        x}_{2\perp}}{|{\vec z}_{\perp} - {\vec x}_{2\perp}|^2} \right]
  \notag \\ & \times \, f^{abc} \, \left[ U^{bd}_{{\vec x}_{1\perp}} -
    U^{bd}_{{\vec b}_{2\perp}} \right] \, \left[ U^{ce}_{{\vec
        x}_{2\perp}} - U^{ce}_{{\vec b}_{2\perp}} \right]\, \left(
    V_{{\vec b}_{1 \perp}} \right)_1 \, \left( V_{{\vec b}_{2 \perp}}
    t^e \, t^d \right)_2 \notag \\ & + \frac{i \, g^3}{4 \, \pi^3} \,
  \int d^2 x \, f^{abc} \, U^{bd}_{{\vec b}_{2\perp}} \, \left[
    U^{ce}_{{\vec x}_{\perp}} - U^{ce}_{{\vec b}_{2\perp}} \right] \,
  \left( V_{{\vec b}_{1 \perp}} \right)_1 \, \left( V_{{\vec b}_{2
        \perp}} t^e \, t^d \right)_2 \, \left( \frac{{\vec
        \epsilon}_\perp^{\, \lambda *} \cdot ({\vec z}_{\perp} - {\vec
        x}_{\perp}) }{|{\vec z}_{\perp} - {\vec x}_{\perp}|^2} \,
    \frac{1}{|{\vec x}_{\perp} - {\vec b}_{2\perp}|^2} \right. \notag
  \\ & \left. - \frac{{\vec \epsilon}_\perp^{\, \lambda *} \cdot
      ({\vec z}_{\perp} - {\vec b}_{2\perp}) }{|{\vec z}_{\perp} -
      {\vec b}_{2\perp}|^2} \, \frac{{\vec z}_{\perp} - {\vec
        x}_{\perp}}{|{\vec z}_{\perp} - {\vec x}_{\perp}|^2} \cdot
    \frac{{\vec x}_{\perp} - {\vec b}_{2\perp}}{|{\vec x}_{\perp} -
      {\vec b}_{2\perp}|^2} - \frac{{\vec \epsilon}_\perp^{\, \lambda
        *} \cdot ({\vec z}_{\perp} - {\vec b}_{2\perp}) }{|{\vec
        z}_{\perp} - {\vec b}_{2\perp}|^2} \, \frac{1}{|{\vec
        x}_{\perp} - {\vec b}_{2\perp}|^2} \right) \notag \\ & +
  \frac{i \, g^3}{4 \, \pi^2} \, f^{abc} \, U^{bd}_{{\vec b}_{2\perp}}
  \, \left[ U^{ce}_{{\vec z}_{\perp}} - U^{ce}_{{\vec b}_{2\perp}}
  \right] \, \left( V_{{\vec b}_{1 \perp}} \right)_1 \, \left(
    V_{{\vec b}_{2 \perp}} t^e \, t^d \right)_2 \, \frac{{\vec
      \epsilon}_\perp^{\, \lambda *} \cdot ({\vec z}_{\perp} - {\vec
      b}_{2\perp}) }{|{\vec z}_{\perp} - {\vec b}_{2\perp}|^2} \, \ln
  \frac{1}{|{\vec z}_{\perp} - {\vec b}_{2\perp}| \, \Lambda}.
\end{align}
This is our final result for the sum of the $D$ graphs. Using
\eq{eq:E} one can easily obtain the sum of the $E$ graphs as well from
\eq{eq:Dall_coord}. Again this result can also be obtained directly
from \eq{eq:ABCsum_coord}.


\subsection{Cross-checks}

To test our main results for the order-$g^3$ gluon production
amplitude in Eqs.~\eqref{eq:ABCsum_coord} and \eqref{eq:Dall_coord}
let us run a couple of cross checks.

First of all, if there is no shock wave, there should be no gluon
production. A simple calculation shows that if one puts all $U=1$ and
all $V=1$ then
\begin{align}
  \label{eq:no_int}
  \sum\limits_{i=1}^{7} A_i = 0 , \ \ \ \sideset{}{'}
  \sum\limits_{i=1}^{12} B_i = 0, \ \ \ \sideset{}{'}
  \sum\limits_{i=1}^{12} C_i =0, \ \ \ \sideset{}{'}
  \sum\limits_{i=1}^{6} D_i = 0, \ \ \ \sideset{}{'}
  \sum\limits_{i=1}^{6} E_i =0,
\end{align}
as expected.

Another issue is gauge invariance. To test our results for gauge
invariance, the order-$g^3$ gluon production amplitude can be
calculated in different sub-gauges of the light-cone gauge, other than
the PV sub-gauge in which the results in Eqs.~\eqref{eq:ABCsum_coord}
and \eqref{eq:Dall_coord} were obtained. In Appendix~\ref{B} we show
that, while the calculations of the order-$g^3$ gluon production
amplitude in the ${\vec A}_\perp (x^- \to + \infty) = 0$ or ${\vec
  A}_\perp (x^- \to - \infty) = 0$ sub-gauges are more involved than
the PV sub-gauge calculation, the end result is the same in all three
gauges for the unprimed sum of all $A$, $B$ and $C$ graphs, and for
the unprimed sum of $D$ and $E$ graphs. Gauge-invariance also holds if
we use retarded gluon ``propagators''. Note that obtaining commutators
of fundamental SU($N_c$) generators for the $B$ and $C$ graphs by
moving gluons across the cut was shown to be a legitimate trick only
in the PV sub-gauge of the light-cone gauge: it is likely that to
re-derive the result from \eq{eq:ABCsum_coord} in the ${\vec A}_\perp
(x^- \to + \infty) = 0$ and ${\vec A}_\perp (x^- \to - \infty) = 0$
sub-gauges of the light-cone gauge, the gluon should also be moved
across the cut in the diagrams like the first two graphs in
\fig{shock_graphs}, complicating the analysis. Hence,
strictly-speaking, we have not shown that our result in
\eq{eq:ABCsum_coord} (with primed sums) is gauge-invariant: while it
may still be gauge-invariant, at the moment we can think of it as a
gauge-invariant amplitude (resulting from the unprimed sums) with the
terms added (in the PV sub-gauge) casting it in a simpler form for
obtaining the gluon production cross section.


\section{Outlook}
\label{sec:outlook}

This paper is the first step in our project to calculate the first
saturation correction in the projectile wave function to the $p+A$
classical gluon production cross section. In this paper we calculated
the order-$g^3$ gluon production amplitude. The main results are given
in transverse coordinate space in Eqs.~\eqref{eq:ABCsum_coord} and
\eqref{eq:Dall_coord}. Transverse momentum space amplitude is given in
Eqs.~\eqref{eq:ABCsum} and \eqref{eq:Dall2}. These results constitute
the first saturation correction to the leading-order gluon production
amplitude in \eq{eq:1gluon} and in \eq{eq:1gluon_coord} respectively.

To complete this project and find the order-$\as^3$ contribution to
the classical gluon production cross section one needs to calculate
the order-$g^5$ amplitude, again involving only two projectile
nucleons (see \fig{xsect}). Similar to the $D$-graphs above, in
calculating the order-$g^5$ amplitude we will be able to simplify the
color algebra by employing the fact that only one of the two nucleons
involved emits a gluon in the complex conjugate amplitude, as shown in
the diagram (ii) of \fig{xsect}. Still, this appears to be a rather
challenging calculation, and it is presently left for future work.


\section*{Acknowledgments}

The authors are grateful to Ian Balitsky for encouraging
discussions. This material is based upon work supported by the
U.S. Department of Energy, Office of Science, Office of Nuclear
Physics under Award Number DE-SC0004286. \\


\appendix

\section{Moving gluons across the cut}
\renewcommand{\theequation}{A\arabic{equation}}
  \setcounter{equation}{0}
\label{A}

The purpose of this Appendix is to show that (a) the inclusion of the
type-(iii) diagrams from \fig{xsect} into the calculation can be
easily accomplished by using retarded gluon Green functions instead of
Feynman propagators; and that (b) the contributions of $B$ and $C$
graphs (shown in \fig{Bgraphs}) to the ${\cal O} (g^3)$ amplitude
along with the contributions of similar diagrams to the ${\cal O}
(g^5)$ amplitude can all be efficiently found by calculating a subset
of those graphs with color-commutators instead of the usual
fundamental color factors. While these statements appear unrelated,
they seem to be hard to disentangle for the $B$ and $C$ types of
graphs.

\begin{figure}[ht]
\begin{center}
\includegraphics[width=0.9 \textwidth]{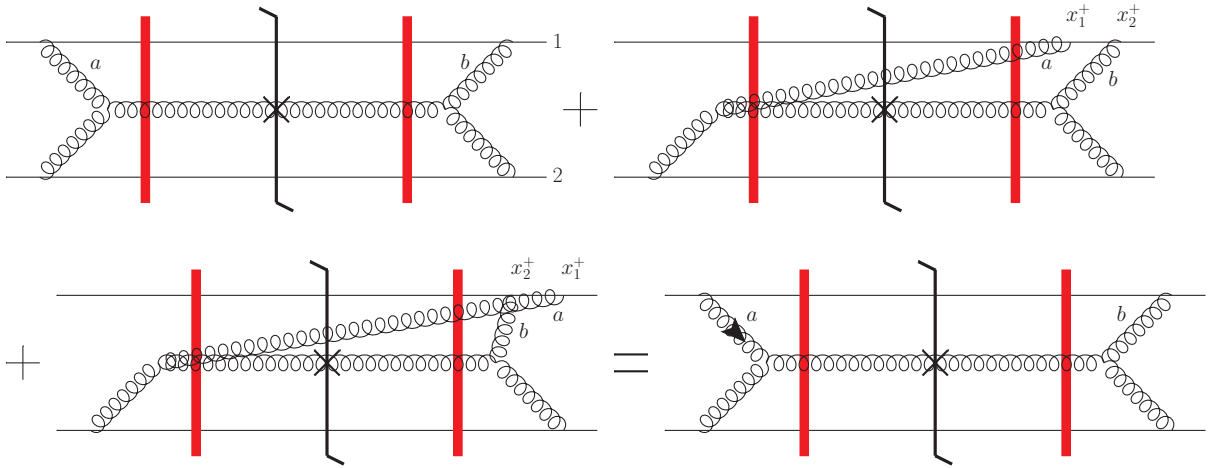} 
\caption{Diagrams contributing to the retarded Green function for the
  gluon propagator in the square of the $A_1$ graph. The retarded
  gluon Green function is labeled by the arrow in the last diagram
  representing the net result.}
\label{retardation1}
\end{center}
\end{figure}

Let us begin with contributions to the cross section coming from the
squares of the $A$-graphs (defined in \fig{Agraphs}). It is
straightforward to see by an explicit diagram analysis that the
argument depicted in \fig{retardation} applies in general, for the
amplitude and the complex conjugate amplitude contributions coming
from any two $A$-graphs. We illustrate this point in
\fig{retardation1} by considering a slightly more involved example
than that shown in \fig{retardation}, namely by studying the square of
diagram $A_1$. The argument of \eq{eq:retarded} applies here as well,
after the following two observations. First of all, the interactions
of the gluon that we moved across the cut in the second and third
diagrams (to the left of the equal sign) of \fig{retardation1} with
the shock wave cancel, as the gluon has the same transverse coordinate
on both sides of the cut such that $U \, U^\dagger = 1$: hence we can
treat this gluon as a free gluon propagator in the first graph, and as
a free cut propagator in the second and third graphs. Second of all,
in all three diagrams to the left of the equal sign of
\fig{retardation1} the interactions of quark $1$ with the shock wave
cancel (for a similar reason to the gluon's, $V \, V^\dagger = 1$),
leading to the same color factor on this quark line, tr~$[t^a \, t^b]
= \delta^{ab}/2$, for all three graphs. Since the longitudinal
coordinate integral ranges of the second and third graphs in
\fig{retardation1}, $x_1^+ > x_2^+$ and $x_2^+ > x_1^+$, complement
each other giving independent $x_1^+$ and $x_2^+$ integrals from
$-\infty$ to $0$ each, we see that the sum of the three diagrams in
\fig{retardation1} leads to a retarded gluon Green function denoted by
an arrow in the last diagram in \fig{retardation1}, just like it was
demonstrated in \eq{eq:retarded}.

Other contributions to the amplitude squared made out of $A$ graphs
(both in the amplitude and in the complex conjugate amplitude) can be
analyzed in a similar manner on a case-by-case basis, leading to the
same result: all internal gluon lines should be represented by
retarded Green functions. Note that to prove the use of retarded
propagators for the gluon lines in the diagrams like $A_2$ one has to
also employ the fact that the ``$+$'' component of the gluon momentum
is conserved in the interaction with the shock wave.

Now we have to analyze the contributions to the amplitude squared
coming from $B$ and $C$ graphs, along with the interference terms
between the $A$ and $B$ or $C$ diagrams. The main tool here, in
addition to \eq{eq:retarded}, would be the cancellation of retarded
Green functions from \fig{cancellation} (for the gluon exchanged
between the quarks either before or after the shock wave both in the
amplitude and the complex conjugate amplitude). Indeed one can proceed
with the diagrammatic arguments, similar to how we outlined the
analysis for the square of the sum of $A$ diagrams above. While we
performed this diagram-by-diagram check, it is rather lengthy and we
are not going to present it here due to a very high number of diagrams
one has to consider. Instead, a somewhat more compact approach to the
problem is based on a formal argument which we present below.

Following \cite{Balitsky:2004rr} we denote gluon fields in the
amplitude by $A_\mu^a$, while the gluon fields in the complex
conjugate amplitude are labeled by ${\cal A}_\mu^a$. These fields are
not the background classical fields of the target, but rather
``quantum'' fields whose Wick contractions give us the $s$-channel
gluon propagators used in calculating diagrams $A$, $B$, $C$, etc. By
analogy to \eqref{eq:Wfund} we can define the Wilson line representing
quark propagation over a finite $x^+$-interval in the amplitude by
\begin{align}
  \label{eq:Wfund_finite}
  {V_{{\vec b}_\perp, b^-} [x_2^+, x_1^+]}_A = \mbox{P} \exp \left\{ i
    \, g \, \int\limits_{x_1^+}^{x_2^+} d x^+ t^a \, A^{a \, -} (x^+,
    b^-, {\vec b}_\perp) \right\}.
\end{align}
The corresponding Wilson line in the complex conjugate amplitude would
be ${V^\dagger_{{\vec b}_\perp, b^-} [x_2^+, x_1^+]}_{\cal A}$, where
${\cal A}$ in the subscript denotes the fact that now one uses the
field ${\cal A}_\mu^a$ in the exponent of \eq{eq:Wfund_finite}.

Further we define gluon propagators as contractions
\begin{subequations}
\begin{align}
  \label{eq:props}
  & \contraction{}{A}{_\mu^a (x) \, }{A} A_\mu^a (x) \, A_\nu^b (y)
  \equiv D^{ab}_{F \, \mu\nu} (x-y) = \langle 0 | \mbox{T} A_\mu^a (x)
  \, A_\nu^b (y) | 0 \rangle = \int \frac{d^4 k}{(2 \pi)^4} \, e^{- i
    k \cdot (x-y)} \, \frac{- i \, \delta^{ab} \, D_{\mu\nu} (k)}{k^2
    + i \epsilon}  \\
  & \contraction{}{{\cal A}}{_\mu^a (x) \, }{{\cal A}} {\cal A}_\mu^a
  (x) \, {\cal A}_\nu^b (y) \equiv {\cal D}^{ab}_{F \, \mu\nu} (x-y) =
  \langle 0 | \overline{\mbox{T}} {\cal A}_\mu^a (x) \, {\cal A}_\nu^b
  (y) | 0 \rangle =
  \left[ D^{ab}_{F \, \mu\nu} (x-y) \right]^* \\
  & \contraction{}{A}{_\mu^a (x) \, }{{\cal A}} A_\mu^a (x) \, {\cal
    A}_\nu^b (y) \equiv D^{ab}_{\mu\nu} (x-y) = \langle 0 | A_\mu^a
  (x) \, {\cal A}_\nu^b (y) | 0 \rangle = - \int \frac{d^4 k}{(2
    \pi)^4} \, e^{- i k \cdot (x-y)} \, \delta^{ab} \, D_{\mu\nu} (k)
  \, (2 \pi) \, \delta (k^2) \, \theta (-k^+) \\
  & \contraction{}{{\cal A}}{_\mu^a (x) \, }{A} {\cal A}_\mu^a (x) \,
  A_\nu^b (y) = D^{ab}_{\mu\nu} (y-x) = \langle 0 | {\cal A}_\mu^a (x)
  \, A_\nu^b (y) | 0 \rangle ,
\end{align}
\end{subequations}
where $D_{\mu\nu} (k)$ is given by \eq{eq:DmunuPV} while T and
$\overline{\mbox{T}}$ denote time- and anti-time-ordering
respectively. (As in the calculation leading to Eqs.~\eqref{eq:ABCsum}
and \eqref{eq:Dall2} we will use the PV sub-gauge of the light-cone
gauge.) Since only the upper ``$-$'' components of the gluon fields
$A_\mu^a$ and ${\cal A}_\mu^a$ enter the Wilson lines such as
\eqref{eq:Wfund_finite}, we will suppress the Lorentz indices
below. That is, we will use
\begin{align}
  \label{eq:suppr}
  D^{ab}_F (x-y) \equiv D^{ab \, - - }_{F} (x-y), \ \ \ {\cal
    D}^{ab}_F (x-y) \equiv {\cal D}^{ab \, - - }_{F} (x-y), \ \ \
  D^{ab} (x-y) \equiv D^{ab \, - - } (x-y).
\end{align}
Note that (cf. \eq{eq:retarded})
\begin{subequations}\label{eq:ret}
\begin{align}
  D^{ab}_F (x-y) - D^{ab} (x-y) & = D^{ab}_{ret} (x-y) \equiv \int
  \frac{d^4 k}{(2 \pi)^4} \, e^{- i k \cdot (x-y)} \, \frac{- i \,
    \delta^{ab} \, D^{--} (k)}{k^2
    + i \epsilon k^+}, \label{eq:ret1} \\
  {\cal D}^{ab}_F (x-y) - D^{ab} (y-x) & = \int \frac{d^4 k}{(2
    \pi)^4} \, e^{i k \cdot (x-y)} \, \frac{i \, \delta^{ab} \, D^{--}
    (k)}{k^2 - i \epsilon k^+} = - D^{ab}_{ret} (x-y), \label{eq:ret2}
\end{align}
\end{subequations}
with $D^{ab}_{ret} (x-y)$ denoting the retarded gluon Green function.
Indeed we are setting up the well-known Keldysh--Schwinger formalism
\cite{Keldysh:1964ud,Schwinger:1960qe}, which was used in small-$x$
physics in \cite{Balitsky:2004rr,Gelis:2008rw,Mueller:2012bn}.

Eqs.~\eqref{eq:ret} is a formal expression behind the cancellations
like those shown in \fig{cancellation}. To see this we concentrate on
the Wilson lines representing quark propagators before the shock wave,
both in the amplitude and in the complex conjugate amplitude. (For the
cancellation it does not matter which interactions take place after
the scattering in the shock wave.) The contribution of the diagrams in
\fig{cancellation} is proportional to
\begin{align}
  \label{eq:Vlines}
  \langle 0 | \mbox{T}_A \overline{\mbox{T}}_{\cal A} \left\{
    {V^\dagger_{{\vec b}_{1\perp}, b_1^-} [0, -\infty]}_{\cal A} \,
    {V_{{\vec b}_{1\perp}, b_1^-} [0, -\infty]}_{A} \otimes
    {V^\dagger_{{\vec b}_{2\perp}, b_2^-} [0, -\infty]}_{\cal A} \,
    {V_{{\vec b}_{2\perp}, b_2^-} [0, -\infty]}_{A} \right\} | 0
  \rangle,
\end{align}
where the $\otimes$ sign indicates that the color indices of each
product $V^\dagger \, V$ are not contracted with each other, and are
fixed at some arbitrary values. Time-ordering for $A_\mu^a$-fields and
anti-time-ordering for ${\cal A}_\mu^a$-fields are denoted by T$_A$
and $\overline{\mbox{T}}_{\cal A}$ correspondingly. For the classical
gluon field diagrams we need contractions between the gluon fields
$A_\mu^a$ and ${\cal A}_\mu^a$ connecting the two projectile quarks:
that is, we need contractions of the fields at $({\vec b}_{1\perp},
b_1^-)$ and the fields at $({\vec b}_{2\perp}, b_2^-)$. Indeed the
diagrams in \fig{cancellation} are examples of such
contractions. Expanding \eq{eq:Vlines} to the first order in the
fields in each $V^\dagger \, V$, and analyzing only the
$\contraction{}{b}{_1 \, }{b} b_1 \, b_2$ contractions, we get
\begin{align}
  \label{eq:contr1}
  & - g^2 \, \langle 0 | \int\limits_{-\infty}^0 d x^+ \, t^a \,
  \left[ A^{a \, -} (x^+, b_1^-, {\vec b}_{1\perp}) - {\cal A}^{a \,
      -} (x^+, b_1^-, {\vec b}_{1\perp}) \right] \otimes
  \int\limits_{-\infty}^0 d y^+ \, t^b \, \left[ A^{b \, -} (y^+,
    b_2^-, {\vec b}_{2\perp}) - {\cal A}^{b \, -} (y^+, b_2^-, {\vec
      b}_{2\perp}) \right] | 0 \rangle \notag \\ & = - g^2 \, t^a
  \otimes t^b \, \int\limits_{-\infty}^0 d x^+ \, dy^+ \, \left[
    D_F^{ab} (x-y) + {\cal D}_F^{ab} (x-y) - D^{ab} (x-y) - D^{ab}
    (y-x) \right] \notag \\ & = - g^2 \, t^a \otimes t^b \,
  \int\limits_{-\infty}^0 d x^+ \, dy^+ \, \left[ D_{ret}^{ab} (x-y) -
    D_{ret}^{ab} (x-y) \right] =0.
\end{align}
This is the formal proof of the cancellation shown in
\fig{cancellation}. Note that before canceling, all the correlators
assembled into the retarded Green functions. We have also employed an
abbreviated notation by suppressing $({\vec b}_{1\perp}, b_1^-)$ and
$({\vec b}_{2\perp}, b_2^-)$ in the arguments of the correlators.

A somewhat more involved calculation demonstrates that expanding
\eq{eq:Vlines} to the second order in the fields in each $V^\dagger \,
V$ and concentrating on the $\contraction{}{b}{_1 \, }{b} b_1 \, b_2$
contractions again, we would also get zero. This means that adding an
extra gluon exchanged between the projectile quarks on either side of
either diagram of \fig{cancellation} (as long as the gluon is both
emitted and absorbed before the scattering in the shock wave) would
still generate canceling diagrams. Two-gluon exchange is the highest
order necessary in the classical field analysis
\cite{Kovchegov:1997pc}.

\begin{figure}[ht]
\begin{center}
\includegraphics[width=0.9 \textwidth]{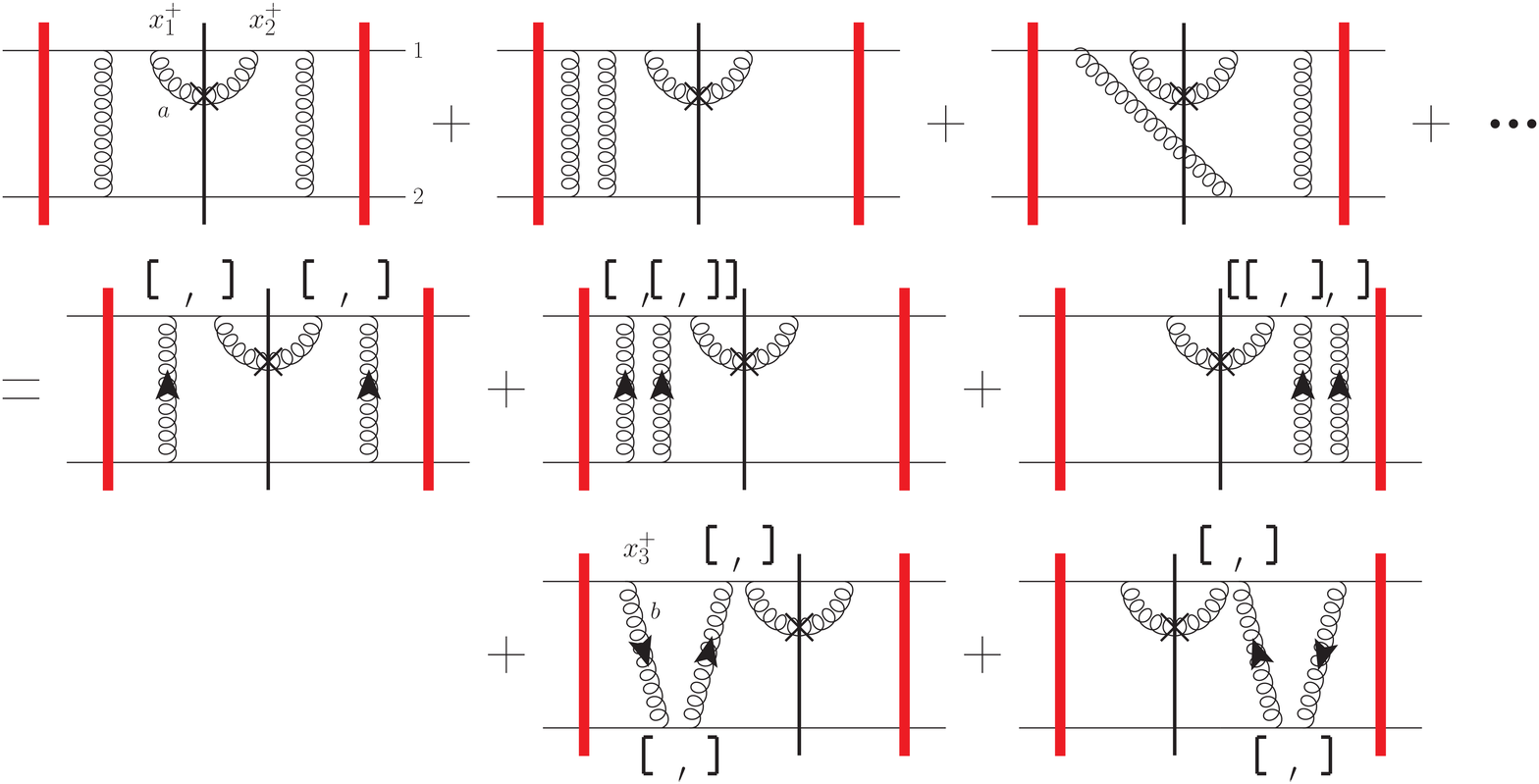} 
\caption{The sum of the diagrams complementing the square of the graph
  $B_9$ and $B_{10}$. The arrows on the gluon lines denote the
  time-flow direction for the retarded propagators, while the brackets
  denote the color matrix commutator. Both time-orderings of the
  quark-gluon vertices on the Wilson line representing quark $2$ are
  implied in the last two diagrams diagrams in the second line. The
  first gluon exchanged between the projectile quarks in the first
  diagram of the last line has the emission time $x_3^+ \in [0,
  +\infty]$ with the color matrix $t^b$ placed to the left of the
  commutator in the amplitude (with similar implication for the last
  diagram).}
\label{B9_org}
\end{center}
\end{figure}

Let us apply the Keldysh-Schwinger formalism described above to the
analysis of the squares of $B$ diagrams. We will work out the square
of the diagrams $B_9 + B_{10}$ (as defined in \fig{Bgraphs}), along
with all the other similar graphs obtained from the square of $B_9 +
B_{10}$ by moving gluons across the cut: contributions of other
diagrams can be found by analogy. A sample of diagrams we need to sum
up is given in the first line of \fig{B9_org}. Namely, we fix the
produced gluon to be emitted at light-cone time $x_1^+$ in the
amplitude and at $x_2^+$ in the complex conjugate amplitude, and sum
up all the graphs where the other two gluons are exchanged between the
two projectile quarks, as long as the gluons are emitted and/or
absorbed after the interaction with the shock wave in the amplitude
and in the complex conjugate amplitude (that is, for $x^+
>0$). Clearly the sample diagrams in the first line of \fig{B9_org}
represent only a small subset of all the graphs that need to be
included.

Instead of summing the diagrams we notice that due to the lack of
gluon emissions and/or absorptions at $x^+ < 0$, the interactions of
the projectile quarks with the shock wave cancel. Hence the
contribution of all the $B_9 + B_{10}$ squared type of diagrams can be
written as proportional to
\begin{align}
  \label{eq:B910sum}
  \langle 0 | \mbox{T}_A \overline{\mbox{T}}_{\cal A} \left\{
    \mbox{tr} \left[ {V^\dagger_{{\vec b}_{1\perp}, b_1^-} [x_2^+ ,
        0]}_{\cal A} \, t^a \, {V^\dagger_{{\vec b}_{1\perp}, b_1^-}
        [+\infty, x_2^+]}_{\cal A} \, {V_{{\vec b}_{1\perp}, b_1^-}
        [+\infty, x_1^+]}_{A} \, t^a \, {V_{{\vec b}_{1\perp}, b_1^-}
        [x_1^+ , 0]}_{A} \right] \right. \notag \\ \times \left. \,
    \mbox{tr} \left[ {V^\dagger_{{\vec b}_{2\perp}, b_2^-} [+\infty,
        0]}_{\cal A} \, {V_{{\vec b}_{2\perp}, b_2^-} [+\infty,
        0]}_{A} \right] \right\} | 0 \rangle
\end{align}
with the expression \eqref{eq:B910sum} containing all the gluon
exchanges between the projectile quarks.

First let us rewrite \eq{eq:B910sum} as
\begin{align}
  \label{eq:B910sum2}
  \langle 0 | \mbox{T}_A \overline{\mbox{T}}_{\cal A} & \left\{
    \mbox{tr} \left[ {V^\dagger_{{\vec b}_{1\perp}, b_1^-} [x_2^+ ,
        0]}_{\cal A} \, t^a \, {V_{{\vec b}_{1\perp}, b_1^-} [x_2^+ ,
        0]}_{\cal A} \! \left( {V^\dagger_{{\vec b}_{1\perp}, b_1^-}
          [+\infty, 0]}_{\cal A} \, {V_{{\vec b}_{1\perp}, b_1^-}
          [+\infty, 0]}_{A} \right) \! {V^\dagger_{{\vec b}_{1\perp},
          b_1^-} [x_1^+ , 0]}_{A} t^a \, {V_{{\vec b}_{1\perp}, b_1^-}
        [x_1^+ , 0]}_{A} \right] \right. \notag \\ & \times \left. \,
    \mbox{tr} \left[ {V^\dagger_{{\vec b}_{2\perp}, b_2^-} [+\infty,
        0]}_{\cal A} \, {V_{{\vec b}_{2\perp}, b_2^-} [+\infty,
        0]}_{A} \right] \right\} | 0 \rangle ,
\end{align}
where the parenthesis around the $V^\dagger \, V$ factor inside the
first trace are placed there just to emphasize this term. 

Once again we are interested in contractions in \eq{eq:B910sum2}
connecting the $b_1$ and $b_2$ lines. By analogy to the analysis of
\eq{eq:Vlines} one can conclude that two contractions between the
fields in the second trace and the fields in the parenthesis inside
the first trace cancel. We are left with contractions between
${V_{b_1}^\dagger [x_2^+ , 0]}_{\cal A} \, t^a \, {V_{b_1} [x_2^+ ,
  0]}_{\cal A}$ from the first trace and the second trace, tr~$\left[
  {V^\dagger_{b_{2}} [+\infty, 0]}_{\cal A} \, {V_{b_2}[+\infty,
    0]}_{A} \right]$, along with the contractions between
${V_{b_1}^\dagger [x_1^+ , 0]}_{A} \, t^a \, {V_{b_1} [x_1^+ ,
  0]}_{A}$ from the first trace and the same second trace,
tr~$\left[{V^\dagger_{b_{2}} [+\infty, 0]}_{\cal A} \,
  {V_{b_2}[+\infty, 0]}_{A} \right]$. We may also have one contraction
between the expression in the parenthesis of \eq{eq:B910sum2} and the
second trace, combined with another contraction either between
${V_{b_1}^\dagger [x_2^+ , 0]}_{\cal A} \, t^a \, {V_{b_1} [x_2^+ ,
  0]}_{\cal A}$ from the first trace and the second trace or between
${V_{b_1}^\dagger [x_1^+ , 0]}_{A} \, t^a \, {V_{b_1} [x_1^+ ,
  0]}_{A}$ from the first trace and the same second trace. This is
exactly the answer illustrated in the second and third lines of
\fig{B9_org}. First of all, since the $b_2$ quark line brings in a
factor of tr~$\left[{V^\dagger_{b_{2}} [+\infty, 0]}_{\cal A} \,
  {V_{b_2}[+\infty, 0]}_{A} \right]$, for each contraction of $A^-
(b_1)$ coming from anywhere in the first trace in \eq{eq:B910sum2}
with $A^- (b_2)$ from the second trace there exists a contraction of
$A^- (b_1)$ with ${\cal A}^- (b_2)$; moreover, the second contraction
comes in with a negative relative sign (due to hermitean conjugation
in $V^\dagger_{\cal A}$ as compared to $V_A$), completing the original
contraction to a retarded Green function, in accordance with
\eq{eq:ret1}. Similarly, for each contraction between ${\cal A}^-
(b_1)$ with ${\cal A}^- (b_2)$ there exists a contraction of ${\cal
  A}^- (b_1)$ with ${A}^- (b_2)$, which comes in with the relative
negative sign, again generating a retarded Green function, but now via
\eq{eq:ret2}. Let us stress again that since we are constructing gluon
production cross section corresponding to the classical field, in this
case we only need two contractions between the $b_1$ and $b_2$ lines,
corresponding to the two gluons exchanged in the diagrams of
\fig{B9_org}. Hence our conclusions should be understood as valid for
up to two gluon exchanges, but not necessarily beyond. The second
point we need to make is that contractions with, say,
${V_{b_1}^\dagger [x_1^+ , 0]}_{A} \, t^a \, {V_{b_1} [x_1^+ ,
  0]}_{A}$ imply that the gluon in the amplitude can interact with the
quark $b_1$ only at light-cone times $0 < x^+ < x_1^+$, that is, to
the left of the emitted gluon in the diagrams of
\fig{B9_org}. Moreover, since
\begin{align}
  \label{eq:Fierz}
  {V_{b_1}^\dagger [x_1^+ , 0]}_{A} \, t^a \, {V_{b_1} [x_1^+ ,
    0]}_{A} = {U^{ab}_{b_1} [x_1^+ , 0]}_{A} \, t^b, 
\end{align}
the interaction is identical to that with a gluon Wilson line
connecting the points $(0, b_1^-, {\vec b}_{1 \perp})$ and $(x_1^+,
b_1^-, {\vec b}_{1 \perp})$ along the $x^+$ light-cone direction. In
passing we note that this is similar to the modification of the quark
source current by a gluon exchange in the classical perturbative
solution of Yang-Mills equations constructed in
\cite{Kovchegov:1997ke} (see Eqs.~(8) there). To conclude we see that
contractions between ${V_{b_1}^\dagger [x_1^+ , 0]}_{A} \, t^a \,
{V_{b_1} [x_1^+ , 0]}_{A}$ and tr~$\left[{V^\dagger_{b_{2}} [+\infty,
    0]}_{\cal A} \, {V_{b_2}[+\infty, 0]}_{A} \right]$ lead to the
diagram to the left of the cut in the first panel of the second row of
\fig{B9_org} (for one contraction) and to the second panel of the
second row in \fig{B9_org} (for two contractions). The gluon
propagators become retarded Green functions, and the color factors are
replaced by commutators. Similar arguments for contractions between
${V_{b_1}^\dagger [x_2^+ , 0]}_{\cal A} \, t^a \, {V_{b_1} [x_2^+ ,
  0]}_{\cal A}$ and tr~$\left[{V^\dagger_{b_{2}} [+\infty, 0]}_{\cal
    A} \, {V_{b_2}[+\infty, 0]}_{A} \right]$ give the right-hand side
of the first diagram in the second line of \fig{B9_org} along with the
last diagram in that line. (Let us stress that in the last two graphs
in \fig{B9_org} we implicitly include a ``crossed'' contribution,
where the quark-gluon vertices on quark line $2$ are interchanged.) A
contraction between ${V_{b_1}^\dagger [x_1^+ , 0]}_{A} \, t^a \,
{V_{b_1} [x_1^+ , 0]}_{A}$ and tr~$\left[{V^\dagger_{b_{2}} [+\infty,
    0]}_{\cal A} \, {V_{b_2}[+\infty, 0]}_{A} \right]$ along with
another contraction between ${V^\dagger_{b_1} [+\infty, 0]}_{\cal A}
\, {V_{b_1} [+\infty, 0]}_{A}$ and tr~$\left[{V^\dagger_{b_{2}}
    [+\infty, 0]}_{\cal A} \, {V_{b_2}[+\infty, 0]}_{A} \right]$ give
the first diagram in the last row of \fig{B9_org} (more precisely, the
part of the diagram to the left of the cut), with the last graph in
this last row being a mirror image with respect to the final-state
cut. Note that these last two diagrams in \fig{B9_org} are zero, since
the contribution of quark-$2$ line is proportional to a color trace of
a commutator. We see that to calculate the contribution to the cross
section of the square of $B_9 + B_{10} + \ldots$, where the ellipsis
represent the multitude of other graph in this class, we only need to
calculate the square of $B_9$ with the retarded gluon Green function
and with the color commutator instead of the standard fundamental
color factor, along with the interference of the ${\cal O} (g^5)$
diagrams (containing a double commutator or two commutators, as shown
in \fig{B9_org}) with the leading-order (${\cal O} (g)$) gluon
production amplitude, shown in the last four panels of
\fig{B9_org}. We have thus completed a demonstration of the original
claim that to facilitate the calculation one can use retarded gluon
propagators, calculate only a sub-set of $B$ (and $C$) graphs with the
commutators, along with calculating only the diagrams of the (i) and
(ii) types by the classification of \fig{xsect}.

A complete proof of these results involves analyzing all the possible
contributions to the amplitude squared. While still tedious, our
formalism developed above makes the proof much more straightforward by
reducing the vast number of diagrams one needs to consider to a much
smaller number of Wilson line correlators like that in
\eq{eq:B910sum}. Let us illustrate our technique by a couple more
examples.

\begin{figure}[ht]
\begin{center}
\includegraphics[width=0.9 \textwidth]{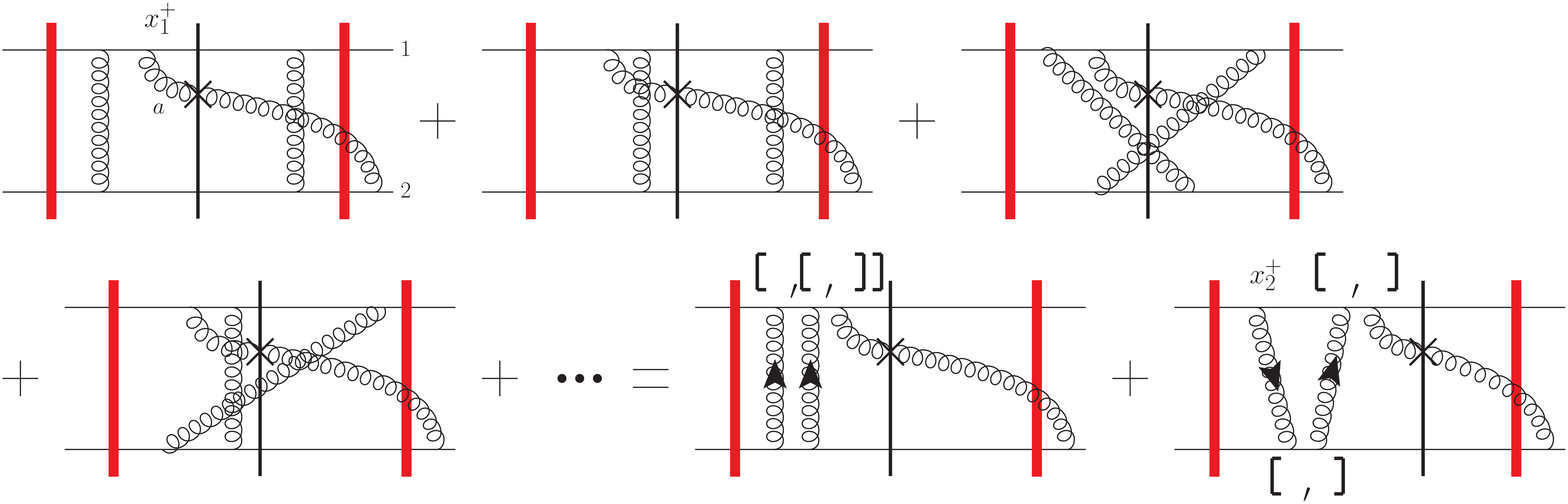} 
\caption{The sum of the diagrams complementing the interference of the
  graphs $B_9$ and $B_{10}$ with $C_5$. The arrows on the gluon lines
  denote the time-flow direction for the retarded propagators, while
  the brackets denote the color matrix commutator. In the last diagram
  the gluon emission time is $x_2^+ \in [0, +\infty]$.}
\label{C5_org}
\end{center}
\end{figure}

First consider the interference of the $B_9 + B_{10} + \ldots$ type of
graphs with $C_5$ (the upside-down reflection of $B_5$ from
\fig{Bgraphs}). This set of graphs is depicted in \fig{C5_org}, where
we only consider gluon exchanges between the projectile quarks at $x^+
>0$ on the either side of the cut. Noticing that the shock wave
interactions with the quark $1$ cancel we conclude that the
contribution of all the graphs of the type shown in \fig{C5_org} is
proportional to
\begin{align}
  \label{eq:C5sum}
  \langle 0 | \mbox{T}_A \overline{\mbox{T}}_{\cal A} \left\{
    \mbox{tr} \left[ {V^\dagger_{{\vec b}_{1\perp}, b_1^-} [+\infty ,
        0]}_{\cal A} \, {V_{{\vec b}_{1\perp}, b_1^-} [+\infty,
        x_1^+]}_{A} \, t^a \, {V_{{\vec b}_{1\perp}, b_1^-} [x_1^+ ,
        0]}_{A} \right] \otimes {V^\dagger_{{\vec b}_{2\perp}, b_2^-}
      [+\infty, 0]}_{\cal A} \, {V_{{\vec b}_{2\perp}, b_2^-}
      [+\infty, 0]}_{A} \right\} | 0 \rangle .
\end{align}
We rewrite \eq{eq:C5sum} as
\begin{align}
  \label{eq:C5sum2}
  \langle 0 | \mbox{T}_A \overline{\mbox{T}}_{\cal A} \left\{
    \mbox{tr} \left[ \left( {V^\dagger_{{\vec b}_{1\perp}, b_1^-}
          [+\infty , 0]}_{\cal A} \, {V_{{\vec b}_{1\perp}, b_1^-}
          [+\infty, 0]}_{A} \right) \, {V^\dagger_{{\vec b}_{1\perp},
          b_1^-} [x_1^+, 0]}_{A} \, t^a \, {V_{{\vec b}_{1\perp},
          b_1^-} [x_1^+ , 0]}_{A} \right] \right. \notag \\ \otimes \,
  \left. {V^\dagger_{{\vec b}_{2\perp}, b_2^-} [+\infty, 0]}_{\cal A}
    \, {V_{{\vec b}_{2\perp}, b_2^-} [+\infty, 0]}_{A} \right\} | 0
  \rangle .
\end{align}
If the two contractions are between ${V^\dagger_{b_1} [+\infty ,
  0]}_{\cal A} \, {V_{b_1} [+\infty, 0]}_{A}$ and ${V^\dagger_{b_2}
  [+\infty, 0]}_{\cal A} \, {V_{b_2} [+\infty, 0]}_{A}$, then they
cancel just like they did in the analysis of \eq{eq:Vlines}. We are
left with two options: one may have two contractions between
${V^\dagger_{b_1} [x_1^+, 0]}_{A} \, t^a \, {V_{b_1} [x_1^+ , 0]}_{A}$
and ${V^\dagger_{b_2} [+\infty, 0]}_{\cal A} \, {V_{b_2} [+\infty,
  0]}_{A}$, giving the answer shown in the first diagram after the
equal sign in \fig{C5_org}, or one contraction between
${V^\dagger_{b_1} [x_1^+, 0]}_{A} \, t^a \, {V_{b_1} [x_1^+ , 0]}_{A}$
and ${V^\dagger_{b_2} [+\infty, 0]}_{\cal A} \, {V_{b_2} [+\infty,
  0]}_{A}$ and another contraction between ${V^\dagger_{b_1} [+\infty
  , 0]}_{\cal A} \, {V_{b_1} [+\infty, 0]}_{A}$ and ${V^\dagger_{b_2}
  [+\infty, 0]}_{\cal A} \, {V_{b_2} [+\infty, 0]}_{A}$. The latter
case gives the last diagram in \fig{C5_org}. Interestingly the first
diagram after the equal sign in \fig{C5_org} is zero, since the trace
of the commutator we get in the color space of quark $1$ is zero. Note
that now the final answer in \fig{C5_org} is completely absorbed into
the interference between the ${\cal O} (g^5)$ diagrams with the ${\cal
  O} (g)$ diagram: hence the diagram $C_5$ is absorbed into the ${\cal
  O} (g^5)$ gluon production amplitude, and is not included into our
${\cal O} (g^3)$ result presented in this work.

\begin{figure}[ht]
\begin{center}
\includegraphics[width=0.9 \textwidth]{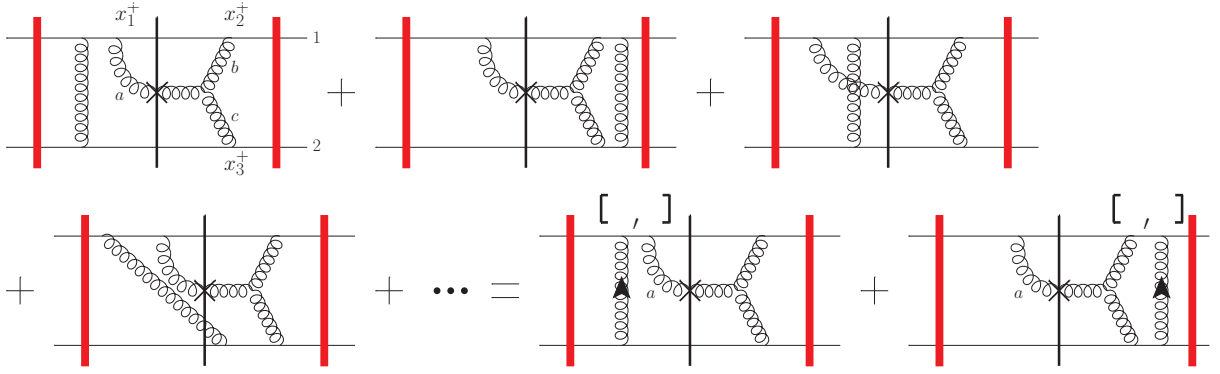} 
\caption{The sum of the diagrams complementing the interference of the
  graphs $B_9$ and $B_{10}$ with $A_5$. The arrows on the gluon line
  denote the time-flow direction for the retarded propagators, while
  the brackets denote the color matrix commutator.}
\label{A5_org}
\end{center}
\end{figure}

Our final example demonstrating our formal technique involves the
interference between the same $B_9 + B_{10} + \ldots$ type of graphs
with $A_5$. It is illustrated in \fig{A5_org}. Similar to the above we
can see that the contribution of the diagrams from \fig{A5_org} is
proportional to
\begin{align}
  \label{eq:A5_sum}
  f^{abc} \, \langle 0 | \mbox{T}_A \overline{\mbox{T}}_{\cal A}
  \left\{ \mbox{tr} \left[ {V^\dagger_{{\vec b}_{1\perp}, b_1^-}
        [x_2^+, 0]}_{\cal A} \, t^b \, {V^\dagger_{{\vec b}_{1\perp},
          b_1^-} [+\infty, x_2^+]}_{\cal A} \, {V_{{\vec b}_{1\perp},
          b_1^-} [+\infty, x_1^+]}_{A} \, t^a \, {V_{{\vec
            b}_{1\perp}, b_1^-} [x_1^+ , 0]}_{A} \right]
  \right. \nonumber \\ \times \, \left.  \mbox{tr} \left[
      {V^\dagger_{{\vec b}_{2\perp}, b_2^-} [x_3^+, 0]}_{\cal A} \,
      t^c \, {V^\dagger_{{\vec b}_{2\perp}, b_2^-} [+\infty ,
        x_3^+]}_{\cal A} \, {V_{{\vec b}_{2\perp}, b_2^-} [+\infty,
        0]}_{A} \right] \right\} | 0 \rangle.
\end{align}
We rewrite this as 
\begin{align}
  \label{eq:A5_sum2}
  & f^{abc} \, \langle 0 | \mbox{T}_A \overline{\mbox{T}}_{\cal A}
  \left\{ \mbox{tr} \left[ {V^\dagger_{{\vec b}_{1\perp}, b_1^-}
        [x_2^+, 0]}_{\cal A} \, t^b \, {V_{{\vec b}_{1\perp}, b_1^-}
        [x_2^+, 0]}_{\cal A} \, \left( {V^\dagger_{{\vec b}_{1\perp},
            b_1^-} [+\infty, 0]}_{\cal A} \, {V_{{\vec b}_{1\perp},
            b_1^-} [+\infty, 0]}_{A} \right) \right. \right. \notag \\
  & \times \left. \left. {V^\dagger_{{\vec b}_{1\perp}, b_1^-} [x_1^+,
        0]}_{A} \, t^a \, {V_{{\vec b}_{1\perp}, b_1^-} [x_1^+ ,
        0]}_{A} \right] \, \mbox{tr} \left[ {V^\dagger_{{\vec
            b}_{2\perp}, b_2^-} [x_3^+, 0]}_{\cal A} \, t^c \,
      {V_{{\vec b}_{2\perp}, b_2^-} [x_3^+, 0]}_{\cal A} \left(
        {V^\dagger_{{\vec b}_{2\perp}, b_2^-} [+\infty , 0]}_{\cal A}
        \, {V_{{\vec b}_{2\perp}, b_2^-} [+\infty, 0]}_{A} \right)
    \right] \right\} | 0 \rangle .
\end{align}
An important difference now is that we are looking at only one
contraction, corresponding to the single gluon exchange between the
projectile quarks in \fig{A5_org}. Similar to the above contractions
between $V^\dagger V$'s in parenthesis cancel. Single contractions
involving ${V^\dagger_{b_2} [x_3^+, 0]}_{\cal A} \, t^c \, {V_{b_2}
  [x_3^+, 0]}_{\cal A}$ in the second trace are zero: they are
proportional to a trace of a commutator. We are left with contractions
between ${V^\dagger_{b_1} [x_2^+, 0]}_{\cal A} \, t^b \, {V_{b_1}
  [x_2^+, 0]}_{\cal A}$ and ${V^\dagger_{b_2} [+\infty , 0]}_{\cal A}
\, {V_{b_2} [+\infty, 0]}_{A}$ and between ${V^\dagger_{b_1} [x_1^+,
  0]}_{A} \, t^a \, {V_{b_1} [x_1^+ , 0]}_{A}$ and ${V^\dagger_{b_2}
  [+\infty , 0]}_{\cal A} \, {V_{b_2} [+\infty, 0]}_{A}$: this is
exactly the answer given in the last two diagrams of \fig{A5_org}.

Note that one can repeat the above argument after moving the gluon
line with color $b$ (or $c$) in the first graph of \fig{A5_org} across
the cut, getting the same answer with two commutator terms. Adding
such contributions to the final result from \fig{A5_org} would turn
the propagators of the gluon lines $b$ and $c$ into retarded Green
functions.

The rest of the proof would be a more or less straightforward
repetition of the above examples, which we verified but are not going
to present here. In the end the propagators of all the internal gluon
lines become retarded Green functions. The $B$ and $C$ graphs
contribute as a subset of these graphs with color commutators.

\begin{figure}[ht]
\begin{center}
\includegraphics[width=0.9 \textwidth]{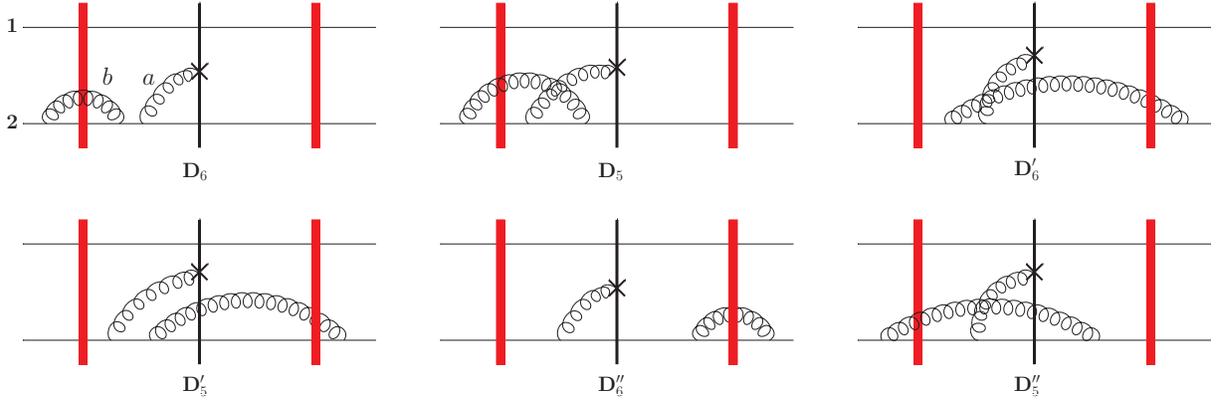} 
\caption{This figure illustrates the rules for calculating diagrams
  $D_5$ and $D_6$ from \fig{Dgraphs} along with the other graphs which
  need to be included. The emissions coming from quark $1$ (see
  e.g. \fig{Dsquared}) in the complex conjugate amplitude are not
  shown as they are not relevant for the calculation at hand.}
\label{Dgraphs56}
\end{center}
\end{figure}

This Appendix presents details of the correspondence between the
``tree-level'' Feynman diagrams and classical gluon fields. It is
possible that an even more compact proof of our main results can be
constructed: however, at the moment it is not known to the authors.

Our techniques presented above can be applied to the $D$ and $E$
graphs as well. However, in those cases a simple diagrammatic
summation is possible. The $D$ diagrams which need to be added
together are shown in \fig{Dgraphs56}. (The emissions off of quark
$1$, such as those in \fig{Dsquared}, are not shown explicitly in
\fig{Dgraphs56} since they are not needed for our analysis.) In the
end one obtains
\begin{align}
  \label{eq:Dsum}
  D_5 + D'_5 + D''_5 + D_6 + D'_6 + D''_6 = D_6 (\mbox{with retarded
    propagators and with} \ [t^a,t^b] \ \mbox{instead of} \ t^a \,
  t^b),
\end{align}
again in agreement with the main claims of this Appendix. The proof
for the $E$ graphs is obtained by a simple swap of the projectile
quarks $1$ and $2$.


\section{On Sub-Gauge Invariance}
\renewcommand{\theequation}{B\arabic{equation}}
  \setcounter{equation}{0}
\label{B}

In this Appendix we demonstrate that up to order $g^3$ the gluon
production amplitude does not depend on the particular sub-gauge used.
Using
\begin{align}
  \notag \frac{1}{l^+ -i \epsilon}-\frac{1}{l^+ +i \epsilon}= 2 \pi i
  \delta(l^+),
\end{align}
we can write the gluon propagator in the ${\vec A}_\perp (x^- \to +
\infty) = 0$ sub-gauge of the light-cone gauge as
\begin{align}
  \label{eq:prop}
  \frac{-i}{l^2+i \epsilon}\left( g_{\mu \nu} - \frac{\eta_\mu
      l_\nu}{l^+ - i \epsilon} - \frac{\eta_\nu l_\mu}{l^+ + i
      \epsilon}\right) = \frac{-i}{l^2+i \epsilon}\left( g_{\mu \nu}
    -\mbox{PV} \left(\frac{1}{l^+} \right) (\eta_\mu l_\nu + \eta_\nu
    l_\mu)\right) + \delta(l^+)\frac{\pi}{l_\perp^2} (\eta_\mu l_\nu -
  \eta_\nu l_\mu).
\end{align}

Here we have split up the gluon propagator in the ${\vec A}_\perp (x^-
\to + \infty) = 0$ sub-gauge into two parts: one is equivalent to the
PV gauge propagator, the other is proportional to $\delta(l^+)$. The
gluon propagator in the ${\vec A}_\perp (x^- \to - \infty) = 0$
sub-gauge can be written in a similar form, the only difference being
the sign of the $\delta(l^+)$ term. The $\delta(l^+)$ component gives
rise to extra terms for a given diagram when compared with the PV
gauge. Due to gauge invariance of the amplitude it is expected that
all these extra terms from all of the diagrams cancel out. This does
end up being the case but the cancellations are not trivial. One needs
to consider more diagrams than simply those shown in the main text as
classes $A$, $B$, $C$, $D$ and $E$.

There are three types of new contributions in the ${\vec A}_\perp (x^-
\to + \infty) = 0$ sub-gauge calculation of gluon production (as
compared to the PV sub-gauge calculation): most of the $A$-graphs
change (by $\Delta A_i$) in the new gauge along with some of the $D$
and $E$ graphs (by $\Delta D_i$ and $\Delta E_i$ respectively), some
of the $B$ and $C$ graphs acquire non-eikonal contributions which are
still leading-order in the ${\vec A}_\perp (x^- \to + \infty) = 0$
sub-gauge (we will refer to those contributions as ``pinched'') and
there are new diagrams which can be identified as the shock-wave
interaction corrections. All of these contributions will be described
in detail below. These contributions are necessary when dealing with
both the ${\vec A}_\perp (x^- \to + \infty) = 0$ and ${\vec A}_\perp
(x^- \to - \infty) = 0$ sub-gauges of the light-cone gauge. Through
the rest of this section we will be working in the ${\vec A}_\perp
(x^- \to + \infty) = 0$ sub-gauge. As will be shown, all of these new
contributions can be written in terms of some sort of a ``gauge
rotation'' acting on the order-$g$ single gluon emission diagrams from
\fig{1gluon}. Since gauge invariance is valid separately for gluon
emissions from quark $1$ and quark $2$, we will only perform the
calculation explicitly for gluon emission from quark $1$.

\begin{figure}[ht]
\begin{center}
\includegraphics[width=0.3 \textwidth]{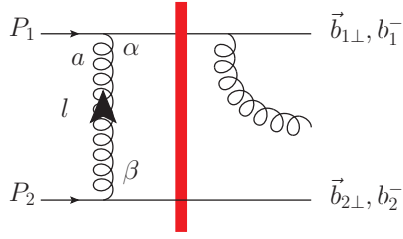} 
\caption{Diagram B$_3$, an example of a diagram with a pinched contribution.}
\label{pinch}
\end{center}
\end{figure}

Pinched contributions originate from terms in the $B$ and $C$ diagrams
which, while zero in the PV sub-gauge, are not zero in the ${\vec
  A}_\perp (x^- \to + \infty) = 0$ sub-gauge. In order to see these
extra contributions one cannot just treat the quark lines as Wilson
lines, as was done in the rest of the paper; instead one must treat
them as high energy quarks with large $P_1^+$ and $P_2^+$
momenta. Here we examine diagram B$_3$ depicted in detail in
\fig{pinch}, to show how this pinched contribution arises.

For the following calculation we only consider the left side of
\fig{pinch}. That is, for the quark lines we only include terms up to
the $\gamma^+$ from the shock wave. Here we show that this diagram can
be written in terms of two components, the PV--sub-gauge contribution
which comes from treating the quark lines as Wilson lines, and a
``gauge rotation''. The left (before the shock wave) side of the
diagram gives
\begin{align}
  \int \frac{d^4 l}{(2 \pi)^4} e^{i l \cdot (b_1 - b_2)} \left\{
    \frac{-i}{l^2+i \epsilon} \left[ g_{\alpha \beta} - \mbox{PV}
      \left( \frac{1}{l^+}\right) (\eta_\alpha l_\beta + \eta_\beta
      l_\alpha)\right] - \frac{\pi}{l_\perp^2} \delta (l^+)
    (\eta_\alpha l_\beta - \eta_\beta l_\alpha)\right\} \notag \\
  \times \left[\gamma^+ \frac{i (\sh P_1 + \sh l)}{(P_1+l)^2+i
      \epsilon}i g (t^a)_1 \gamma^\alpha u_\sigma (P_1)\right]
  \left[\gamma^+ \frac{i (\sh P_2 - \sh l)}{(P_2 -l)^2+i \epsilon}i g
    (t^a)_2 \gamma^\beta u_\rho (P_2)\right]. \label{B3orig}
\end{align}
Here $b_1^{\mu} = (0, b_1^-, {\vec b}_{1\perp})$ and $b_2^{\mu} = (0,
b_2^-, {\vec b}_{2\perp})$ denote the intersection points of the
projectile quarks trajectories with the shock wave. We also assume
that the quarks are massless. We rewrite \eqref{B3orig} more compactly
as
\begin{align}
\label{B3original}
\left( g^2 (t^a)_1 (t^a)_2 \right) \int \frac{d^4 l}{(2 \pi)^4} e^{i l
  \cdot (b_1 - b_2)} \Gamma_1^\alpha \Gamma_2^\beta \left(
  \frac{-i}{l^2+i \epsilon} \left[ g_{\alpha \beta} - \mbox{PV} \left(
      \frac{1}{l^+}\right) (\eta_\alpha l_\beta + \eta_\beta
    l_\alpha)\right] - \frac{\pi}{l_\perp^2} \delta (l^+) (\eta_\alpha
  l_\beta - \eta_\beta l_\alpha)\right),
\end{align}
where we have defined the following symbols:
\begin{align}
  \notag \Gamma_1^\alpha & = \left[\gamma^+ \frac{ (\sh P_1 + \sh
      l)}{(P_1+l)^2+i \epsilon} \gamma^\alpha u_\sigma (P_1)\right],
  \\ \notag \Gamma_2^\beta & = \left[\gamma^+ \frac{ (\sh P_2 - \sh
      l)}{(P_2 -l)^2+i \epsilon} \gamma^\beta u_\rho (P_2)\right].
\end{align}

Before continuing with the remainder of the calculation it is useful
to evaluate the following in the high-energy limit,
\begin{align}
  \eta_\alpha \Gamma_1^\alpha & = \left[\gamma^+ \frac{ (\sh P_1 + \sh
      l)}{(P_1+l)^2+i \epsilon} \gamma^+ u_\sigma (P_1)\right] \notag
  \\ & \approx \left[\gamma^+ \gamma^- P_1^+ \gamma^+ u_\sigma
    (P_1)\right] \frac{1}{2 P_1^+ l^- +i \epsilon}
  \notag \\  & = \left[\gamma^+ u_\sigma (P_1)\right] \frac{1}{ l^- +i \epsilon}, \notag \\
  l_\alpha \Gamma_1^\alpha & = \left[\gamma^+ \frac{ (\sh P_1 + \sh
      l)}{(P_1+l)^2+i \epsilon} \sh l u_\sigma (P_1)\right] \notag \\
  & \approx \left[\gamma^+ \frac{ (\sh P_1 + \sh l)}{(P_1+l)^2+i
      \epsilon} (\sh P_1 + \sh l) u_\sigma (P_1)\right] \notag \\ & =
  \left[\gamma^+ u_\sigma (P_1)\right]. \notag
\end{align}
Using similar tricks for $\eta_\beta \Gamma_2^\beta$ and $l_\beta
\Gamma_2^\beta$ we have
\begin{align}
  \notag \eta_\beta \Gamma_2^\beta & \approx \left[\gamma^+ u_\rho
    (P_2)\right] \frac{-1}{ l^- -i \epsilon}, \\
  \notag l_\beta \Gamma_2^\beta & \approx \left[\gamma^+ u_\rho
    (P_2)\right](-1).
\end{align}

Let us first evaluate \eq{B3original} by treating the quark lines as
Wilson lines. This is equivalent to taking only the upper ``$--$''
component of the gluon propagator (lowercase $\alpha = +, \beta = +$)
and then taking the high energy limit, i.e., taking $P_1^+, P_2^+$
large. The term proportional to $\delta(l^+)$ is exactly zero for
$\alpha = +, \beta = +$. We end up solely with the PV--sub-gauge
contribution of the form
\begin{align} 
  \notag \left( g^2 (t^a)_1 (t^a)_2 \right) & \int \frac{d^4 l}{(2
    \pi)^4} e^{i l \cdot (b_1 - b_2)} \; \eta_\alpha \Gamma_1^\alpha
  \; \eta_\beta \Gamma_2^\beta
  \frac{i}{l^2+i \epsilon} \mbox{PV} \left( \frac{1}{l^+}\right) (2 l^-) \\
  = & \left[ \gamma^+ u_\sigma(P_1)\right]_1 \left[ \gamma^+ u_\rho
    (P_2)\right]_2 \left( i g^2 (t^a)_1 (t^a)_2 \right) \int \frac{d^4
    l}{(2 \pi)^4} e^{i l \cdot (b_1 - b_2)} \frac{-1}{l^2+i \epsilon}
  \mbox{PV} \left( \frac{1}{l^+} \right) \mbox{PV} \left(
    \frac{2}{l^-} \right). \label{B3wilson}
\end{align}

Now lets evaluate \eq{B3original} using a complete treatment of the
quark lines. This allows us to accurately calculate the term
proportional to $\delta(l^+)$. To do this we first evaluate the
following expressions:
\begin{align}
  \left( \eta_\alpha l_\beta + \eta_\beta l_\alpha \right)
  \Gamma_1^\alpha \Gamma_2^\beta & = \left[ \gamma^+
    u_\sigma(P_1)\right] \left[ \gamma^+ u_\rho (P_2)\right] \left(-
    \frac{1}{l^- +i \epsilon}- \frac{1}{l^- -i \epsilon} \right)
  \notag \\ & = \left[ \gamma^+ u_\sigma(P_1)\right] \left[ \gamma^+
    u_\rho (P_2)\right]
  \mbox{PV} \left( \frac{-2}{l^-} \right), \notag \\
  \left( \eta_\alpha l_\beta - \eta_\beta l_\alpha \right)
  \Gamma_1^\alpha \Gamma_2^\beta & = \left[ \gamma^+
    u_\sigma(P_1)\right] \left[ \gamma^+ u_\rho (P_2)\right] \left(-
    \frac{1}{l^- +i \epsilon}+ \frac{1}{l^- -i \epsilon} \right)
  \notag \\ & = \left[ \gamma^+ u_\sigma(P_1)\right]
  \left[ \gamma^+ u_\rho (P_2)\right] 2 \pi i \delta(l^-), \notag \\
  g_{\alpha \beta} \Gamma_1^\alpha \Gamma_2^\beta & = 0. \notag
\end{align}
With these results in hand we evaluate \eq{B3original} arriving at
\begin{align} 
  \left[ \gamma^+ u_\sigma(P_1)\right]_1 \left[ \gamma^+ u_\rho
    (P_2)\right]_2 \left( i g^2 (t^a)_1 (t^a)_2 \right) \int \frac{d^4
    l}{(2 \pi)^4} e^{i l \cdot (b_1 - b_2)} \left[ \frac{-1}{l^2+i
      \epsilon} \mbox{PV} \left( \frac{1}{l^+} \right) \mbox{PV}
    \left( \frac{2}{l^-} \right) - \frac{2 \pi^2}{l_\perp^2} \delta
    (l^+) \delta (l^-)\right]. \label{B5wilson}
\end{align}

The first term in the square brackets in the integrand of
\eqref{B5wilson} corresponds to the PV part of the gluon
propagator. Notice how this matches the result ones gets from treating
the quark lines as Wilson lines, shown in \eq{B3wilson}. This is
important because this justifies the treatment of these diagrams in
the PV gauge used in the main text and in Appendix~\ref{A}. The second
term is the gauge-dependent term proportional to $\delta(l^+)$. Notice
how it also contains a $\delta(l^-)$, which embodies the ``pinching''
of the singularities in the $l^-$-integral: this is why we refer to
such terms as the ``pinched'' contributions. This term was hidden
before when we just took the ``$--$'' component of the gluon
propagator. We can see now that the naive eikonal approximation used
in arriving at \eq{B3wilson} caused us to miss this term.

\begin{figure}[ht]
\begin{center}
\includegraphics[width=0.35 \textwidth]{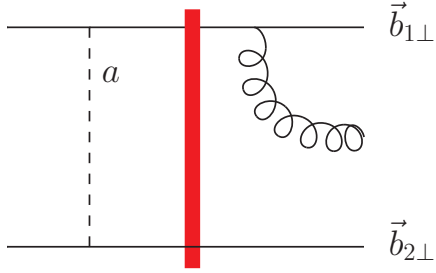} 
\caption{The pinched contribution associated with the diagram $B_3$.}
\label{rotation}
\end{center}
\end{figure}

Now that we have separated the sub-gauge-dependent part of the diagram
from the PV part we can analyze these separately. The PV sub-gauge
contribution (the first term in the square brackets of
\eqref{B5wilson}) leads to the expression for the diagrams $B_3$ given
in \eq{B3} of the main text. The second term in the square brackets of
\eqref{B5wilson} yields, after integration over $l$,
\begin{align}
\label{eqrotation}
\left[ \gamma^+ u_\sigma(P_1)\right]_1 \left[ \gamma^+ u_\rho
  (P_2)\right]_2 \left(- \frac{ i}{4 \pi} g^2 (t^a)_1 (t^a)_2
  \ln{\frac{1}{|{\vec b}_{1\perp} - {\vec b}_{2\perp} | \, \Lambda}}
\right).
\end{align}
We will refer to such terms as ``gauge rotations''.  It will be
denoted graphically by a dashed line with the end points at the
transverse positions whose difference is inside the logarithm (for
\eq{eqrotation} these would be ${\vec b}_{1\perp}$ and ${\vec b}_{2
  \perp}$). If the end point is attached to a quark it would give a
fundamental generator, if it is attached to a gluon it would give an
adjoint generator. Both generators have the same color. The graphical
representation of the sub-gauge-dependent part of $B_3$, algebraically
given in \eq{eqrotation}, is shown in \fig{rotation}. This graphical
notation is used for the rest of this section.

The pinched diagram contribution arises only when the gluon is exchanged
between the projectile quark either before any other interaction takes
place or when such an exchange happens after all the interactions took
place. Analyzing all the $B$-graphs in \fig{Bgraphs}, we conclude that
only 4 of those have the pinched contributions when they are
calculated in the ${\vec A}_\perp (x^- \to + \infty) = 0$ (or ${\vec
  A}_\perp (x^- \to - \infty) = 0$) sub-gauge. These are $B_1$, $B_3$,
$B_5$ and $B_{10}$. The pinched contributions coming from these
diagrams are denoted here as $\Delta B_1,\Delta B_3,\Delta B_5$ and
$\Delta B_{10}$. All of them are shown in \fig{all} below. (Indeed
there are also $\Delta C_1,\Delta C_3,\Delta C_5$ and $\Delta C_{10}$:
these are important when checking gauge-invariance of the amplitude
with the gluon emitted off of quark 2.)

\begin{figure}[ht]
\begin{center}
\includegraphics[width=0.9 \textwidth]{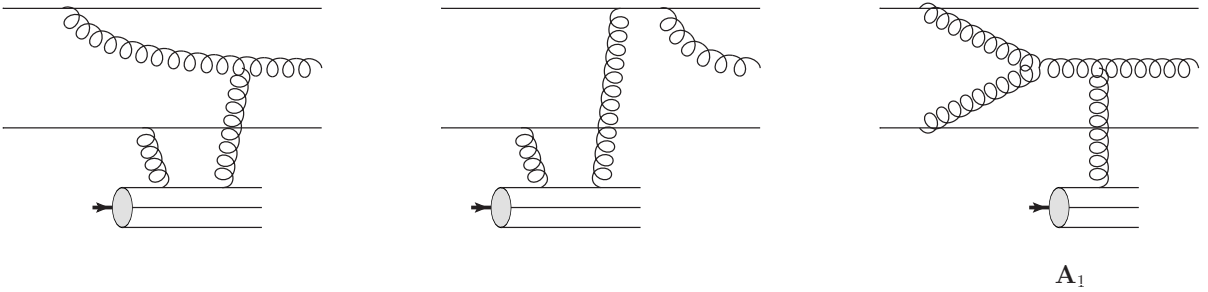} 
\caption{Two left panels: examples of shock-wave correction diagrams
  which need to be considered in the gauges explored in this
  Appendix. Right panel: a diagram in the $A_1$ class which is of the
  same order in the coupling as the two other diagrams in this
  figure. The composite object at the bottom of the diagrams is a
  nucleon in the target nucleus. }
\label{shock_graphs}
\end{center}
\end{figure}

Now let us focus on the shock-wave corrections. The diagrams which we
have in mind are shown in the two left panels of \fig{shock_graphs}
(cf. \cite{Kovchegov:1998bi}). These diagrams do not belong to the
$A$, $B$, $C$, $D$ and $E$ diagram types considered in the main text:
the two-gluon interaction with the target nucleon in those graphs does
{\sl not} leave the nucleon in a color-singlet state, which is
indicated by the nucleon breaking up in \fig{shock_graphs}. One
concludes that these diagrams are higher-order corrections to the
Glauber--Mueller scattering in the target, which is limited by two
gluons per nucleon \cite{Mueller:1989st}. However, his does not allow
one to simply neglect these graphs, since this correction is enhanced
by a power of $A_1^{1/3}$ (in the cross section) due to the extra
projectile nucleon involved in the interaction. In the right panel of
\fig{shock_graphs} we show a diagram in the $A_1$ class, with the
interaction with the target limited to a single-gluon exchange with
one of the target nucleons. Clearly all three diagrams in
\fig{shock_graphs} are of the same order in our power counting: they
are all order-$g^5$ and have the same projectile and target nucleons
participating in the interaction. Hence the first two graphs in
\fig{shock_graphs}, along with other similar diagrams involving more
multiple rescatterings in the target, are of the same order as the
$A$, $B$, $C$, $D$ and $E$ diagrams and have to be included in the
analysis. In fact one may worry why we did not include them in the
main-text calculation in the PV sub-gauge of the light-cone gauge:
below we justify neglecting these diagrams in the PV sub-gauge
calculation performed in the main text.

\begin{figure}[ht]
\begin{center}
\includegraphics[width=0.9 \textwidth]{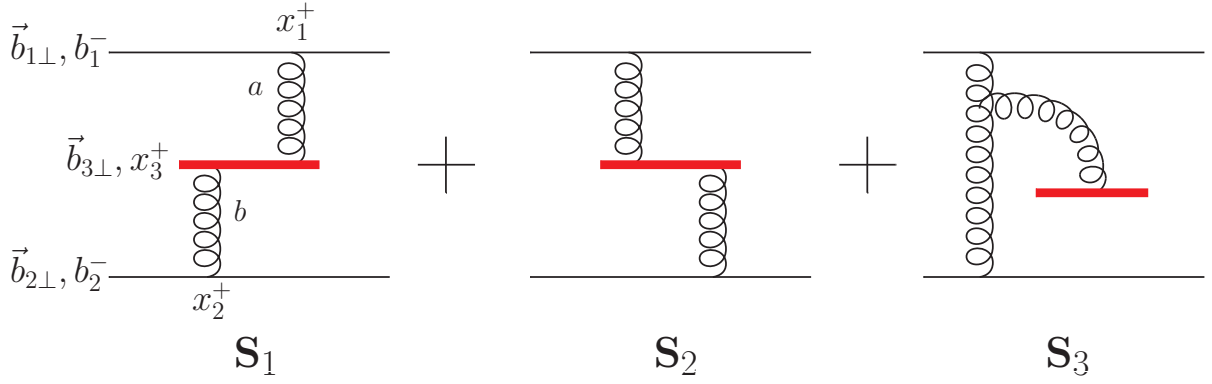} 
\caption{The diagrams correcting the interaction of two projectile
  quarks (top two straight horizontal lines) with a quark coming from
  a single nucleon in the shock wave (the thick red line).}
\label{Sdiag}
\end{center}
\end{figure}

To resum the corrections of the type shown in the left two diagrams of
\fig{shock_graphs}, we have to include corrections like this for
either one of the many nucleons involved in the shock-wave
interaction. We first consider scattering of two projectile quarks
(coming from two different nucleons) on a shock-wave target. The
corrections to the scattering on a single target nucleon are shown in
\fig{Sdiag} and labeled $S_1$, $S_2$ and $S_3$. (Our analysis will be
similarly valid for scattering of the gluons coming from the
projectile nucleons on the target.) Note that the diagram $S_3$ in
\fig{Sdiag} also has the standard eikonal contribution where the gluon
exchanged between the projectile quark lines is long-lived in the
$s$-channel: those types of contributions are included in the analysis
of diagrams $A-E$ in the main text and will be discarded here.

In the $S$ diagrams shown in \fig{Sdiag}, quarks $1$ and $2$ are
eikonal quarks traveling in the $x^+$ direction. Quark $3$, labeled by
a thick red line, originates in a nucleon from the target and thus has
a large $P^-$ momentum and is traveling in the $x^-$ direction. While
the evaluation of these diagrams is explicitly done for quarks,
similar results exist for the case where quark(s) $1$ and/or $2$
are/is replaced by (a) gluon(s). For this calculation we will just
focus on the quark case. The diagrams in \fig{Sdiag} should be
understood as being inside the shock wave; they can take place at any
point in the shock wave. Since we are summing over all possible
diagrams, in a given target charge distribution there will be diagrams
where a given target nucleon has diagrams $S_1$, $S_2$ and $S_3$
associated with it. We have to calculate and sum up all three diagrams
associated with each nucleon in the target. As an example we calculate
diagram S$_1$ explicitly.

\begin{figure}[ht]
\begin{center}
\includegraphics[width=0.4 \textwidth]{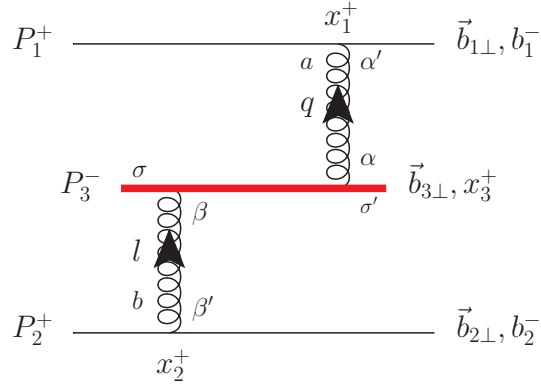} 
\caption{Diagram $S_1$, a correction to the interaction with a single
  nucleon in the target shock-wave. The thick red line denotes a quark
  in the target nucleon.}
\label{S1}
\end{center}
\end{figure}

This diagram $S_1$ is shown in detail in \fig{S1}. As usual, we will
treat the projectile quarks as the eikonal lines along the $x^+$
light-cone, while treating the target quark as a regular quark with a
large $P_3^-$ component of its momentum. (Note that we normalize the
target quark by multiplying its spinor matrix element by $1/(2
P_3^-)$.) Fourier-transforming into coordinate space we obtain the
following contribution:
\begin{align}
  S_1 = g^4 \, (t^a)_1 \, (t^b)_2 \, (t^a t^b)_3 & \int \frac{d^4
    l}{(2 \pi)^4}\frac{d q^- d^2 q}{(2 \pi)^3} e^{ - i l^+ (b_1^- -
    b_2^-) -i q^- (x^+_1 - x^+_3)-i l^- (x^+_3 - x^+_2) + i \vec
    q_\perp \cdot (\vec b_{1 \perp} - \vec b_{3 \perp})+i \vec l_\perp
    \cdot (\vec b_{3 \perp} - \vec b_{2 \perp})}
  \notag \\
  & \frac{1}{2 P^-_3} \, \bar u_{\sigma'} (P_3 + l - q)
  \gamma^{\alpha} \frac{i (\sh P_3 + \sh l)}{(P_3 + l)^2 + i \epsilon}
  \gamma^{\beta} u_{\sigma} (P_3) \, g^{\alpha' -} \, g^{\beta' -}
  \notag \\
  & \left[ \frac{-i}{l^2+i \epsilon}\left( g_{\beta' \beta}
      -\mbox{PV}\left(\frac{1}{l^+}\right) (\eta_{\beta'} l_\beta +
      \eta_\beta l_{\beta'})\right) + \delta(l^+)\frac{\pi}{l_\perp^2}
    (\eta_{\beta'} l_\beta - \eta_\beta l_{\beta'}) \right]
  \notag \\
  & \left[ \frac{-i}{q^2+i \epsilon}\left( g_{\alpha \alpha'}
      -\mbox{PV}\left(\frac{1}{q^+} \right) (\eta_\alpha q_{\alpha'} +
      \eta_{\alpha'} q_\alpha)\right) +
    \delta(q^+)\frac{\pi}{q_\perp^2} (\eta_\alpha q_{\alpha'} -
    \eta_{\alpha'} q_\alpha) \right]. \label{S11}
\end{align}
In arriving at \eq{S11} we have used the on-shell condition for the
outgoing target quark,
\begin{align}
\label{on-shell}
  l^+ - q^+ = \frac{(\vec l_\perp - \vec q_\perp)^2}{2 P^-_3},
\end{align}
to eliminate the $q^+$ integral. Since quarks $1$ and $2$ are
completely eikonal we replaced their spinor matrix elements by
$g^{\alpha' -} \, g^{\beta' -}$. In the target quark spinor matrix
element from \eq{S11} one can only have $\alpha = +, \perp$ and $\beta
= +, \perp$ since the gluon propagators vanish for either $\alpha = -$
or $\beta = -$ due to the gauge choice. The leading contribution in
$P^-_3$ is given by the $\alpha, \beta = \perp$ component. Naively one
would expect such contribution to be sub-eikonal, suppressed by a
power of $P_3^-$. As one could see below, this is indeed the case, but
only until one integrates over $l^+$ picking the pole either at
$l^+=0$ or at $q^+=0$. The residues at such poles generate enhancement
(due to pinching of the poles in the product of propagators in
\eqref{S11}), making the end contribution leading order. Since we are
interested only in this pinched pole contribution, below we will
evaluate all the expressions keeping the $l^+ \approx 0$ approximation
in mind.

Keeping the $\alpha, \beta = \perp$ component allows us to write the
product of the two gluon propagators in \eqref{S11} as
\begin{align}
  \left( \frac{i}{l^2 + i \epsilon} \mbox{PV}\frac{1}{l^+} +
    \frac{\pi}{l_\perp^2}\delta(l^+) \right) \left( \frac{i}{q^2 + i
      \epsilon} \mbox{PV}\frac{1}{q^+} - \frac{\pi}{q_\perp^2}
    \delta(q^+) \right) \, \vec q_{\perp \alpha} \, \vec l_{\perp
    \beta} . \label{Sprop}
\end{align}
Multiplying the spinor matrix element of quark $3$ from \eqref{S11} by
$\vec q_{\perp \alpha} \, \vec l_{\perp \beta}$ from the above
expression yields
\begin{align}
  & \frac{1}{2 P^-_3} \, \bar u_{\sigma'} (P_3 + l - q) \ \sh {\vec
    q}_\perp \frac{i (\sh P_3 + \sh l)}{(P_3 + l)^2 + i \epsilon} \,
  \sh {\vec l}_\perp u_{\sigma} (P_3)
  \notag \\
  & = \frac{1}{2 P^-_3} \frac{i}{(P_3 + l)^2 + i \epsilon} \bar
  u_{\sigma'} (P_3 + l - q) \ \sh {\vec q}_\perp \left( \gamma^- \sh
    {\vec l}_\perp l^+ - l_\perp^2 \right) u_{\sigma}
  (P_3), \notag \\
  & = \frac{1}{2 P^-_3} \frac{i}{(P_3 + l)^2 + i \epsilon} \left[ 2
    P^-_3 l^+ \left( \vec l_\perp \cdot \vec q_\perp - i \sigma \;
      \vec l_\perp \times \vec q_\perp \right) - l_\perp^2 \left( \vec
      q_\perp \cdot (\vec q_\perp - \vec l_\perp) - i \sigma \; \vec
      l_\perp \times \vec q_\perp \right) \right]
  \notag \\
  & \approx \frac{i}{2 P^-_3} \left[ \vec l_\perp \cdot \vec q_\perp -
    i \sigma \; \vec l_\perp \times \vec q_\perp - l_\perp^2 \left(
      q_\perp^2- 2 \, \vec l_\perp \cdot \vec q_\perp \right)
    \frac{1}{(P_3 + l)^2 + i \epsilon} \right] \notag \\ & \approx i
  \, \frac{l^+ - q^+}{(\vec l_\perp - \vec q_\perp)^2} \left[ \vec
    l_\perp \cdot \vec q_\perp - i \sigma \; \vec l_\perp \times \vec
    q_\perp - l_\perp^2 \left( q_\perp^2- 2 \, \vec l_\perp \cdot \vec
      q_\perp \right) \frac{1}{2 P_3^- l^+ - l_\perp^2 + i \epsilon}
  \right].\label{S12}
\end{align}
Here we have for simplicity assumed that $P_3^\mu = (0, P_3^-,
0_\perp)$ and defined $\sh {\vec q}_\perp = - {\vec q}_\perp \cdot
{\vec \gamma}_\perp$.  To evaluate the spinor matrix elements we
assumed that the spinors are chosen in the Lepage--Brodsky basis
\cite{Lepage:1980fj} (for the ``$-$'' moving quark) and used the table
of spinor matrix elements in \cite{Lepage:1980fj} (see also
\cite{KovchegovLevin}) with $+ \leftrightarrow -$ to write
\begin{align}
  \bar u_{\sigma'} (P_3 + l - q) \ \sh {\vec q}_\perp \gamma^- \sh {\vec
    l}_\perp u_{\sigma} (P_3) & = 2 P^-_3 \left( \vec l_\perp \cdot
    \vec q_\perp - i \sigma \; \vec l_\perp \times \vec q_\perp
  \right)
  \notag \\
  \bar u_{\sigma'} (P_3 + l -q) \ \sh {\vec q}_\perp u_{\sigma} (P_3) &
  = \vec q_\perp \cdot (\vec q_\perp - \vec l_\perp) - i \sigma \;
  \vec l_\perp \times \vec q_\perp . \notag
\end{align}
We have also assumed that $l^+ \ll l_\perp$ in \eqref{S12} as we are
only after the contribution at the $l^+ =0$ (or $q^+=0$) pole. In the
last step of \eq{S12} we have used the on-shell condition
\eqref{on-shell} for quark $3$.

Combining Eqs.~\eqref{S12}, \eqref{S11} and \eqref{Sprop} we have
\begin{align}
  S_1 = - g^4 \, (t^a)_1 \, (t^b)_2 \, (t^a t^b)_3 & \int \frac{d^4
    l}{(2 \pi)^4}\frac{d q^- d^2 q}{(2 \pi)^3} e^{- i l^+ (b_1^- -
    b_2^-) -i q^- (x^+_1 - x^+_3)-i l^- (x^+_3 - x^+_2) + i \vec
    q_\perp \cdot (\vec b_{1 \perp} - \vec b_{3 \perp} )+i \vec
    l_\perp \cdot (\vec b_{3 \perp}- \vec b_{2 \perp})}
  \notag \\
  \times & \frac{1}{(\vec l_\perp - \vec q_\perp)^2} \left[ \vec
    l_\perp \cdot \vec q_\perp - i \sigma \; \vec l_\perp \times \vec
    q_\perp - l_\perp^2 \left( q_\perp^2- 2 \vec l_\perp \cdot \vec
      q_\perp \right) \frac{1}{2 P_3^- l^+ - l_\perp^2 + i
      \epsilon}\right]
  \notag \\
  \times & \left[ \frac{i}{l^2 + i \epsilon} \frac{1}{q^2 + i
      \epsilon}\left(
      \mbox{PV}\frac{1}{q^+}-\mbox{PV}\frac{1}{l^+}\right) +
    \frac{\pi}{l_\perp^2 q_\perp^2}(\delta(l^+)+\delta(q^+))\right]
  \notag
\end{align}
Integrating over $l^-, q^-$ and $l^+$, keeping in mind that $q^+$ is
set by the on shell condition \eqref{on-shell}, we arrive at the final
result
\begin{align}
  \notag S_1 = \int & \frac{d^2 l}{(2 \pi)^2}\frac{d^2 q}{(2 \pi)^2}
  e^{i \vec q_\perp \cdot (\vec b_{1 \perp} - \vec b_{3 \perp})+i \vec
    l_\perp \cdot (\vec b_{3 \perp} - \vec b_{2 \perp})}
  \delta( x^+_1 - x^+_3) \, \delta( x^+_2 - x^+_3) \notag \\
  & \times \left[ -\frac{1}{2} g^4 (t^a)_1 (t^b)_2 (t^at^b )_3
    \frac{1}{l_\perp^2 q_\perp^2}+ \frac{1}{2} g^4 (t^a)_1 (t^b)_2
    (t^a t^b )_3 \frac{i}{l_\perp^2 q_\perp^2 (\vec l - \vec q)^2}
    \left( 2 \sigma \; \vec l_\perp \times \vec q_\perp + i
      (q_\perp^2-l_\perp^2)\right) \right].
\end{align}

Using similar techniques to calculate the other two graphs in
\fig{Sdiag} we have in total
\begin{subequations}
\begin{align}
  S_1 = & \int \frac{d^2 l}{(2 \pi)^2}\frac{d^2 q}{(2 \pi)^2} e^{i
    \vec q_\perp \cdot (\vec b_{1 \perp} - \vec b_{3 \perp})+i \vec
    l_\perp \cdot (\vec b_{3 \perp} - \vec b_{2 \perp})}
  \delta( x^+_1 - x^+_3) \delta( x^+_2 - x^+_3) \notag \\
  & \left[-\frac{1}{2} g^4 (t^a)_1 (t^b)_2 (t^at^b )_3
    \frac{1}{l_\perp^2 q_\perp^2} + \frac{1}{2} g^4 (t^a)_1 (t^b)_2
    (t^a t^b )_3 \frac{i}{l_\perp^2 q_\perp^2 (\vec l_\perp - \vec
      q_\perp)^2} \left( 2 \sigma \; \vec l_\perp \times \vec q_\perp
      + i (q_\perp^2-l_\perp^2)
    \right) \right] \\
  S_2 = & \int \frac{d^2 l}{(2 \pi)^2}\frac{d^2 q}{(2 \pi)^2} e^{i
    \vec q_\perp \cdot (\vec b_{1 \perp} - \vec b_{3 \perp})+i \vec
    l_\perp \cdot (\vec b_{3 \perp} - \vec b_{2 \perp})}
  \delta( x^+_1 - x^+_3) \delta( x^+_2 - x^+_3) \notag \\
  & \left[ -\frac{1}{2} g^4 (t^a)_1 (t^b)_2 (t^bt^a )_3
    \frac{1}{l_\perp^2 q_\perp^2} - \frac{1}{2} g^4 (t^a)_1 (t^b)_2
    (t^b t^a )_3 \frac{i}{l_\perp^2 q_\perp^2 (\vec l_\perp - \vec
      q_\perp)^2} \left( 2 \sigma \; \vec l_\perp \times \vec q_\perp
      + i (q_\perp^2-l_\perp^2)\right) \right] \\
  {\tilde S}_3 = & \int \frac{d^2 l}{(2 \pi)^2}\frac{d^2 q}{(2 \pi)^2}
  e^{i \vec q_\perp \cdot (\vec b_{1 \perp} - \vec b_{3 \perp})+i \vec
    l_\perp \cdot (\vec b_{3 \perp} - \vec b_{2 \perp})}
  \delta( x^+_1 - x^+_3) \delta( x^+_2 - x^+_3) \notag \\
  & \left[ 
    - g^4 (t^a)_1 (t^b)_2 [t^a, t^b ]_3 \frac{i \sigma \;
      \vec l_\perp \times \vec q_\perp}{l_\perp^2 q_\perp^2 (\vec
      l_\perp - \vec q_\perp)^2} \right] ,
\end{align}
\end{subequations}
where we have dropped the eikonal term in $S_3$ labeling the
remainder ${\tilde S}_3$.

\begin{figure}[b]
\begin{center}
\includegraphics[width=0.9 \textwidth]{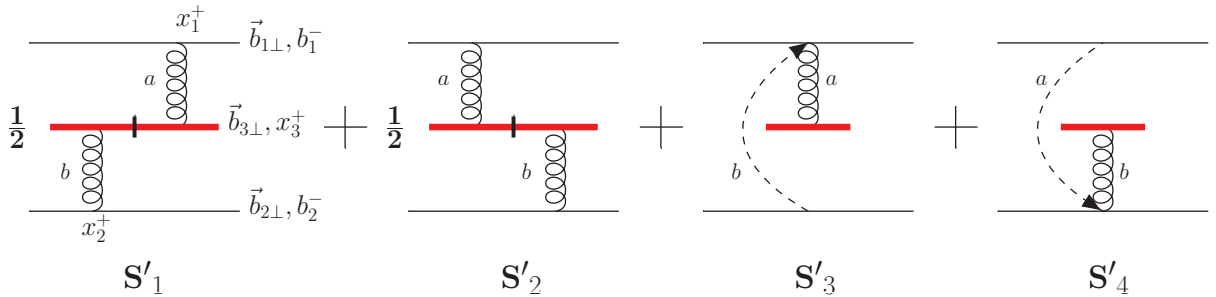} 
\caption{Diagrammatic representation of the four contributions
  contained in Eqs.~\eqref{Sprimes}. The short line crossing the
  (thick red) target quark in the left two diagrams indicates that the
  quark is on mass-shell.  
  The dashed line represents an instantaneous gluon exchange. The
  arrow on the dashed line entering a quark-gluon vertex denotes a
  commutator between the fundamental color generators (with the color
  matrix of the dashed line in the first position in the commutator).
}
\label{S'diag}
\end{center}
\end{figure}

Summing all of these terms up we arrive at
\begin{align}
\label{shockwavetotal}
S_1 + S_2 + {\tilde S}_3 =& \int \frac{d^2 l}{(2 \pi)^2}\frac{d^2
  q}{(2 \pi)^2} e^{i \vec q_\perp \cdot (\vec b_{1 \perp} - \vec b_{3
    \perp})+i \vec l_\perp \cdot (\vec b_{3 \perp} - \vec b_{2
    \perp})}
\delta( x^+_1 - x^+_3) \delta( x^+_2 - x^+_3) \notag \\
& \left[-\frac{1}{2} g^4 (t^a)_1 (t^b)_2 \{ t^a,t^b \}_3 \frac{1}{
    l_\perp^2 q_\perp^2}+ \frac{1}{2} g^4 (t^a)_1 (t^b)_2 [t^a,t^b]_3
  \left( \frac{1}{q_\perp^2 }-\frac{1}{l_\perp^2} \right)
  \frac{1}{(\vec l_\perp - \vec q_\perp)^2} \right].
\end{align}
Notice how the polarization ($\sigma$) dependence of the target quark
associated with the $S$ diagrams goes away when one adds the diagrams
together. As we Fourier-transform \eq{shockwavetotal} over transverse
momenta it is convenient to split up the resulting equation into the
following four pieces, which allow for an intuitive diagrammatic
interpretation:
\begin{subequations}\label{Sprimes}
\begin{align}
  S'_1 = & \frac{1}{2} \left[ \frac{-i}{2 \pi} g^2 (t^a)_1 (t^a)_3 \ln
    \left( \frac{1}{|\vec b_{1 \perp} - \vec b_{3 \perp}| \Lambda}
    \right) \right] \left[ \frac{-i}{2 \pi} g^2 (t^b)_2 (t^b)_3 \ln
    \left( \frac{1}{|\vec b_{2 \perp} - \vec b_{3 \perp}|
        \Lambda}\right) \right]
  \delta( x^+_1 - x^+_3) \delta( x^+_2 - x^+_3) \label{Sprimesa} \\
  S'_2 = & \frac{1}{2} \left[ \frac{-i}{2 \pi} g^2 (t^b)_2 (t^b)_3 \ln
    \left( \frac{1}{|\vec b_{2 \perp} - \vec b_{3 \perp}|
        \Lambda}\right) \right] \left[ \frac{-i}{2 \pi} g^2 (t^a)_1
    (t^a)_3 \ln \left( \frac{1}{|\vec b_{1 \perp} - \vec b_{3 \perp}|
        \Lambda} \right) \right]
  \delta( x^+_1 - x^+_3) \delta( x^+_2 - x^+_3) \label{Sprimesb} \\
  S'_3 = & \left[ \frac{-i}{2 \pi} g^2 (t^a)_3 \ln \left(
      \frac{1}{|\vec b_{1 \perp} - \vec b_{3 \perp}| \Lambda} \right)
  \right] \left[ \frac{-i}{4 \pi} g^2 [t^b, t^a]_1 (t^b)_2 \ln \left(
      \frac{1}{|\vec b_{1 \perp} - \vec b_{2 \perp}| \Lambda}\right)
  \right]
  \delta( x^+_1 - x^+_3) \delta( x^+_2 - x^+_3) \label{Sprimesc} \\
  S'_4 = & \left[ \frac{-i}{2 \pi} g^2 (t^b)_3 \ln \left(
      \frac{1}{|\vec b_{2 \perp} - \vec b_{3 \perp}| \Lambda} \right)
  \right] \left[ \frac{-i}{4 \pi} g^2 (t^a)_1 [t^a, t^b]_2 \ln \left(
      \frac{1}{|\vec b_{1 \perp} - \vec b_{2 \perp}| \Lambda}\right)
  \right] \delta( x^+_1 - x^+_3) \delta( x^+_2 -
  x^+_3). \label{Sprimesd}
\end{align}
\end{subequations}
The diagrams representing these four expressions are shown in
\fig{S'diag}. Notice that these satisfy the condition
\begin{align}
\notag
  S_1 + S_2 + {\tilde S}_3 = \sum\limits_{i=1}^4 S'_i.
\end{align}

Let us go over the physical significance behind the four contributions
in Eqs.~\eqref{Sprimes}. The $\delta( x^+_1 - x^+_3) \delta( x^+_2 -
x^+_3)$ term means all of these interactions are instantaneous, they
occur at a single $x^+$ position. (In the end of the calculation all
the $x^+$ coordinates are integrated out leading to the GM
exponentiation of the result.) All of the diagrams in \fig{S'diag} are
various corrections to the eikonal scattering seen in the MV model or
GM approximation. Diagrams $S'_1$ and $S'_2$ consist of the eikonal
scattering of the two quarks off the same target nucleon, each of them
scattering in a classical field of the target quark, since the quark
line is put on mass shell between the rescatterings (as indicated by a
``cut'' through the line). Indeed, each term in the square brackets of
Eqs.~\eqref{Sprimesa} and \eqref{Sprimesb} represents a single
$t$-channel gluon exchange between the projectile quarks at ${\vec
  b}_{1\perp}$ and ${\vec b}_{2\perp}$ and the target quark at ${\vec
  b}_{3\perp}$. The contributions in \eqref{Sprimesa} and
\eqref{Sprimesb} come in with a factor of $1/2$ shown explicitly in
the two left diagrams of \fig{S'diag}. The only difference between
$S'_1$ and $S'_2$ is the color factor associated with the target
quark.

Note that the contributions $S'_1$ and $S'_2$ also occur in the PV
sub-gauge of the light-cone gauge. One can see this by noticing that
they would remain if one drops the delta-function parts of the gluon
propagators in \eq{S11} (and in a similar calculation for
$S_2$). Therefore, these contributions are sub-gauge invariant. In
addition, these terms end up canceling out when one considers the
amplitude squared (since bringing either of the $t$-channel gluons in
these diagrams across the cut gives an overall minus sign
\cite{Kovchegov:1998bi}), meaning that these diagrams do not effect
the gluon production cross section. This is why they are not included
in the calculation in the main text leading to
Eqs.~\eqref{eq:ABCsum_coord} and \eqref{eq:Dall_coord}.

\begin{figure}[ht]
\begin{center}
\includegraphics[width=0.9 \textwidth]{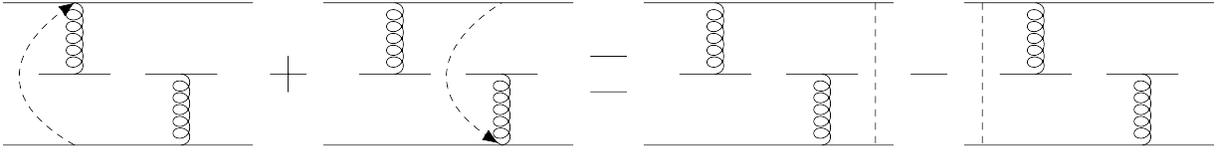} 
\caption{Summing up the gauge rotations for a given target nucleon
  distribution.}
\label{cellproof}
\end{center}
\end{figure}

The other two contributions, $S'_3$ and $S'_4$, which arise in the
${\vec A}_\perp (x^- \to + \infty) = 0$ (and, up to a sign, in ${\vec
  A}_\perp (x^- \to - \infty) = 0$) sub-gauge of the light-cone gauge
but not in the PV sub-gauge, correspond to one of the projectile quark
lines scattering in a classical field of the target while being
color-rotated by the field of another projectile quark (see
e.g. \cite{Kovchegov:1997pc} for the color-rotation terminology). The
first square brackets in each of Eqs.~\eqref{Sprimesc} and
\eqref{Sprimesd} contain a single $t$-channel gluon exchange
corresponding to the classical field of the target quark. The second
pair of square brackets in Eqs.~\eqref{Sprimesc} and \eqref{Sprimesd}
contain the ``gauge rotation'' illustrated by the dashed line in
\fig{S'diag} using the notation defined in \eq{eqrotation} and
\fig{rotation}. The arrow at the end of the dashed line indicates a
commutator of fundamental color matrices for the dashed line ``gluon''
and for the ``true'' gluon entering the quark-gluon vertex: the first
term in the commutator contains the color matrix for the dashed line
placed to the left of the color matrix of the ``true'' gluon.

\begin{figure}[ht]
\begin{center}
\includegraphics[width=0.7 \textwidth]{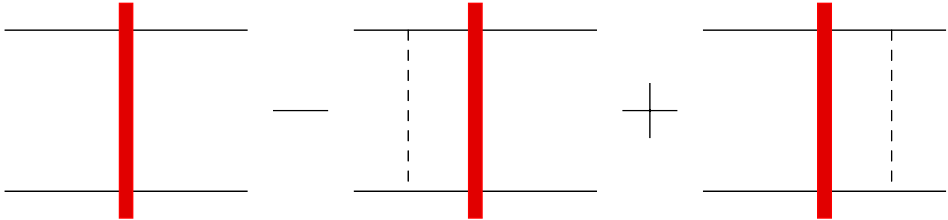} 
\caption{Shock wave with order-$g^2$ corrections.}
\label{shockwave}
\end{center}
\end{figure}

To figure out how corrections from the diagrams $S'_3$ and $S'_4$
contribute to the amplitude with multiple rescatterings we need to add
up all possible gauge rotations for multiple interactions with the
target nucleons. The case of the two projectile quarks scattering on
two target nucleons is shown in \fig{cellproof}. On the left of the
equality in \fig{cellproof} we have shown two contributions arising
due to the corrections like those shown in the right two diagrams of
\fig{S'diag}: one for the first nucleons and one for the second
nucleon. Analyzing the sum on the left we see that the gauge rotations
inside the shock wave cancel out and we only end up with rotations on
the outside of the charge distribution. The final result on the right
of \fig{shockwave} gives an order-$g^2$ correction to the shock waves
with the dashed line without an arrow defined as in \eq{eqrotation}
and \fig{rotation}. Repeating this argument for any number of target
nucleons, by including both corrections from the right two diagrams of
\fig{S'diag} for each nucleon, we arrive at the same conclusion: the
net result of all such corrections is equal to a diagram with a dashed
line to the right of the shock wave minus the diagram with the dashed
line to the left of the shock wave. We see that in the sub-gauge of
interest each shock wave interaction is accompanied by the order-$g^2$
corrections as shown in \fig{shockwave}. The dashed line contributions
are given by the expression in the parenthesis of \eq{eqrotation},
since all the $x^+$ coordinates are integrated out in going from
$S'_3$ and $S'_4$ in \eq{Sprimes} to the contribution to the
scattering amplitude.

Using this result it is straightforward to deduce the contributions of
gauge rotations to the gluon production amplitude. One simply has to
take an order-$g$ gluon production amplitude for the scattering of two
projectile quarks on the target and add all possible dashed lines to
it (connecting a pair of $s$-channel lines) immediately to the left
and to the right of the shock wave (with the appropriate signs). The
resulting corrections to the single gluon production amplitude are
depicted in \fig{all} below under the category (II): they are labeled
$\Delta S_i$.

\begin{figure}[ht]
\begin{center}
\includegraphics[width=0.9 \textwidth]{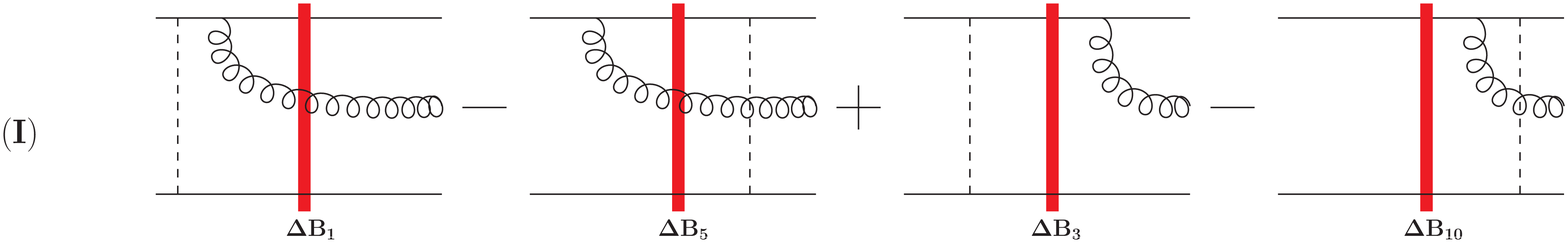} \\[5mm]
\includegraphics[width=0.9 \textwidth]{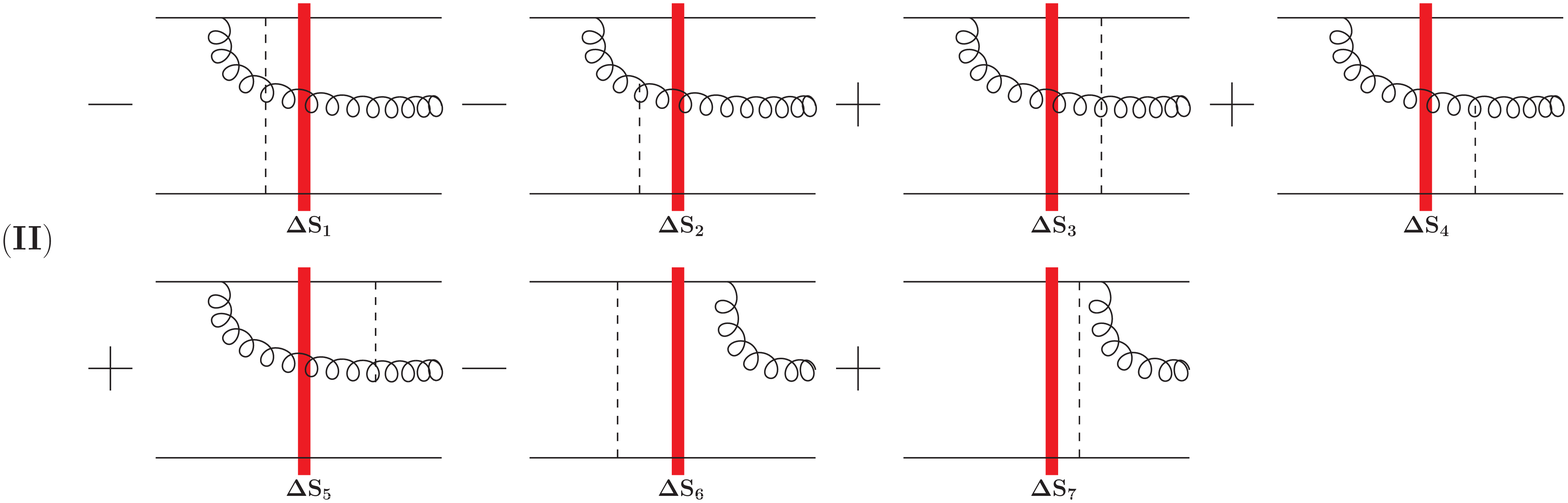} \\[5mm]
\includegraphics[width=0.9 \textwidth]{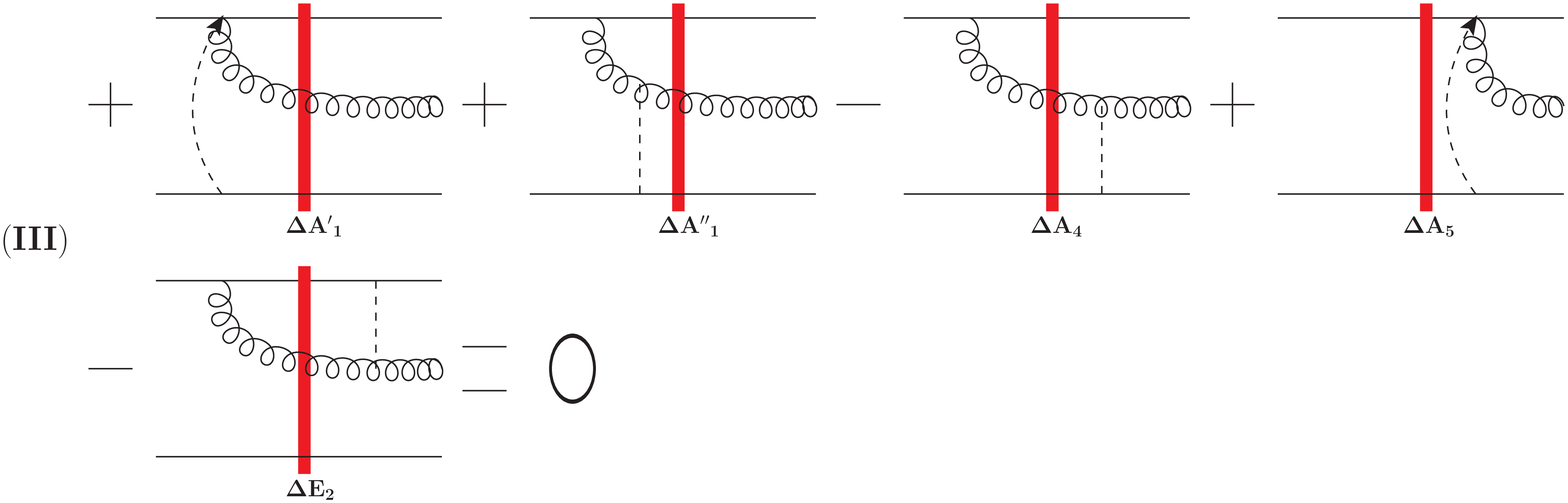}
\caption{The various gauge-dependent corrections (with the produced
  gluon emitted by quark $1$): (I) Pinched contributions; (II) shock
  wave corrections; (III) corrections to diagrams of type A and
  E. Each of the labels $\Delta A_i$, $\Delta B_i$, $\Delta S_i$, and
  $\Delta E_2$ denote both the diagram above it and the sign in front
  of the diagram. In the PV gauge all of these contributions are
  zero. Notice how the sum of all of these is zero, demonstrating
  sub-gauge invariance.}
\label{all}
\end{center}
\end{figure}

The last contributions we have to consider are the gauge dependent
parts of the $A$, $D$ and $E$ diagrams. Calculating these is
straightforward although time-consuming, just repeating the
calculation done in the main text of the paper but this time including
the part of the gluon propagators proportional to $\delta(l^+)$ (see
\eq{eq:prop}). It is interesting to note that for an ``instantaneous''
term such as this, it does not matter whether one uses Feynman
propagators or retarded gluon Green functions: hence our demonstration
of the sub-gauge invariance also applies if one uses retarded gluon
Green functions, as is done in the main text.

With this in mind all of the gauge-dependent terms of diagrams $A$ and
$E$ (corresponding to a gluon emission from quark $1$) are shown in
\fig{all} and are labeled $\Delta A_i$ and $\Delta E_2$. These terms
were obtained by an explicit calculation. Note that $\Delta A_i = A_i
({\vec A}_\perp (x^- \to + \infty) = 0 \ \mbox{sub-gauge}) - A_i
(\mbox{PV sub-gauge})$, with the same definition for $\Delta
E_2$. Also shown in \fig{all} are ``pinched'' contributions $\Delta
B_i$. Note that in \fig{all} the labels $\Delta A_i$, $\Delta B_i$,
$\Delta S_i$, and $\Delta E_2$ include both the diagram above each of
them and the sign in front of the diagram. All of the gauge-dependent
contributions to Feynman diagrams at the order $g^3$ shown in
\fig{all}, that is, the ``pinched'' contributions, shock wave
corrections, and gauge-dependent diagram corrections cancel out,
preserving gauge invariance. More precisely we have
\begin{align}
  \Delta A'_1+\Delta B_2+\Delta S_1=0;
  \notag \\ \Delta A''_1+\Delta S_2=0;
  \notag \\ \Delta A_4 + \Delta S_4=0;
  \notag \\ \Delta A_5 + \Delta B_{10} + \Delta S_7 = 0;
  \notag \\ \Delta E_2 + \Delta S_5=0;
  \notag \\ \Delta B_5 + \Delta S_3 = 0;
  \notag \\ \Delta B_3 + \Delta S_6 = 0.
\label{final}
\end{align}
Hence our result (the sum of all the $A$, $B$ and $C$ graphs and the
sum of all the $D$ and $E$ graphs) is independent of the choice of
sub-gauge in the light-cone gauge.



\begin{thebibliography}{10}

\bibitem{Gribov:1984tu}
L.~V. Gribov, E.~M. Levin, and M.~G. Ryskin, {\it {Semihard Processes in QCD}},
   {\em Phys. Rept.} {\bf 100} (1983) 1--150.

\bibitem{Iancu:2003xm}
E.~Iancu and R.~Venugopalan, {\it The color glass condensate and high energy
  scattering in {QCD}},  \href{http://xxx.lanl.gov/abs/hep-ph/0303204}{{\tt
  hep-ph/0303204}}.

\bibitem{Jalilian-Marian:2005jf}
J.~Jalilian-Marian and Y.~V. Kovchegov, {\it Saturation physics and deuteron
  gold collisions at {RHIC}},  {\em Prog. Part. Nucl. Phys.} {\bf 56} (2006)
  104--231, [\href{http://xxx.lanl.gov/abs/hep-ph/0505052}{{\tt
  hep-ph/0505052}}].

\bibitem{Weigert:2005us}
H.~Weigert, {\it Evolution at small {$x_{bj}$: The Color Glass Condensate}},
  {\em Prog. Part. Nucl. Phys.} {\bf 55} (2005) 461--565,
  [\href{http://xxx.lanl.gov/abs/hep-ph/0501087}{{\tt hep-ph/0501087}}].

\bibitem{Gelis:2010nm}
F.~Gelis, E.~Iancu, J.~Jalilian-Marian, and R.~Venugopalan, {\it {The Color
  Glass Condensate}},  {\em Ann.Rev.Nucl.Part.Sci.} {\bf 60} (2010) 463--489,
  [\href{http://xxx.lanl.gov/abs/1002.0333}{{\tt arXiv:1002.0333}}].

\bibitem{Albacete:2014fwa}
J.~L. Albacete and C.~Marquet, {\it {Gluon saturation and initial conditions
  for relativistic heavy ion collisions}},  {\em Prog.Part.Nucl.Phys.} {\bf 76}
  (2014) 1--42, [\href{http://xxx.lanl.gov/abs/1401.4866}{{\tt
  arXiv:1401.4866}}].

\bibitem{Balitsky:2001gj}
I.~Balitsky, {\it {High-energy QCD and Wilson lines}},
  \href{http://xxx.lanl.gov/abs/hep-ph/0101042}{{\tt hep-ph/0101042}}.

\bibitem{KovchegovLevin}
Y.~V. Kovchegov and E.~Levin, {\em Quantum Chromodynamics at High Energy}.
\newblock Cambridge University Press, 2012.

\bibitem{Kovner:1995ts}
A.~Kovner, L.~D. McLerran, and H.~Weigert, {\it Gluon production at high
  transverse momentum in the mclerran-venugopalan model of nuclear structure
  functions},  {\em Phys. Rev.} {\bf D52} (1995) 3809--3814,
  [\href{http://xxx.lanl.gov/abs/hep-ph/9505320}{{\tt hep-ph/9505320}}].

\bibitem{Kovner:1995ja}
A.~Kovner, L.~D. McLerran, and H.~Weigert, {\it Gluon production from
  non{A}belian {W}eizsacker-{W}illiams fields in nucleus-nucleus collisions},
  {\em Phys. Rev.} {\bf D52} (1995) 6231--6237,
  [\href{http://xxx.lanl.gov/abs/hep-ph/9502289}{{\tt hep-ph/9502289}}].

\bibitem{Kovchegov:1996ty}
Y.~V. Kovchegov, {\it Non-abelian {Weizs\"{a}cker-Williams} field and a two-
  dimensional effective color charge density for a very large nucleus},  {\em
  Phys. Rev.} {\bf D54} (1996) 5463--5469,
  [\href{http://xxx.lanl.gov/abs/hep-ph/9605446}{{\tt hep-ph/9605446}}].

\bibitem{Kovchegov:1997ke}
Y.~V. Kovchegov and D.~H. Rischke, {\it Classical gluon radiation in
  ultrarelativistic nucleus nucleus collisions},  {\em Phys. Rev.} {\bf C56}
  (1997) 1084--1094, [\href{http://xxx.lanl.gov/abs/hep-ph/9704201}{{\tt
  hep-ph/9704201}}].

\bibitem{Jalilian-Marian:1997xn}
J.~Jalilian-Marian, A.~Kovner, L.~D. McLerran, and H.~Weigert, {\it The
  intrinsic glue distribution at very small x},  {\em Phys. Rev.} {\bf D55}
  (1997) 5414--5428, [\href{http://xxx.lanl.gov/abs/hep-ph/9606337}{{\tt
  hep-ph/9606337}}].

\bibitem{McLerran:1994vd}
L.~D. McLerran and R.~Venugopalan, {\it Green's functions in the color field of
  a large nucleus},  {\em Phys. Rev.} {\bf D50} (1994) 2225--2233,
  [\href{http://xxx.lanl.gov/abs/hep-ph/9402335}{{\tt hep-ph/9402335}}].

\bibitem{McLerran:1993ka}
L.~D. McLerran and R.~Venugopalan, {\it Gluon distribution functions for very
  large nuclei at small transverse momentum},  {\em Phys. Rev.} {\bf D49}
  (1994) 3352--3355, [\href{http://xxx.lanl.gov/abs/hep-ph/9311205}{{\tt
  hep-ph/9311205}}].

\bibitem{McLerran:1993ni}
L.~D. McLerran and R.~Venugopalan, {\it Computing quark and gluon distribution
  functions for very large nuclei},  {\em Phys. Rev.} {\bf D49} (1994)
  2233--2241, [\href{http://xxx.lanl.gov/abs/hep-ph/9309289}{{\tt
  hep-ph/9309289}}].

\bibitem{Kovchegov:1997pc}
Y.~V. Kovchegov, {\it {Quantum structure of the non-Abelian
  Weizs\"{a}cker-Williams field for a very large nucleus}},  {\em Phys. Rev.}
  {\bf D55} (1997) 5445--5455,
  [\href{http://xxx.lanl.gov/abs/hep-ph/9701229}{{\tt hep-ph/9701229}}].

\bibitem{Krasnitz:1998ns}
A.~Krasnitz and R.~Venugopalan, {\it Non-perturbative computation of gluon
  mini-jet production in nuclear collisions at very high energies},  {\em Nucl.
  Phys.} {\bf B557} (1999) 237,
  [\href{http://xxx.lanl.gov/abs/hep-ph/9809433}{{\tt hep-ph/9809433}}].

\bibitem{Krasnitz:1999wc}
A.~Krasnitz and R.~Venugopalan, {\it The initial energy density of gluons
  produced in very high energy nuclear collisions},  {\em Phys. Rev. Lett.}
  {\bf 84} (2000) 4309--4312,
  [\href{http://xxx.lanl.gov/abs/hep-ph/9909203}{{\tt hep-ph/9909203}}].

\bibitem{Krasnitz:2002mn}
A.~Krasnitz, Y.~Nara, and R.~Venugopalan, {\it {Gluon production in the color
  glass condensate model of collisions of ultrarelativistic finite nuclei}},
  {\em Nucl. Phys.} {\bf A717} (2003) 268--290,
  [\href{http://xxx.lanl.gov/abs/hep-ph/0209269}{{\tt hep-ph/0209269}}].

\bibitem{Krasnitz:2003jw}
A.~Krasnitz, Y.~Nara, and R.~Venugopalan, {\it Classical gluodynamics of high
  energy nuclear collisions: An erratum and an update},  {\em Nucl. Phys.} {\bf
  A727} (2003) 427--436, [\href{http://xxx.lanl.gov/abs/hep-ph/0305112}{{\tt
  hep-ph/0305112}}].

\bibitem{Krasnitz:2003nv}
A.~Krasnitz, Y.~Nara, and R.~Venugopalan, {\it Probing a color glass condensate
  in high energy heavy ion collisions},  {\em Braz. J. Phys.} {\bf 33} (2003)
  223--230.

\bibitem{Lappi:2003bi}
T.~Lappi, {\it Production of gluons in the classical field model for heavy ion
  collisions},  {\em Phys. Rev.} {\bf C67} (2003) 054903,
  [\href{http://xxx.lanl.gov/abs/hep-ph/0303076}{{\tt hep-ph/0303076}}].

\bibitem{Blaizot:2010kh}
J.~P. Blaizot, T.~Lappi, and Y.~Mehtar-Tani, {\it {On the gluon spectrum in the
  glasma}},  {\em Nucl. Phys.} {\bf A846} (2010) 63--82,
  [\href{http://xxx.lanl.gov/abs/1005.0955}{{\tt arXiv:1005.0955}}].

\bibitem{Kuraev:1977fs}
E.~A. Kuraev, L.~N. Lipatov, and V.~S. Fadin, {\it {The Pomeranchuk
  singlularity in non-Abelian gauge theories}},  {\em Sov. Phys. JETP} {\bf 45}
  (1977) 199--204.

\bibitem{Balitsky:1978ic}
I.~Balitsky and L.~Lipatov, {\it {The Pomeranchuk Singularity in Quantum
  Chromodynamics}},  {\em Sov.J.Nucl.Phys.} {\bf 28} (1978) 822--829.

\bibitem{Balitsky:1996ub}
I.~Balitsky, {\it Operator expansion for high-energy scattering},  {\em Nucl.
  Phys.} {\bf B463} (1996) 99--160,
  [\href{http://xxx.lanl.gov/abs/hep-ph/9509348}{{\tt hep-ph/9509348}}].

\bibitem{Balitsky:1998ya}
I.~Balitsky, {\it Factorization and high-energy effective action},  {\em Phys.
  Rev.} {\bf D60} (1999) 014020,
  [\href{http://xxx.lanl.gov/abs/hep-ph/9812311}{{\tt hep-ph/9812311}}].

\bibitem{Kovchegov:1999yj}
Y.~V. Kovchegov, {\it Small-x {$F_2$} structure function of a nucleus including
  multiple pomeron exchanges},  {\em Phys. Rev.} {\bf D60} (1999) 034008,
  [\href{http://xxx.lanl.gov/abs/hep-ph/9901281}{{\tt hep-ph/9901281}}].

\bibitem{Kovchegov:1999ua}
Y.~V. Kovchegov, {\it Unitarization of the {BFKL} pomeron on a nucleus},  {\em
  Phys. Rev.} {\bf D61} (2000) 074018,
  [\href{http://xxx.lanl.gov/abs/hep-ph/9905214}{{\tt hep-ph/9905214}}].

\bibitem{Jalilian-Marian:1997dw}
J.~Jalilian-Marian, A.~Kovner, and H.~Weigert, {\it The {Wilson}
  renormalization group for low x physics: Gluon evolution at finite parton
  density},  {\em Phys. Rev.} {\bf D59} (1998) 014015,
  [\href{http://xxx.lanl.gov/abs/hep-ph/9709432}{{\tt hep-ph/9709432}}].

\bibitem{Jalilian-Marian:1997gr}
J.~Jalilian-Marian, A.~Kovner, A.~Leonidov, and H.~Weigert, {\it The {Wilson}
  renormalization group for low x physics: Towards the high density regime},
  {\em Phys. Rev.} {\bf D59} (1998) 014014,
  [\href{http://xxx.lanl.gov/abs/hep-ph/9706377}{{\tt hep-ph/9706377}}].

\bibitem{Iancu:2001ad}
E.~Iancu, A.~Leonidov, and L.~D. McLerran, {\it {The renormalization group
  equation for the color glass condensate}},  {\em Phys. Lett.} {\bf B510}
  (2001) 133--144.

\bibitem{Iancu:2000hn}
E.~Iancu, A.~Leonidov, and L.~D. McLerran, {\it Nonlinear gluon evolution in
  the color glass condensate. {I}},  {\em Nucl. Phys.} {\bf A692} (2001)
  583--645, [\href{http://xxx.lanl.gov/abs/hep-ph/0011241}{{\tt
  hep-ph/0011241}}].

\bibitem{Gardi:2006rp}
E.~Gardi, J.~Kuokkanen, K.~Rummukainen, and H.~Weigert, {\it Running coupling
  and power corrections in nonlinear evolution at the high-energy limit},  {\em
  Nucl. Phys.} {\bf A784} (2007) 282--340,
  [\href{http://xxx.lanl.gov/abs/hep-ph/0609087}{{\tt hep-ph/0609087}}].

\bibitem{Balitsky:2006wa}
I.~I. Balitsky, {\it {Quark Contribution to the Small-$x$ Evolution of Color
  Dipole}},  {\em Phys. Rev. D} {\bf 75} (2007) 014001,
  [\href{http://xxx.lanl.gov/abs/hep-ph/0609105}{{\tt hep-ph/0609105}}].

\bibitem{Kovchegov:2006vj}
Y.~Kovchegov and H.~Weigert, {\it {Triumvirate of Running Couplings in
  Small-$x$ Evolution}},  {\em Nucl. Phys. {\bf A}} {\bf 784} (2007) 188--226,
  [\href{http://xxx.lanl.gov/abs/hep-ph/0609090}{{\tt hep-ph/0609090}}].

\bibitem{Albacete:2007yr}
J.~L. Albacete and Y.~V. Kovchegov, {\it Solving high energy evolution equation
  including running coupling corrections},  {\em Phys. Rev.} {\bf D75} (2007)
  125021, [\href{http://xxx.lanl.gov/abs/0704.0612}{{\tt 0704.0612}}].

\bibitem{Horowitz:2010yg}
W.~A. Horowitz and Y.~V. Kovchegov, {\it {Running Coupling Corrections to High
  Energy Inclusive Gluon Production}},  {\em Nucl. Phys.} {\bf A849} (2011)
  72--97, [\href{http://xxx.lanl.gov/abs/1009.0545}{{\tt arXiv:1009.0545}}].

\bibitem{Fadin:1975cb}
V.~S. Fadin, E.~Kuraev, and L.~Lipatov, {\it {On the Pomeranchuk Singularity in
  Asymptotically Free Theories}},  {\em Phys.Lett.} {\bf B60} (1975) 50--52.

\bibitem{Kovchegov:1998bi}
Y.~V. Kovchegov and A.~H. Mueller, {\it Gluon production in current nucleus and
  nucleon nucleus collisions in a quasi-classical approximation},  {\em Nucl.
  Phys.} {\bf B529} (1998) 451--479,
  [\href{http://xxx.lanl.gov/abs/hep-ph/9802440}{{\tt hep-ph/9802440}}].

\bibitem{Kopeliovich:1998nw}
B.~Z. Kopeliovich, A.~V. Tarasov, and A.~Schafer, {\it Bremsstrahlung of a
  quark propagating through a nucleus},  {\em Phys. Rev.} {\bf C59} (1999)
  1609--1619, [\href{http://xxx.lanl.gov/abs/hep-ph/9808378}{{\tt
  hep-ph/9808378}}].

\bibitem{Dumitru:2001ux}
A.~Dumitru and L.~D. McLerran, {\it How protons shatter colored glass},  {\em
  Nucl. Phys.} {\bf A700} (2002) 492--508,
  [\href{http://xxx.lanl.gov/abs/hep-ph/0105268}{{\tt hep-ph/0105268}}].

\bibitem{Kovchegov:2000hz}
Y.~V. Kovchegov, {\it Classical initial conditions for ultrarelativistic heavy
  ion collisions},  {\em Nucl. Phys.} {\bf A692} (2001) 557--582,
  [\href{http://xxx.lanl.gov/abs/hep-ph/0011252}{{\tt hep-ph/0011252}}].

\bibitem{Blaizot:2008yb}
J.-P. Blaizot and Y.~Mehtar-Tani, {\it {The Classical field created in early
  stages of high energy nucleus-nucleus collisions}},  {\em Nucl.Phys.} {\bf
  A818} (2009) 97--119, [\href{http://xxx.lanl.gov/abs/0806.1422}{{\tt
  arXiv:0806.1422}}].

\bibitem{Kharzeev:2001gp}
D.~Kharzeev and E.~Levin, {\it {Manifestations of high density QCD in the first
  RHIC data}},  {\em Phys. Lett.} {\bf B523} (2001) 79--87,
  [\href{http://xxx.lanl.gov/abs/nucl-th/0108006}{{\tt nucl-th/0108006}}].

\bibitem{Kharzeev:2000ph}
D.~Kharzeev and M.~Nardi, {\it {Hadron production in nuclear collisions at RHIC
  and high density QCD}},  {\em Phys. Lett.} {\bf B507} (2001) 121--128,
  [\href{http://xxx.lanl.gov/abs/nucl-th/0012025}{{\tt nucl-th/0012025}}].

\bibitem{ALbacete:2010ad}
J.~L. Albacete and A.~Dumitru, {\it {A model for gluon production in heavy-ion
  collisions at the LHC with rcBK unintegrated gluon densities}},
  \href{http://xxx.lanl.gov/abs/1011.5161}{{\tt arXiv:1011.5161}}.

\bibitem{Kovchegov:2001sc}
Y.~V. Kovchegov and K.~Tuchin, {\it Inclusive gluon production in dis at high
  parton density},  {\em Phys. Rev.} {\bf D65} (2002) 074026,
  [\href{http://xxx.lanl.gov/abs/hep-ph/0111362}{{\tt hep-ph/0111362}}].

\bibitem{Kharzeev:2003wz}
D.~Kharzeev, Y.~V. Kovchegov, and K.~Tuchin, {\it Cronin effect and high-p(t)
  suppression in p a collisions},  {\em Phys. Rev.} {\bf D68} (2003) 094013,
  [\href{http://xxx.lanl.gov/abs/hep-ph/0307037}{{\tt hep-ph/0307037}}].

\bibitem{Mueller:1989st}
A.~H. Mueller, {\it {Small x Behavior and Parton Saturation: A QCD Model}},
  {\em Nucl. Phys.} {\bf B335} (1990) 115.

\bibitem{Balitsky:2004rr}
I.~Balitsky, {\it {Scattering of shock waves in QCD}},  {\em Phys. Rev.} {\bf
  D70} (2004) 114030, [\href{http://xxx.lanl.gov/abs/hep-ph/0409314}{{\tt
  hep-ph/0409314}}].

\bibitem{Belitsky:2002sm}
A.~V. Belitsky, X.~Ji, and F.~Yuan, {\it {Final state interactions and gauge
  invariant parton distributions}},  {\em Nucl.Phys.} {\bf B656} (2003)
  165--198, [\href{http://xxx.lanl.gov/abs/hep-ph/0208038}{{\tt
  hep-ph/0208038}}].

\bibitem{Liou:2012xy}
T.~Liou, {\it {Color-neutral heavy particle production in nucleus-nucleus
  collisions in the quasi-classical approximation}},  {\em Nucl.Phys.} {\bf
  A897} (2013) 122--140, [\href{http://xxx.lanl.gov/abs/1206.6123}{{\tt
  arXiv:1206.6123}}].

\bibitem{Liou:2013qya}
T.~Liou, A.~Mueller, and B.~Wu, {\it {Radiative $p_\bot$-broadening of
  high-energy quarks and gluons in QCD matter}},  {\em Nucl.Phys.} {\bf A916}
  (2013) 102--125, [\href{http://xxx.lanl.gov/abs/1304.7677}{{\tt
  arXiv:1304.7677}}].

\bibitem{Lepage:1980fj}
G.~P. Lepage and S.~J. Brodsky, {\it Exclusive processes in perturbative
  quantum chromodynamics},  {\em Phys. Rev.} {\bf D22} (1980) 2157.

\bibitem{Keldysh:1964ud}
L.~Keldysh, {\it {Diagram technique for nonequilibrium processes}},  {\em
  Zh.Eksp.Teor.Fiz.} {\bf 47} (1964) 1515--1527.

\bibitem{Schwinger:1960qe}
J.~S. Schwinger, {\it {Brownian motion of a quantum oscillator}},  {\em
  J.Math.Phys.} {\bf 2} (1961) 407--432.

\bibitem{Gelis:2008rw}
F.~Gelis, T.~Lappi, and R.~Venugopalan, {\it {High energy factorization in
  nucleus-nucleus collisions}},  {\em Phys.Rev.} {\bf D78} (2008) 054019,
  [\href{http://xxx.lanl.gov/abs/0804.2630}{{\tt arXiv:0804.2630}}].

\bibitem{Mueller:2012bn}
A.~Mueller and S.~Munier, {\it {$p_{\perp}$-broadening and production processes
  versus dipole/quadrupole amplitudes at next-to-leading order}},  {\em
  Nucl.Phys.} {\bf A893} (2012) 43--86,
  [\href{http://xxx.lanl.gov/abs/1206.1333}{{\tt arXiv:1206.1333}}].

\end{thebibliography}

\providecommand{\href}[2]{#2}\begingroup\raggedright\endgroup

\end{document}